\documentclass[iop]{emulateapj}
\usepackage[dvips]{color}
\usepackage{graphicx}
\usepackage{apjfonts}
\usepackage{rotating}
\usepackage[USenglish]{babel}
\newcommand{\kms}{${\rm km}\,{\rm s}^{-1}$}
\newcommand{\masyr}{${\rm mas}\,{\rm yr}^{-1}$}
\newcommand{\maspx}{${\rm mas}\,{\rm pixel}^{-1}$}
\newcommand{\pixyr}{${\rm pixel}\,{\rm yr}^{-1}$}

\newcommand{\etal}{et al.\ }
\newcommand{\x}{$\times$}

\usepackage{longtable}
\usepackage[pdfa]{hyperref}
\usepackage{ulem}

\begin{document}

\title{HUBBLE SPACE TELESCOPE PROPER MOTION (HSTPROMO) CATALOGS OF
  GALACTIC GLOBULAR CLUSTERS$^{\ast}$. I. SAMPLE SELECTION, DATA
  REDUCTION AND NGC~7078 RESULTS}

\author{A.\ Bellini\altaffilmark{1},
J.\ Anderson\altaffilmark{1},
R.\ P.\ van der Marel\altaffilmark{1}, 
L.\ L.\ Watkins\altaffilmark{1},
I.\ R.\ King\altaffilmark{2},
P.\ Bianchini\altaffilmark{3},
J.\ Chanam\'{e}\altaffilmark{4},
R.\ Chandar\altaffilmark{5},
A.\ M.\ Cool\altaffilmark{6},
F.\ R.\ Ferraro\altaffilmark{7},
H.\ Ford\altaffilmark{8},
D.\ Massari\altaffilmark{7}}
\altaffiltext{$\ast$}{Based on proprietary and
    archival observations with the NASA/ESA Hubble Space Telescope,
    obtained at the Space Telescope Science Institute, which is
    operated by AURA, Inc., under NASA contract NAS 5-26555.}
\altaffiltext{1}{Space Telescope Science Institute, 3700 San Martin
  Drive, Baltimore, 21218, MD, USA}
\altaffiltext{2}{Department of Astronomy, University of Washington,
  Box 351580, Seattle, WA 98195, USA}
\altaffiltext{3}{Max Planck Institute for Astronomy, K\"onigstuhl 17,
  69117 Heidelberg, Germany}
\altaffiltext{4}{Istituto de Astrof\'{i}sica,
  Pontificia Universidad Cat\'{o}lica de Chile, Av. Vicu\~{n}a
  Mackenna 4860, Macul 782-0436, Santiago, Chile}
\altaffiltext{5}{Department of Physics and Astronomy, The University
  of Toledo, 2801 West Bancroft Street, Toledo, 43606, OH, USA}
\altaffiltext{6}{Department of Physics and Astronomy, San Francisco
  State University, 1600 Holloway Avenue, San Francisco, CA 94132,
  USA}
\altaffiltext{7}{Dipartimento di Fisica e Astronomia, Universit\`{a} di
  Bologna, via Ranzani 1, 40127 Bologna, Italy}
\altaffiltext{8}{Department of Physics and Astronomy, The Johns
  Hopkins University, 3400 North Charles Street, Baltimore, 21218, MD,
  USA}

\email{bellini@stsci.edu}

\begin{abstract}
We present the first study of high-precision internal proper motions
(PMs) in a large sample of globular clusters, based on \textit{Hubble
  Space Telescope} (\textit{HST}) data obtained over the past decade
with the ACS/WFC, ACS/HRC, and WFC3/UVIS instruments. We determine PMs
for over 1.3 million stars in the central regions of 22 clusters, with
a median number of $\sim$60,000 stars per cluster. These PMs have the
potential to significantly advance our understanding of the internal
kinematics of globular clusters by extending past line-of-sight (LOS)
velocity measurements to two- or three-dimensional velocities, lower
stellar masses, and larger sample sizes. We describe the reduction
pipeline that we developed to derive homogeneous PMs from the very
heterogeneous archival data. We demonstrate the quality of the
measurements through extensive Monte-Carlo simulations. We also
discuss the PM errors introduced by various systematic effects, and
the techniques that we have developed to correct or remove them to the
extent possible. We provide in electronic form the catalog for
NGC~7078 (M~15), which consists of 77,837 stars in the central
$2\farcm4$. We validate the catalog by comparison with existing PM
measurements and LOS velocities, and use it to study the dependence of
the velocity dispersion on radius, stellar magnitude (or mass) along
the main sequence, and direction in the plane of the sky
(radial/tangential).  Subsequent papers in this series will explore a
range of applications in globular-cluster science, and will also
present the PM catalogs for the other sample clusters.
\end{abstract}


\maketitle

\section{Introduction}
\label{sec:1}

Globular clusters (GCs) are the oldest surviving stellar systems in
galaxies. As such, they provide valuable information on the earliest
phases of galactic evolution, and have been the target of numerous
studies during the past century.  Measures of the stellar motions in
GCs, for instance, allow us to constrain the structure, formation, and
dynamical evolution of these ancient stellar systems, and in turn,
that of the Milky Way itself.

Almost all of what is known about the internal motions within GCs is
based on spectroscopic line-of-sight (LOS) velocity measurements.
Observations of the kinematics of GCs have come a long way since,
e.g., Illingworth (1976) measured the velocity dispersions of 10
clusters using the broadening of absorption lines in integrated-light
spectra and Da Costa et al.\ (1977) measured the velocities for 11
stars in NGC~6397.  The largest published samples today have
velocities for a few thousand stars (e.g., Gebhardt et al.\ 2000;
Malavolta et al.\ 2014, Massari et al.\ 2014).

Despite the major improvements provided by LOS-based studies on our
understanding of the dynamics of GCs, there are some intrinsic
limitations. First of all, the need for spectroscopy implies that only
the brighter (more massive) stars in a GC can be observed. Moreover,
in the crowded central regions of the cluster core, spectroscopy is
limited by source confusion. Even integral-field spectroscopy is
affected by the shot noise from the brightest sources. Moreover, LOS
measurements are limited to measure only one component of the motion,
and therefore several model-dependent assumptions are required to
infer the three-dimensional structure of GCs.

A significant improvement in data quality is possible with
proper-motion (PM) measurements.  Indeed, PMs have the potential to
provide several advantages over LOS velocity studies:\ (1) No
spectroscopy is required, so the more plentiful fainter stars can be
studied, which yields better statistics on the kinematical quantities
of interest.  (2) Stars are measured individually, in contrast with
integrated-light measurements, which contain a disproportionate
contribution from bright giants.  (3) Two components of velocity are
measured instead of just one.  More importantly, it directly reveals
the velocity-dispersion anisotropy of the cluster, thus removing the
mass-anisotropy degeneracy (Binney \& Mamon 1982).

PMs are small, and difficult to measure with ground-based telescopes,
where they require an enormous effort to achieve only a modest
accuracy, particularly for faint stars in crowded fields (e.g., van
Leeuwen et al.\ 2000; Bellini et al.\ 2009).  On the other hand, the
stable environment of space makes the \textit{Hubble Space Telescope}
(\textit{HST}) an excellent astrometric tool.  Its diffraction-limited
resolution allows it to distinguish and measure positions and fluxes
for stars all the way to the center of most globular clusters.  Apart
from small changes due to breathing, its point spread function and
geometric distortion have been extremely stable over the two decades
since the repair mission.

\textit{HST} has the ability to measure PMs of unmatched quality
compared with any ground-based facility, and even in the most-crowded
central regions of GCs. Our team has developed methods to do this
accurately (e.g., Anderson \& King 2003a; Bellini, Anderson \& Bedin
2011). For instance, for a GC 5 kpc from the Sun, a dispersion of 10
\kms\ corresponds to $\sim 0.42$ \masyr; with a WFC3/UVIS scale of 40
mas$\,$pix$^{-1}$ this gives $\sim 0.1$ pixel over a 10-year time
baseline.  Since our measurement techniques reach a precision of
$\sim$0.01 pixel per single exposure for bright, unsaturated sources,
a tenth of a pixel is easy game, even for rather faint stars, so that
large numbers of proper motions depend only on the availability of
archival data.  To date, detailed \textit{HST} internal PM dynamics of
GCs have been studied for only a handful of clusters:\ NGC~104
(47~Tuc, McLaughlin et al.\ 2006), NGC~7078 (M~15, McNamara et al.\
2003), NGC~6266 (McNamara et al.\ 2011, 2012), and NGC~5139
($\omega$~Cen, Anderson \& van der Marel 2010) -- but a deluge is now
imminent; the project is described by Piotto et al.\ (2014), and the
first result paper has been submitted (Milone et al.\ 2014).

With high-quality PM catalogs it will be possible to address, for a
large number of GCs, many important topics: (1) \textit{Cluster-field
  separation}, for a better identification of bona-fide cluster
members for luminosity- and mass-function analyses and the study of
binaries and exotic stars, and to provide clean samples of targets for
spectroscopic follow-up.  (2) \textit{Internal motions}, to study in
detail the kinematics and the dynamics of GCs in general, and of each
population component in particular (with the aim of looking for fossil
signatures of distinct star-formation events).  (3) \textit{Absolute
  motions}, by estimating an absolute proper-motion zero point using
background galaxies as a reference frame (e.g., the series of papers
starting with Dinescu et al.\ 1997 and continuing as Casetti-Dinescu,
and Bellini et al. 2010 using ground-based observations, and Bedin et
al.\ 2003, Milone et al.\ 2006, and Massari et al.\ 2013 using
\textit{HST}).  Absolute PMs, in conjunction with radial velocities,
allow calculation of Galactic orbits of GCs; At the same time the
orbits that they exhibit are an indicator of the shape of the Galactic
potential.  (4) \textit{Geometric distance}, by comparing the LOS
velocity dispersion with that on the plane of the sky (Rees 1995,
1997). This will provide a scale of GC distances that is independent
of those based on stellar evolution or RR Lyrae stars.  (5)
\textit{Cluster rotation on the plane of the sky}, from the measure of
the stellar velocities as a function of the position angle at
different radial distances (e.g., Anderson \& King
2003b).\footnote{Cluster rotations can also be measured
    spectroscopically, see, e.g., Peterson \& Cudworth (1994);
    Bianchini et al.\ (2013).}  (6) \textit{Energy equipartition},
from the analysis of stellar velocity dispersion as a function of the
stellar mass (e.g., Trenti \& van der Marel 2013).  (7) \textit{Mass
  segregation}, by studying the stellar velocity dispersion as a
function of the distance from the cluster center for different stellar
masses.  (8) \textit{(An)isotropy}, by comparing tangential and radial
components of the stellar motion. (9) \textit{Full three-dimensional
  cluster dynamics}, when also LOS velocities are known.  The
availability of all the three components of the motion will directly
constrain the three-dimensional velocity and phase-space distribution
functions.  (10) \textit{Constraints on the presence of an
  intermediate-mass black hole}, by looking for both fast-moving
individual stars and for a sudden increase in the
velocity-dispersion-profile near the center (e.g., van der Marel \&
Anderson 2010).

\begin{table*}[th!]
\begin{center}
\label{tab:1}
\begin{tabular}{ccccccccc}
\multicolumn{9}{c}{\textsc{Table 1}}\\
\multicolumn{9}{c}{\textsc{Globular Clusters and their Parameters}}\\
\hline\hline
Cluster ID&R.A.$^\triangleright$&Dec.$^\triangleright$&$D_\odot$$^\ast$&[Fe/H]$^\ast$&$E(B-V)$$^\ast$&$\sigma_{V_{\rm LOS}}$$^\ast$&
$r_{\rm c}$$^\ast$&$r_{\rm h}$$^\ast$\\
   &$(^{\rm h}$:$^{\rm m}$:$^{\rm s})$&$(^{\circ}$:$^\prime$:$^{\prime\prime})$&kpc& & &
km$\,{\rm s}^{-1}$&$^\prime$ &$^\prime$ \\
\hline
NGC~104 (47~Tuc)       &00:24:05.71&$-$72:04:52.7& 4.5&$-$0.72&0.04&11.0$\pm$0.3&0.36&3.17\\
NGC~288                &00:52:45.24&$-$26:34:57.4& 8.9&$-$1.32&0.03& 2.9$\pm$0.3&1.35&2.23\\
NGC~362                &01:03:14.26&$-$70:50:55.6& 8.6&$-$1.26&0.05& 6.4$\pm$0.3&0.18&0.82\\
NGC~1851               &05:14:06.76&$-$40:02:47.6&12.1&$-$1.18&0.02&10.4$\pm$0.5&0.09&0.51\\
NGC~2808               &09:12:03.10&$-$64:51:48.6& 9.6&$-$1.14&0.22&13.4$\pm$1.2&0.25&0.80\\
NGC~5139 ($\omega$~Cen)&13:26:47.24$^\diamond$&$-$47:28:46.45$^\diamond$& 5.2&$-$1.53&0.12&16.8$\pm$0.3&2.37&5.00\\
NGC~5904 (M~5)         &15:18:33.22&$+$02:04:51.7& 7.5&$-$1.29&0.03& 5.5$\pm$0.4&0.44&1.77\\
NGC~5927               &15:28:00.69&$-$50:40:22.9& 7.7&$-$0.49&0.45& 8.8$^\dagger$&0.42&1.10\\
NGC~6266 (M~62)        &17:01:12.78$^\ddagger$&$-$30:06:46.0$^\ddagger$& 6.8&$-$1.18&0.47&14.3$\pm$0.4&0.22&0.92\\
NGC~6341 (M~92)        &17:17:07.39&$+$43:08:09.4& 8.3&$-$2.31&0.02& 6.0$\pm$0.4&0.26&1.02\\
NGC~6362               &17:31:54.99&$-$67:02:54.0& 7.6&$-$0.99&0.09& 2.8$\pm$0.4&1.13&2.05\\
NGC~6388               &17:36:17.23&$-$44:44:07.8& 9.9&$-$0.55&0.37&18.9$\pm$0.8&0.12&0.52\\
NGC~6397               &17:40:42.09&$-$53:40:27.6& 2.3&$-$2.02&0.18& 4.5$\pm$0.2&0.05&2.90\\
NGC~6441               &17:50:13.06&$-$37:03:05.2&11.6&$-$0.46&0.47&18.0$\pm$0.2&0.13&0.57\\
NGC~6535               &18:03:50.51&$-$00:17:51.5& 6.8&$-$1.79&0.34& 2.4$\pm$0.5&0.36&0.85\\
NGC~6624               &18:23:40.51&$-$30:21:39.7& 7.9&$-$0.44&0.28& 5.4$\pm$0.5&0.06&0.82\\
NGC~6656 (M~22)        &18:36:23.94&$-$23:54:17.1& 3.2&$-$1.70&0.34& 7.8$\pm$0.3&1.33&3.36\\
NGC~6681 (M~70)        &18:43:12.76&$-$32:17:31.6& 9.0&$-$1.62&0.07& 5.2$\pm$0.5&0.03&0.71\\
NGC~6715 (M~54)        &18:55:03.33&$-$30:28:47.5&26.5&$-$1.49&0.15&10.5$\pm$0.3&0.09&0.82\\
NGC~6752               &19:10:52.11&$-$59:59:04.4& 4.0&$-$1.54&0.04& 4.9$\pm$0.4&0.17&1.91\\
NGC~7078 (M~15)        &21:29:58.33&$+$12:10:01.2&10.4&$-$2.37&0.10&13.5$\pm$0.9&0.14&1.00\\
NGC~7099 (M~30)        &21:40:22.12&$-$23:10:47.5& 8.1&$-$2.27&0.03& 5.5$\pm$0.4&0.06&1.03\\
\hline\hline
\multicolumn{9}{l}{\small{$\triangleright$ From Goldsbury et al.\ (2010),  unless stated otherwise.}}\\
\multicolumn{9}{l}{\small{$\ast$ From Harris 1996 (2010 edition), unless stated 
otherwise. $D_\odot$ is the GC distance from the Sun.}}\\
\multicolumn{9}{l}{\small{$\diamond$ From Anderson \& van der Marel (2010).}}\\
\multicolumn{9}{l}{\small{$\dagger$ From Gnedin et al.\ (2002)}.}\\
\multicolumn{9}{l}{\small{$\ddagger$ From Beccari et al.\ (2006)}.}\\
\end{tabular}
\end{center}
\end{table*}

Unfortunately, \textit{HST} has executed only a very limited number of
programs specifically aimed at the study of internal PM dynamics of
GCs. Even so, many GCs have been observed with \textit{HST} for dozens
of different studies, and several of these clusters have been observed
on multiple occasions.  Motivated by the enormous scientific potential
offered by high-precision PM measurements of stars in GCs, we started
a project to derive high-precision PM catalogs for all GCs with
suitable multi-epoch image material in the \textit{HST} archive. This
project is part of -- and uses techniques developed in the context of
-- the \textit{HST} proper-motion (HSTPROMO)
collaboration\footnote{For details see the HSTPROMO home page at
  \url{http://www.stsci.edu/~marel/hstpromo.html}.}, a set of
\textit{HST} projects aimed at improving our dynamical understanding
of stars, clusters, and galaxies in the nearby Universe through the
measurement and interpretation of PMs (e.g., van der Marel et al.\
2013).

The paper is organized as follows. In Section~\ref{s:samplesel} we
present the sample of GCs and data sets used for our study. In
Sections~\ref{s:data red}, \ref{s:master frame} and \ref{s:pm} we
describe our detailed procedures for raw data reduction, astrometry,
and PM measurements, respectively.  In Section~\ref{s:sim} we test the
accuracy of our procedures on simulated data. Section~\ref{s:sys}
describes the effects of systematic errors and how we mitigate their
effects.  In Section~\ref{s:cat and res} we discuss some of the
kinematical quantities implied by the catalog of PMs for the GC
NGC~7078 (M~15). Conclusions are presented in
Section~\ref{s:conclusions}. Appendices present tables (available
electronically) with listings of the \textit{HST} data sets we used
for each cluster, and with the NGC~7078 PM catalog.

This is the first of a series of several papers.  Future papers in
this series will present the PM catalogs for the other GCs in our
sample, will discuss the kinematical quantities they imply for these
GCs, and will address many of the scientific topics listed above.

\section{Sample Selection}
\label{s:samplesel}

This work is based on archival \textit{HST} images taken with three
different cameras:\ (1) the Ultraviolet-Visible channel of the
Wide-Field Camera 3 (WFC3/UVIS); (2) the Wide-Field Channel of the
Advanced Camera for Surveys (ACS/WFC); and (3) the High-Resolution
Channel of ACS (ACS/HRC).

The physical characteristics of these cameras are as follows:\ the
WFC3/UVIS camera is made up of two $4096\times2048$-pixel chips, with
a pixel-scale of about 40 \maspx; ACS/WFC has the same number of
resolution elements as the WFC3/UVIS , but it has a larger sampling of
50 \maspx; ACS/HRC is the \textit{HST} instrument with the finest
resolution, being about 25 \maspx, and it is made up of a single chip
of 1024 pixels on each side.

Wide-Field Planetary Camera 2 (WFPC2) exposures were not taken into
account because, despite the larger time baseline they can generally
provide, there would be only a marginal increase in PM accuracy, due
primarily to the larger pixel size (larger position uncertainties) and
the smaller dynamical range of the WFPC2 chips (fewer well-measured
stars), particularly in the crowded cores, which is the focus of this
study.

Ten GCs were specifically observed with \textit{HST} by some of us to
study their internal motions, namely:
\begin{itemize}
\item{NGC~362, NGC~6624, NGC~6681, NGC~7078, NGC~7099 (GO-10401,
  PI:\ R.~Chandar);}
\item{NGC~2808, NGC~6341, NGC~6752 (GO-10335 and GO-11801, PI:\ H.~Ford);}
\item{NGC~6266, (GO-11609, PI:\ J.~Chanam\'{e});}
\item{NGC~6715 (GO-12274, PI:\ R.~P.~van der Marel).}
\end{itemize}
In January 2011 we searched through the \textit{HST} archive to look
for other suitable data and additional GCs, imaged with the three
aforementioned cameras and with a total time baseline of at least 2
years. Twelve GCs were found satisfying these two criteria, and we
successfully submitted an archival \textit{HST} proposal (AR-12845,
PI:\ A.~Bellini) to analyze them. The clusters are: NGC~104, NGC~288,
NGC~1851, NGC~5139, NGC~5904, NGC~5927, NGC~6362, NGC~6388, NGC~6397,
NGC~6441, NGC~6535 and NGC~6656.  A summary of the general properties
for all 22 GCs is given in Table~1.  A complete list of observations
used for our analysis of each cluster can be found in the appendix.

\section{Data Reduction}
\label{s:data red}

\subsection{Measuring Stellar Position and Fluxes in each Exposure}
\label{ss:mag and pos}

This work is based solely on \texttt{\_flt} or \texttt{\_flc} type
images.  These images are produced by the standard \textit{HST}
calibration pipeline \textit{CALWF3} (for WFC3) or \textit{CALACS}
(for ACS).  Images of type \texttt{\_flt} are dark- and
bias-subtracted and flat-fielded, but not resampled (like the
\texttt{\_drz} type images);\ \texttt{\_flc} images are \texttt{\_flt}
exposures that are also charge-transfer-efficiency (CTE) corrected
(see below).  The choice to use non-resampled images is motivated by
the fact that we need to retain information about where exactly a
photon hit the detector in order to minimize systematic errors in the
PMs.

\subsubsection{Charge-Transfer Efficiency Corrections}

Charge-transfer errors arise from the damaging effects of cosmic rays
on the detectors.  CTE losses affect both the shape (and therefore,
position) and the measured flux of stars, and these errors increase
over time (see, e.g., Anderson \& Bedin 2010).  CTE effects are more
severe when the image background is low, e.g. for short-exposures or
when bluer filters are used. It is a crucial step to properly model
and correct these CTE losses if we want to measure high-quality PMs.

The CTE correction for ACS is especially important on exposures taken
after the camera was repaired in 2009 (7 years after its
installation), while CTE damage is only mild or marginal on earlier
exposures. For the WFC of ACS, the CTE correction is already included
in the \textit{CALACS} pipeline (\texttt{\_flc} extension). The
correction is not available for the HRC of ACS, but this is only a
minor issue, as the HRC stopped operating in 2006 and it was not
repaired during the last \textit{HST} Service Mission 4 (SM4).
Moreover, the HRC read-out also has a maximum of 1024 transfers, so
that at its worst its CTE losses are only half as bad as the WFC.

An official CTE correction for WFC3/UVIS has been recently made
available, but it had not been implemented within the WFC3 calibration
pipeline at the time of our reductions.  So we manually corrected each
individual WFC3/UVIS \texttt{\_flt} exposure with the stand-alone CTE
correction routine available on the official UVIS
website\footnote{\url{http://www.stsci.edu/hst/wfc3/tools/cte\_tools}.}
to create \texttt{\_flc} images.

\subsubsection{ACS/WFC}

All ACS/WFC \texttt{\_flc} images were reduced using the
publicly-available \texttt{FORTRAN} program
\texttt{img2xym\_WFC.09x10}, which is described in detail in Anderson
\& King
(2006a).\footnote{\url{http://www.stsci.edu/~jayander/ACSWFC_PSFs/}.}
The program does a single pass of finding and measures each star in
each exposure by fitting a spatially-varying effective point spread
function (PSF), ignoring any contribution from neighbors.

Library PSFs for several filters are provided along with the reduction
software.  To take into account the variation of the PSF across the
Field-of-View (FoV), the library PSFs are made up of an array of $9
\times 10$ PSFs across the detector. At any given location on the
detector, the local PSF is then obtained through a bi-linear
interpolation of the four surrounding library PSFs.

During its $\sim 90$ min. orbital period around the Earth,
\textit{HST} is cyclically heated by the Earth and Sun. As a result,
the focal length changes slightly during each orbit.  This effect,
known as ``telescope breathing'', affects the shape of the PSF in a
non-constant way across the field of view (FoV). To take into account
the time-dependent variations of the PSFs, for each individual
exposure we derived an additional array of up to $5\times5$
perturbation PSFs by modeling the residuals of library-PSF-subtracted
stars across the detector. These perturbation PSFs were then
interpolated into the $9\times10$ array of the library PSFs and added
to them. The final set of PSFs (one set for each exposure) was then
used to fit stellar profiles.

\subsubsection{WFC3/UVIS}

Star positions and fluxes on WFC3/UVIS images were measured with the
software \texttt{img2xym\_wfc3uv}, adapted mostly from
\texttt{img2xym\_WFC.09x10}. Library, spatially-varying PSFs are
available also for this detector (in an array of $7\times 8$ PSFs). As
done for the ACS/WFC, we derived an additional array of perturbation
PSFs for each WFC3/UVIS exposure and combined it with the library PSFs
to fit stellar profiles.  (For a more comprehensive analysis of
spatial and time variations of UVIS PSFs see Sabbi \& Bellini 2013).

\subsubsection{ACS/HRC}

The measurement of stellar fluxes and positions in each ACS/HRC image
was performed by using the publicly available routine
\texttt{img2xym\_HRC} and library PSFs. Because of the small FoV of
HRC, there was no need to create spatially-varying PSFs, and a
constant PSF for each filter is adeguate to properly represent stellar
profiles all across the detector.  We investigated the possibility of
taking into account the time-dependent part of the PSFs but found that
perturbation PSFs were able to provide only a negligible improvement
in modeling stellar profiles.

\subsection{Single-Exposure Catalogs}
\label{ss:single cats}

The \texttt{img2xym}-routine family used here produces a catalog of
positions and fluxes of each measured star in each individual
exposure, together with some other additional quantities and
diagnostics, such as the quality-of-fit (\texttt{QFIT}) parameter,
which tells us how well a source has been fit with the PSF model
(Anderson et al.\ 2008).

Neighbor subtraction was not taken into account, so stars were
measured as they are on the exposures. Our aim is to measure PMs as
precisely as possible, so we decided to focus our attention on
relatively isolated stars, for which positions can be reliably
measured on individual exposures. The positions of blended stars, or
stars for which the profile is impaired by brighter neighbors, would
be affected by systematics in any case (see Section~\ref{ss:qfit+}).

The precision with which we are able to measure positions for
well-exposed stars on a single image is of the order of $\lesssim
0.01$ pixels (See Section~\ref{ss:experr}). This level of precision
can be achieved thanks to the high quality of the carefully-modeled,
fully-empirical PSFs at our disposal.
 
\subsection{Geometric-Distortion Corrections}
\label{ss:gdc}

Stellar positions in each individual exposure were corrected for
geometric distortion using the state-of-the-art solutions available
for ACS/WFC (Anderson \& King 2006a), ACS/HRC (Anderson \& King
2006b), and WFC3/UVIS (Bellini \& Bedin 2009; Bellini, Anderson \&
Bedin 2011). These corrections are able to provide distortion-free
stellar positions with residuals of the order of $\lesssim 0.01$ pixel
(about the same precision offered by the PSF-fitting).  This level of
precision in the distortion solution depends strongly on the adopted
PSFs, and cannot be achieved with simple centroid-type approaches,
with opticts-based PSFs, or even with empirical PSFs that do not
adeguately treat the PSF's spatial variations.

WFC3/UVIS is affected by a chromatic dependence of the geometric
distortion, and the effect is larger for the bluer filters (see, e.g.,
Fig.~6 of Bellini, Anderson \& Bedin 2011).  The problem likely
resides in the fused-silica CCD windows within the optical system,
which refract blue and red photons differently and exhibit a sharp
increase of the refractive index in the ultraviolet regime.

We showed in Bellini, Anderson \& Bedin( 2011) that there are
negligible color-dependent residuals in the UVIS distortion solutions
for filters redward of F275W. A similar chromatic dependence of the
distortion solution might also be present for the bluer filters of
ACS/HRC.  To minimize this subtle systematic effect, we decided to
exclude any exposure taken through filters bluer than F336W for UVIS,
and F330W for HRC.

The bluest filter available for ACS/WFC peaks at 435 nm (F435W), and
no chromatic dependence of the distortion solution has been reported
for this camera. The ACS/WFC, however, experienced a slight change in
the geometric-distortion solution after it was repaired during
SM4. Post-SM4 positional residuals obtained with pre-SM4
geometric-distortion solutions can be of the order of 0.05 pixels, and
therefore need to be corrected. We carefully modeled the post-SM4
deviation of the distortion solution with a look-up table of
residuals.\footnote{\url{http://www.stsci.edu/~jayander/ACSWFC_PSFs/POST-SM4/}.}
The accuracy of the post-SM4 geometric-distortion solutions for the
ACS/WFC are comparable with the pre-SM4 solution, and is of the order
of $\lesssim 0.01$ pixels.

\section{The Master Frame}
\label{s:master frame}

The 22 GCs for which we want to measure PMs all have different
apparent size and core density. Moreover, most of the archival data
come from projects with scientific goals other than high-precision
astrometry. As a result, the data sets at our disposal are extremely
heterogeneous in terms of used cameras, filters, chosen exposure time,
dither strategy, number of exposures and time baseline.

Despite the severe lack of similarity among the data sets, it is
important to be able to measure PMs for all 22 clusters in a
homogeneous and standardized fashion. This eases subsequent analyses
and comparisons of the dynamical properties of each cluster.  To
obtain a homogeneous set of PM catalogs we had to address several
issues.

The first issue concerns the definition of the reference system
(master frame) on which to register the stellar positions. The master
frame needs to be defined in a consistent way for each cluster, and to
have the same properties. Luckily, there is one data set in common
between all but one GC (NGC~6266): GO-10775, PI:\ A.\ Sarajedini. This
data set has been reduced with software tools similar to the ones we
employed here (for more details see Anderson et al.\ 2008). Its
astro-photometric catalogs are publicly
available\footnote{\url{http://www.astro.ufl.edu/~ata/public_hstgc/data-}
   \url{bases.html}.},
and their high quality and reliability are supported by several dozens
of papers.  Moreover, the GO-10775 data were taken in 2006, and
usually lie in between the time baseline of the data sets of each
cluster, thus limiting bias effects in computing PMs.

The GO-10775 catalogs have stellar positions in equatorial units and
in ACS/WFC pixels (rescaled to be exactly 50 \maspx). The pixel-based
reference frame has North up and East to the left, and places the
center of each GC (as defined in Harris 1996) at location (3000,
3000). To better exploit the GO-10775 catalogs as our reference
systems, we applied the following three changes:
\begin{enumerate}
\item{We modified the pixel scale from 50 to 40 \maspx, which is the
  WFC3/UVIS pixel scale, and represents a compromise between the
  ACS/HRC and the ACS/WFC pixel scales).}
\item{We shifted the cluster-center positions to location (5000,
  5000), in order to accommodate all overlapping data sets with
  GO-10775 (which have different pointings and orientations) without
  having to deal with negative coordinates.}
\item{We removed from the GO-10775 catalogs those stars for which the
  position was not well measured, following the prescriptions given in
  Anderson et al. (2008). In addition, we removed stars belonging to
  any of the following cases: (1) saturated stars; (2) stars fainter
  than instrumental magnitude\footnote{The instrumental magnitude is
    defined as $-2.5\times\log({\rm flux})$, where the flux in counts
    is the volume under the PSF that best fits a stellar profile.We
    will use instrumental magnitudes extensively throughout this
    paper, as they offer an immediate sense of the signal-to-noise
    ratio of measured sources. As a reference, a typical \textit{HST}
    central PSF value is $\sim 0.2$ (i.e., 20\% of the source flux is
    in its central pixel):\ this means that saturated stars (central
    pixel $\geq 55\,000$ counts) will have magnitudes brighter than
    instrumental magnitude
    $-2.5\times\log(55000/0.2)=-13.6$. Moreover, stars with
    instrumental magnitude $-$10 will have a signal-to-noise ratio of
    100.} $-5.7$ in either F606W or F814W; (3) stars with positional
  error larger than 5 mas in either coordinate; (4) stars with
  photometric error larger than 0.2 mag in either filter; and (5)
  stars with $o_V$ or $o_I$, i.e. the ratio of neighbor vs.\ star
  light in the aperture greater than 1.}
\end{enumerate}

Although a GO-10775 catalog is available for $\omega$~Cen, we decided
instead to base its reference system on the GO-9442 data set (PI:\ A.\
Cool).  The reason for this is twofold: (1) the GO-9442 field of view
is nine times larger than that of GO-10775, and there are other
projects (such as GO-10252) that overlap with GO-9442 but not with
GO-10775, thus allowing PM measurements at larger radial distances;
and (2) the GO-9442 observation strategy was very similar to that of
GO-10775 in terms of dithering scheme, number of exposures and
exposure time. Only the chosen filters are different, on account of
the different scientific goals. Moreover, data of GO-9442 were reduced
by one of us (J.\ Anderson) with a preliminary version of the same
software used to create the GO-10775 database.  To transform the
GO-9442 catalog into our reference system, we applied the same
aforementioned changes applied to GO-10775 catalogs.

In order to obtain a reference system for NGC~6266, we noted that the
data of GO-10210 were taken following a very similar observing
strategy to that of GO-9442 for $\omega$~Cen. Therefore we reduced
GO-10120 following the prescriptions given in Anderson et al.\ (2008)
to produce a star catalog analogous to those of GO-10775, and we
applied the same three changes as for the GO-10775 data sets.

\begin{figure*}[ht!]
\centering
\includegraphics[width=15cm]{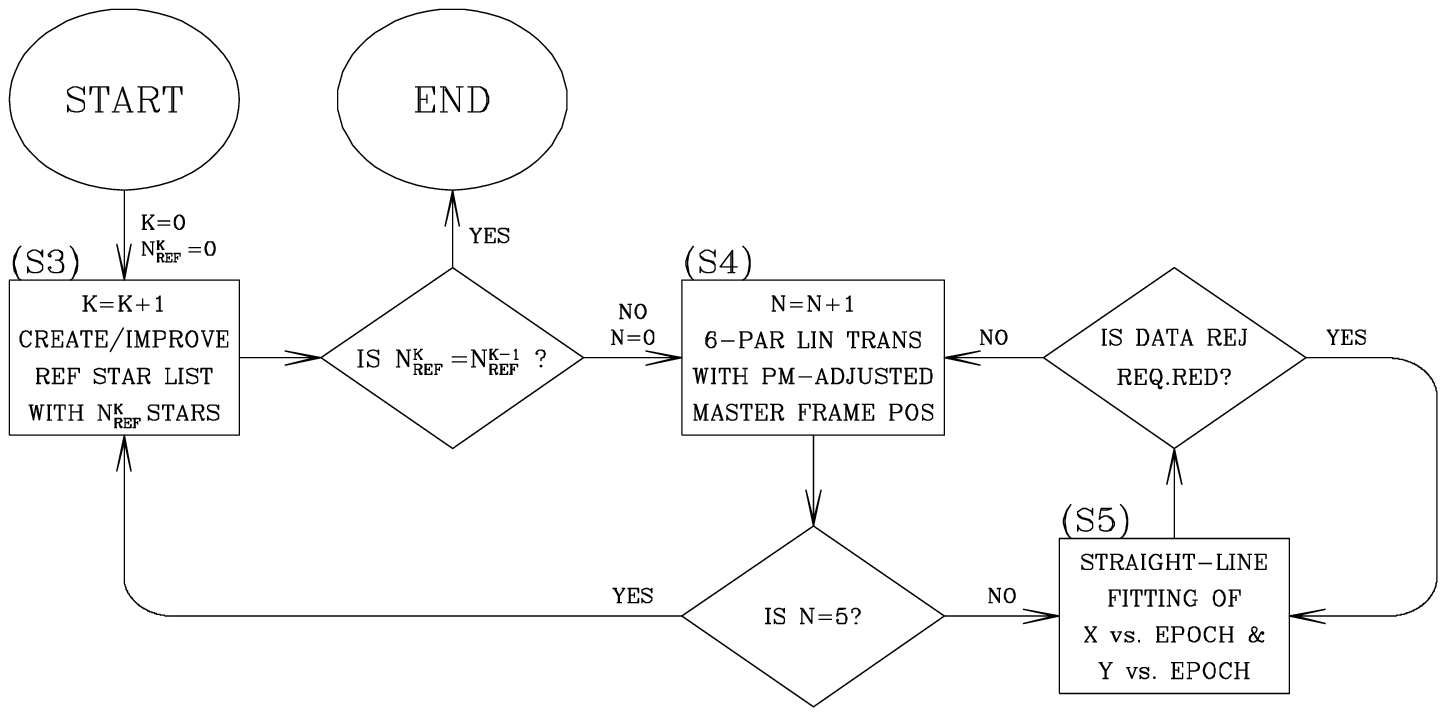}
\caption{Flow chart illustrating the adopted scheme to compute
  PMs. The three main steps discussed in the text are marked as (S3),
  (S4) and (S5). See Sections \ref{ss:ref list}, \ref{ss:master pos}
  and \ref{ss:linfit}, respectively, for details.}
\label{f:flowc}
\end{figure*}

\section{Proper motions}
\label{s:pm}

In the simple situation of repeated observations taken in only two
epochs, one can simply measure the average position of stars within
each epoch, and then obtain PMs as the difference in position between
the second and the first epoch, divided by the time baseline. In
reality, our data sets generally contain a varying number of epochs,
sometimes with one exposure only. Even when there are multiple
exposures within a given epoch (which may span several weeks), stars
are usually measured through different filters and with different
exposure times --and hence different signal-to-noise--, and it is not
trivial to properly determine an average position for them within each
epoch.  Therefore, we decided to treat each individual exposure as a
stand-alone epoch, and to measure PMs by fitting a straight line to
the data in the position versus epoch space (essentially the so-called
\textit{central overlap} method, first proposed by Eichhorn \&
Jefferys 1971).

Our general strategy for measuring PMs can be summarized into five
main steps: (1) measure stellar positions in each individual exposure;
(2) cross-identify the same stars in all the exposures where they can
be found; (3) define a reference network of stars with respect to
which we can compute PMs; (4) transform stellar positions onto a
common reference frame; (5) fit straight lines to the
reference-frame-position-versus-epoch data to obtain PMs.

Steps (3), (4) and (5) are nested into each other, and each of them
requires some iteration in order to reject discrepant observations and
improve the PM measurements. The basic scheme of the iterative process
is summarized in the flow-chart of Fig.~\ref{f:flowc}.  We have
already discussed step (1) in Section \ref{s:data red}; the following
subsections will provide a comprehensive explanation of the subsequent
steps.

\subsection{Linking master-frame to single-catalog stellar positions}
\label{ss:linking}

First of all, each star in the master-frame list needs to be
identified in each individual exposure where it can be found.  The
cross-identification is performed by means of general six-parameter
linear transformations. These allow us to transform stellar positions
as measured in the individual exposures onto the reference system, and
associate them with the closest star of the master frame list.

We are matching up stars that have moved in random directions as time
has passed. To limit the number of mismatches, we considered only
stars from which master-frame matches are within 2.5 pixels
($0\farcs1$). This criterion necessarily limits our ability to measure
the motion of very-fast-moving stars. As an example, let us take the
NGC~5927 data set. The time baselines to the reference dataset
(GO-10775) is 3.87 years for GO-9453 and about 4.38 years for GO-11664
and GO-11729. The fastest motion we can measure for stars present only
in the GO-10775 and GO-9453 data is $\mu=2.5\times40/3.87
{\rm\ mas\ yr}^{-1}=25.84 {\rm\ mas\ yr}^{-1}$. This limit is further
reduced to 22.83 \masyr\ if stars are measured in the GO-11664 and/or
GO-11729 data sets but not in the GO-9453 one (see also Table~A8).
These PMs correspond to $\sim 940$ \kms\ and $830$ \kms\ at the
distance of NGC~5927, but would correspond to smaller velocities for
foreground stars.

At the initial stage, there is no need to fine-tune the linear
transformations, so long as we are able to identify master-frame stars
in each exposure. We will later compute improved transformations to
precisely place single-exposure stellar positions onto the master
frame.

\begin{figure}[ht!]
\centering
\includegraphics[width=\columnwidth]{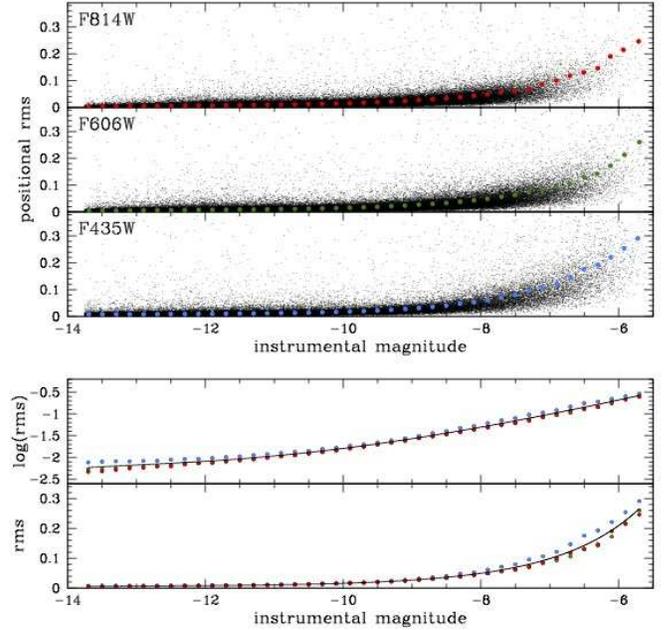}
\caption{Modeling of the expected errors for the ACS/WFC camera. The
  top three panels show the 1D positional RMS as a function of the
  instrumental magnitude for three filters of the central field of
  $\omega$~Cen. We computed the 68.27 percentile of these RMS in bins
  of 0.2 mag, and fitted a 5th-order polynomial to them. The bottom
  two panels show the binned RMS in linear and logarithmic units,
  together with the fitted function.}
\label{f:errormod}
\end{figure}

\subsection{Expected errors}
\label{ss:experr}

Since each exposure corresponds to a stand-alone epoch, we cannot
directly measure stellar positional errors from the RMS of the
residuals around an epoch-averaged position, as in the case of
multiple exposures per epoch. Instead, we need to assign an a-priori
expected error based on some assumptions.

We reduced thousands of \textit{HST} images and found as expected that
there is a general trend of increasing positional RMS as a function of
the instrumental magnitude. This trend is stable over time and has
little dependence on the filter used.  For this reason, we decided to
model this trend for the three \textit{HST} detectors employed here
and assign an expected positional error to each star of each
individual catalog according to its instrumental magnitude.

To model the ACS/WFC expected-error trend, we chose the exposures of
the core of $\omega$~Cen, a moderately-crowded field containing
several thousand stars, and imaged through several dithered exposures
in the F435W, F606W and F814W filters (to sample the available
wavelength coverage).  For each filter, we computed average star
magnitudes and positions, and measured the positional RMS of the
residuals about the mean.  Stars brighter than instrumental magnitude
$\sim -13.7$ are saturated and were not taken into account. Stars
fainter than $\sim-5.7$ are generally close to the shot-noise level
for single-exposure measurements, and define the faint limit of the
model.

The top three panels of Fig.~\ref{f:errormod} show one-dimensional
positional RMS as a function of the instrumental magnitude for F814W,
F606W and F435W from top to bottom.  We divided each sample of points
in bins of 0.2 mag, and computed a 3$\sigma$-clipped 68.27 percentile
of the positional RMS within each bin (full colored circles).  The
bottom two panels of the figure show sampled values of the three
filters, in linear and logarithmic units, as a function of the
instrumental magnitude. The logarithmic units allow one to better
distinguish the sampled values in the bright regime, while the linear
units work better for the faint regime. A least-squares 5th-order
polynomial is fit to the points in the log plane to model the
positional RMS trend. This model provides our expected errors for the
ACS/ WFC camera.

For the ACS/HRC and the WFC3/UVIS cameras we used the central fields
of 47~Tuc and $\omega$~Cen, respectively\footnote{No suitable ACS/HRC
  exposures of the core of $\omega$~Cen have been taken, while the
  core of 47~Tuc was used as ACS/HRC calibration field.}, and followed
the same procedures used for the ACS/WFC camera to model the
positional RMS, and thus the expected errors, as a function of the
instrumental magnitude.  For these two detectors we again modeled the
expected errors using three filters: a blue, an intermediate and a red
filter. As for the ACS/WFC, the intermediate and red filters are the
F606W and the F814W.  As the blue filter for ACS/HRC we chose F475W
instead of the ACS/WFC F435W, because F475W exposures are more
numerous and have longer exposure times.  Because the WFC3/UVIS
detector covers bluer wavelengths than the ACS/WFC, the adopted blue
filter was the F336W (which is also the bluest filter used to compute
PMs). The average modeled curves of the expected errors for the
ACS/HRC and the WFC3/UVIS cameras are very similar to those for the
ACS/WFC shown in Fig.~\ref{f:errormod}.

\subsection{The Reference-Star List}
\label{ss:ref list}

At this stage in the reduction process, we are ready to start
measuring PMs. We want to stress here that we will compute
\textit{relative} and not \textit{absolute} PMs. The main reason is
that the cores of GCs are so dense that the light of a background
galaxy can hardly push itself above the scattered light of the
cluster. (One of the few clusters in which there are enough galaxies
to actually measure absolute PMs is NGC~6681, see Massari et
al.\ 2013.)  Therefore, in general we need to choose a reference set
of objects other than background galaxies against which to measure
motions.  This leaves the cluster stars and the field stars. The
cluster stars have a much tighter PM distribution, so they are the
obvious choice. Our motions will thus be in a frame that moves and
rotates with the cluster.

We want to use only the best-measured, unsaturated master-frame stars
in order to minimize transformation residuals.  Master-frame
magnitudes are rezo-pointed with respect to the deep exposures of
GO-10775, therefore the short-exposure saturation limit in
instrumental magnitudes is about $-16.5$, and the long-exposure limit
is about $-13.5$.  Stars between $-16.5$ and $-13.5$ mag are measured
only in the short exposures. Generally, the best-measured stars lie
within $\sim 3$ mag of the saturation limit. Therefore, in principle
we could consider all stars between instrumental magnitude $-16.5$ and
$-10$ in our reference list.  However, because of the large variety of
exposure times in our data sets, it could be that these bright stars
are too bright (i.e., saturated) in some exposures. We therefore
adopted a compromise by including fainter, less-constrained stars in
the reference list to obtain an adequate number of reference stars for
the transformations by extending the magnitude range of the
reference-list stars to instrumental magnitude $-8$.

The process of creating the reference star list is labeled as (S3) on
the flow chart of Fig.~\ref{f:flowc}.  We start by selecting cluster
members on the basis of their positions on the color-magnitude diagram
(CMD).  To make the selection easier, especially for those clusters
with high reddening foreground values, we corrected the master-frame
photometry for differential reddening as done in Bellini et
al. (2013), following prescriptions given in Milone et al. (2012). A
few field stars will still be included, but once PMs are computed, we
refined our reference-star list by removing from it those stars
with PMs that are inconsistent with the cluster's bulk motion.  This
is an iterative process that ends when, from one iteration to the
next, the number of stars in the reference list stops decreasing,
meaning that we have computed PMs with respect to a list of bona-fide
cluster members that is as genuine as we can hope to obtain.

\begin{figure}[t!]
\centering
\includegraphics[width=\columnwidth]{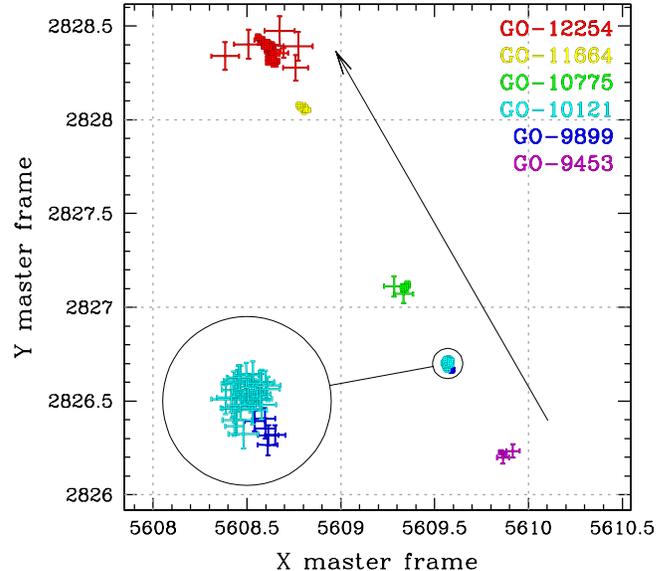}
\caption{Transformed positions of a single star of the NGC~6752 data
  set, taken at six different epochs, as they appear on the reference
  system.  Master-frame pixels are highlighted with dashed lines. Star
  positions and errorbars are color-coded according to their program
  ID. Colors go from violet to green to red, moving from the 2002 to
  2006 to 2011 epochs. A zoomed-in region of GO-9899 and GO-10121
  positions is enclosed for clarity. An arrow shows the motion of the
  star during $\sim 9$ years.}
\label{f:masterpos}
\end{figure}

\subsection{Positions on the Master Frame}
\label{ss:master pos}

For each exposure, we transformed the distortion-corrected positions
of its stars into the master frame using general six-parameter linear
transformations. Only bright, unsaturated reference stars in common
between the single-exposure catalog and the master-frame catalog were
used to compute the transformation parameters (i.e., reference stars
that in the single-exposure catalogs are brighter than instrumental
magnitude $-9.5$).

We chose to restrict the use of common reference stars to the same
amplifier, to limit the impact of uncorrected geometric-distortion and
CTE-mitigation residuals.  The ACS/WFC and WFC3/UVIS cameras have 4
amplifiers each, corresponding to an area of $2048\times2048$
pixels. On the other hand, the ACS/HRC camera has only one amplifier,
therefore this restriction does not apply.

The geometric distortion has a smooth variation across the detectors,
and therefore it can be considered locally flat.  If we were to use
the local-transformation approach (see, e.g., Anderson et al.\ 2009;
Bellini et al.\ 2009), we would have minimized the impact of
uncorrected geometric-distortion residuals. However, the adopted
amplifier-type restriction (a sort of semi-local approach) allows us
to limit these effects.  We will henceforth refer to the PMs thus
obtained as "amplifier-based". This in contrast to "locally-corrected"
PMs, which are discussed in Section 7.3. Both types of PMs are listed
in our catalogs. Which PMs are best depends on the specific scientific
application.

Concerning CTE-correction residuals, y-CTE effects (i.e., trails along
the Y axis of the detector), vary as a function of their distance from
the register. Each amplifier has its own register. To date, there is
no pixel-based x-CTE correction (i.e.,trails along the X axis)
available for \textit{HST}. However, the impact of x-CTE effects is
order of magnitudes smaller than that of y-CTE, and to the first
order, it should be compensated for by our amplifier-based approach.

Since all the stars in our reference list are moving in random
directions with respect to each other with some dispersion, each and
every transformed star position is affected by a systematic error of
${\rm err} \propto \sqrt{\sigma_{\rm ref}/N_{\rm ref}}$, where $N_{\rm
  ref}$ is the total number of reference stars used for the
transformation and $\sigma_{\rm ref}$ their PM dispersion. This
implies that a large number of reference stars is best to minimize
this source of error.  On the other hand, it is not uncommon to have
only a handful of reference stars to use for the transformations,
especially in partially-overlapping data sets, or when the image depth
is very different. A good compromise for the used data sets was found
by rejecting all transformed stars that had less than 75 reference
stars within their amplifier for ACS/WFC and WFC3/UVIS exposures, and
less than 50 for ACS/HRC exposures. In the vast majority of cases, the
typical number of reference stars used for the transformations is
larger than 300.

As mentioned, the reference stars do also move themselves. As a
result, when we transform stellar positions of exposures taken years
apart from the master-frame epoch, we will necessarily have to deal
with larger transformation residuals. These residuals will in turn
translate into larger uncertainties in the transformed positions of
stars. We can bypass this problem by correcting the positions of the
reference stars to correspond to the epoch of the single-exposure
catalog that we want to transform.

Obviously, we need to know the PM of the reference stars to compute
their position adjustments. As a consequence, computing positions on
the master frame is an iterative process.  With improved
transformations we will be able to measure more precise PMs, and with
them obtain even better transformations. We found that 5 iterations
were enough to minimize the transformation residuals.

Once all the stars of all the exposures are transformed into the
master frame, each master-frame star will be characterized by several
slightly different positions, each of them referring to a different
exposure (i.e., a different epoch). In Fig.~\ref{f:masterpos} we
illustrate this concept for a rapidly-moving star in the field of
NGC~6752. On the master frame (the pixels of which are highlighted by
dashed lines), each point represents a transformed single-exposure
position. Errorbars are obtained using expected errors (from
Sectio.~\ref{ss:experr}), sp that larger errorbars refer to shorter
exposure times. For clarity, we color-coded star positions according
to their program number. The epochs of the observations go from 2002
(GO-9453, purple data) to 2011 (GO-12254, red data). We recall that
the master-frame epoch is defined by the GO-10775 observations (in
green). The actual master-frame position of this star lies underneath
the green points (not shown).  Data of GO-9899 and GO-10121 are
separated by less than 3 months, and their position is magnified in
the enclosed circle. An arrow indicates the motion of the star over
$\sim 9$ years.

\begin{figure}[t!]
\centering
\includegraphics[width=\columnwidth]{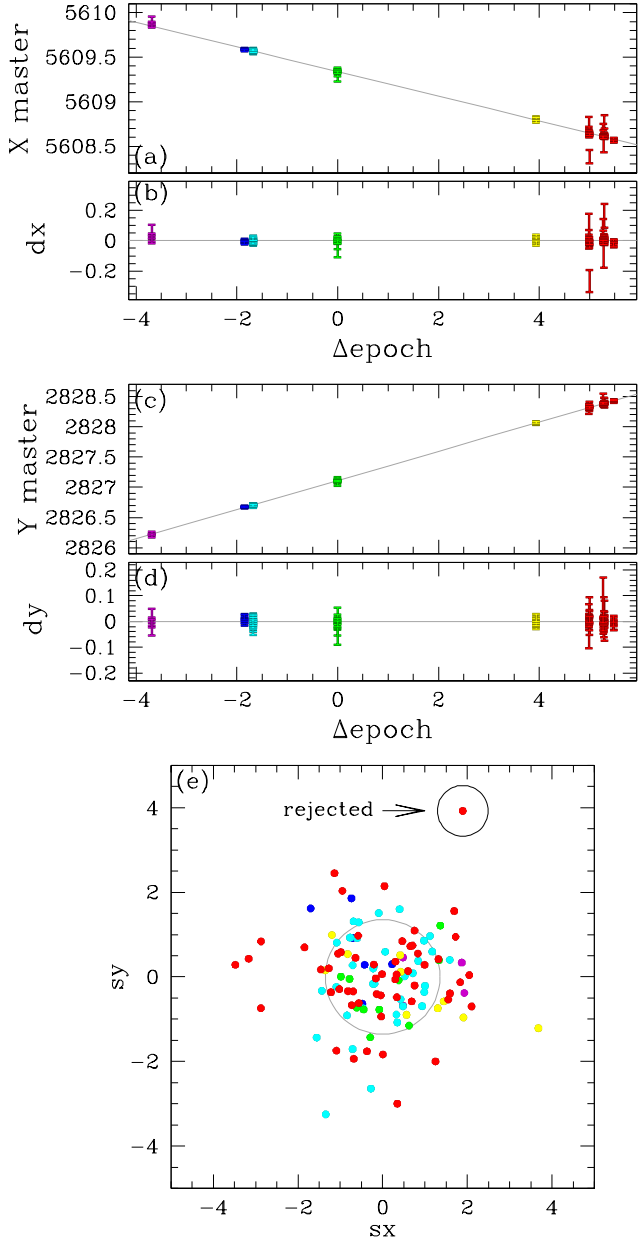}
\caption{Illustrative example of the least-squares straight-line
  fitting procedure. The chosen star is the same shown in
  Fig.~\ref{f:masterpos} (and points are color-coded accordingly).
  Panel (a) shows the X positions versus the epoch of the observations
  with respect to the master-frame epoch, in Julian years. The fitted
  line is marked in grey. The residuals of the fit are in Panel
  (b). Panels (c) and (d) show the same for the Y positions. Panel (e)
  illustrates the adopted rejection criterion. In the normalized and
  rescaled residual plane (sx, sy) (where points resemble a
  two-dimensional Gaussian), we identify the outermost point, and
  check whether its probability of being that far out is inconsistent
  with that of a two-dimensional Gaussian distribution at a confidence
  level of 97.5\%. If not, the data point is rejected (as in the
  example), and the straight-line fitting process is repeated without
  it.}
\label{f:linfit}
\end{figure}

\subsection{Proper-Motion Fitting and Data Rejection}
\label{ss:linfit}

Let us suppose that for a given star we have $N$ total positions in
the master frame. Each position has an associated expected
one-dimensional error and an epoch of observation, and is therefore
characterized by the quadruplet $(x_N,y_N,e_N,t_N)$. To measure the
motion of this star along the X and Y axes, we used a weighted
least-squares to fit a straight line to the data points $(x_N,t_N)$
and $(y_N,t_N)$. We progressively improve the fit by rejecting
outliers or badly-measured observations. This iterative
straight-line-fitting process is marked as (S5) in the flow chart of
Fig.~\ref{f:flowc}.

We require that a star have at least 4 data points, with at least 6
months of time baseline between the second and the second-from-last
point, in order for its PM to be measured. These conditions must be
satisfied at every stage of the fitting/rejection process.

Before starting with the iterative process, we identify and reject
obvious outliers. This task is done by removing one point at a time,
then fitting the straight lines to the remaining $N-1$ points. If the
distance of a removed point from its associated fitted line is larger
than 10 times its expected error, the point is rejected immediately.
Such data points generally come from objects with a cosmic-ray event
within their fitting radius. As a result, the centroid is shifted
toward the cosmic ray, and their measured luminosity is enhanced by
the cosmic-ray counts.

Let us suppose that a star still has $N$ data points after these preliminary
selections.  We fit two weighted straight lines to the points
$(x_N,t_N)$ and $(y_N,t_N)$.  An example of these fits for the same
star used in Fig.~\ref{f:masterpos} is illustrated in
Fig.~\ref{f:linfit}. Data points are color-coded as in
Fig.~\ref{f:masterpos}.  Panel (a) of Fig.~\ref{f:linfit} shows the
fitted line in the X-position versus epoch plane, where the epoch of
each point is expressed relative to the master-frame epoch (T=0, in
years). Panel (c) shows the fit for the Y-position versus epoch.
Panels (b) and (d) show the residuals $(dx_N,dy_N)$ of the points
around the straight-line fits.

To identify and reject the marginal outliers we adopted the
one-point-at-a-time approach as follows:\ We define error-normalized
quantities $dx_N^{\prime}=dx_N/e_N$, $dy_N^{\prime}=dy_N/e_N$, and
their sum in quadrature
$r_N=\sqrt{{dx_N^{\prime}}^2+{dy_N^{\prime}}^2}$. For a Gaussian
distribution, the cumulative probability distribution of $r_N$ is
$P[r_N]=1-\exp{(-r_N^2/2)}$. Alternatively, if the enclosed
probability is $p_N$, then $r_N=\sqrt{-2\times\ln(1-p_N)}$.  For
example, for $p=0.6$ (the reference value we adopted) $r=1.3537$.
This means that in a two-dimensional Gaussian distribution 60\% of the
points should be within 1.3537$\sigma$.  Let the 60$^{\rm th}$
percentile value of $r_N$ of the data points be $M$.  Then, to ensure
that our residuals are consistent with the expected Gaussian, we would
need to multiply all our $e_N$ values by a factor $1.3537/M$. We let
the rescaled, normalized residuals be $(sx_N,sy_N)$\footnote{The
  rescaling can be done in principle using any percentile value. Our
  choice of using $p=0.6$ is motivated by the fact that $p$ needs to
  be small enough so that the distribution is not sensitive to
  outliers, but $p$ also needs to be large enough to guarantee good
  statistics.}.

After the rescaling, to lowest order the cloud of data points should
be consistent with a two-dimensional Gaussian. Panel (e) of
Fig.~\ref{f:linfit} shows the distribution of the normalized and
rescaled residuals $(sx_N,sy_N)$. A circle of radius 1.3537 encloses
60\% of the points (in grey).  We now identify the outermost data
point, at distance $R$. The probability that one data point has such a
high value of $R$ is $P[1/1]=\exp{(-R^2/2)}$.  Since there are $N$
total points in the distribution, the probability of finding 1 data
point out of $N$ with such a high $R$ is $P[1/N]=1-(1-P[1/1])^N$. For
example, if $R=3$ then $P[1/1]\sim 1$\%, and $P[1/3]\sim N\times
P[1/1]$. So, for $N=10$ data points, there is a 10\% chance of having
a $\geq 3 \sigma$ outlier.

We set a confidence threshold $Q$ for accepting data points at
$2.5$\%. If the data point with the highest $R$ has $P[1/N]<Q$, then
the data point is rejected and the straight-line fitting process is
repeated. The iterations stop when all the remaining data points are
consistent with a two-dimensional Gaussian distribution.  At this
point we also compute the errors of the slopes (proper motions) and
intercepts of the fitted lines, and the reduced $\chi^2$ values. We
report PM errors measured in two distinct ways: (1) using the
estimated errors as weights; and (2) using the actual residuals of the
data points around the fitted lines, as described in the
Section~\ref{ss:MC}. It would also be possible to compute PM errors in
a third, independent way, by multiplying the expected errors by the
square root of the reduced $\chi^2$ values, as all these quantities
are included in our PM catalogs.

To summarize, our rejection algorithm works as follows:
\begin{enumerate}
\item{Preliminary rejection of obvious outliers;}
\item{Straight-line fitting to X and Y positions versus epoch;}
\item{Rescaling of normalized residuals to be consistent with a
  two-dimensional Gaussian distribution;}
\item{Checking whenever the outermost data point has $P[1/N]<Q$:}
\begin{description}
\item{YES: reject the outermost data point, return to 2.}
\item{NO: continue.}
\end{description}
\item{Final straight-line fitting with the final set of acceptable
  data points to obtain the final straight-line-fit parameters and
  errors.}
\end{enumerate}

\section{Simulations}
\label{s:sim}

In order to test the performance, accuracy and reliability of our PM
measurements, we carried out two types of simulations.  The first
simulation is based on a series of Monte-Carlo tests that focus on our
ability to reject outliers and obtain accurate values for the PMs and
their errors.  The second simulation tests our PM measurements in an
artificial-star field representing a typical case, with
globular-cluster stars and several field-star components, each of
which has its own spatial density, bulk motion and velocity
dispersion.

\subsection{Single-Star Monte-Carlo Simulations}
\label{ss:MC}

Our Monte-Carlo tests focus on the PM measurement of one single star,
in cases where we have 10, 50 or 200 data points. For each case we run
100$\,$000 random realizations in which data points span a time
baseline of 5 years. Two thirds of the points are at t=0, and the
remaining are either randomly distributed or placed at the ends of the
time baseline ($\pm 2.5$ years). Most of the data points have an
assigned positional displacement that follows a Gaussian distribution
with $\sigma=0.01$ pixel. Five percent of the points are displaced
with a dispersion 10 times larger, to mimic a population of outlier
measurements, while an additional 5\% of the points are misplaced by
up to $\pm 5$ pixels, to mimic possible mismatches.

In each Monte-Carlo run, individual observations were rejected based
on the procedures described in Section~\ref{ss:linfit}, but the
least-square fits for the slope (the PM components $\mu_x$ and
$\mu_y$) and the intercepts (the positions at t=0: $\overline{x}$ and
$\overline{y}$) are computed with weights from the
signal-to-noise-based error estimates from
Section~\ref{ss:experr}. The error estimates from each point are also
used to compute errors in the motions and positions.  For various
reasons (cosmic rays, bad pixels, neighbors, etc.), individual
observations can have errors that are larger than the expected errors,
but not large enough to cause the observation to be rejected. To
estimate the influence of these points on the errors in the
measurements, we determine a residual for every point (using a fit to
the four parameters that excludes that point) and adopt that residual
as the estimate for the error in that determination. We then
redetermine the errors in the slopes and intercepts using the same
procedure as before. Since different observations have different
impact on the slope and intercept determinations, this allow us to
construct a more empirical estimate of the errors in the derived
parameters.

Finally, for each of the three cases we computed the Monte-Carlo RMS
of the measured$-$true residual distribution for each of the derived
quantities (err$_{\overline{x}}$, err$_{\overline{y}}$, err$_{\mu_x}$
and err$_{\mu_y}$), and compared them with the average of the two
different error estimates.  The results are shown in Table~2.  In the
case with 10 points, which resembles those data sets with few
observations, the expected errors tend to underestimate the true
errors, while the residual-based error estimates are more consistent
with the true errors, although slightly larger.  When more data points
are available, both ways of computing the errors are in very good
agreement with the Monte-Carlo RMS.

These results suggest that our fitting, rejection and error-estimation
algorithms are working well. Note that we did not simulate here the
potential of small systematic errors (such as imperfect CTE
corrections) in the bulk of the measurements. In reality, such errors
will always be present at some level. The residual-based PM errors
should therefore generally be more accurate than the PM errors based
on assumed error estimates. The latter propagate only the random error
in individual exposures, and are unable to take into account small but
present systematic errors.

\begin{table}[t!]
\label{t:outputmc}
\begin{center}
\small{
\begin{tabular}{ccccc}
\multicolumn{5}{c}{\textsc{Table 2}}\\
\multicolumn{5}{c}{\textsc{Results of Monte-Carlo simulations$^\dagger$}}\\
\hline
\hline
Type&err$_{\overline{x}}$&err$_{\overline{y}}$&err$_{\mu_x}$&err$_{\mu_y}$\\
\hline
\multicolumn{5}{c}{10 data points}\\
\hline
Monte-Carlo RMS &5.68&5.60&1.61&1.61\\
Average expected errors &5.09&5.13&1.46&1.47\\
Average residual-based  &5.94&5.92&1.71&1.73\\
\hline
\multicolumn{5}{c}{50 data points}\\
\hline
Monte-Carlo RMS &1.89&1.90&0.64&0.64\\
Average expected errors &1.87&1.86&0.63&0.63\\
Average residual-based  &1.90&1.90&0.66&0.66\\
\hline
\multicolumn{5}{c}{200 data points}\\
\hline
Monte-Carlo RMS &0.93&0.93&0.32&0.32\\
Average expected errors &0.92&0.92&0.32&0.32\\
Average residual-based  &0.92&0.93&0.32&0.32\\
\hline\hline
\multicolumn{5}{l}{$^\dagger$ Units of 0.001
  pixels for err$_{\overline{x}}$ and err$_{\overline{y}}$, and 0.001}\\
\multicolumn{5}{l}{\phantom{$^\dagger$ }\pixyr\ for err$_{\mu_x}$ and
  err$_{\mu_y}$.}\\
\end{tabular}}
\end{center}
\end{table}

\subsection{Comprehensive Data Simulations}
\label{ss:sim2}

In order to test the automated procedure of converging on
cluster-member-based PMs, The second simulation concerns the PM
measurement and analysis of a field containing $\sim 19$\,$000$
simulated stars resembling cluster stars, field stars and stars of two
Milky-Way satellite galaxies.  Each star component has its own spatial
density, proper motion and velocity dispersion. We started by setting
up the input master frame catalog, and then we extracted from it
single-exposure catalogs simulating different exposure times, dithers,
roll-angle orientations, cameras and epochs.

\begin{figure*}[ht!]
\centering
\includegraphics[width=18.5cm]{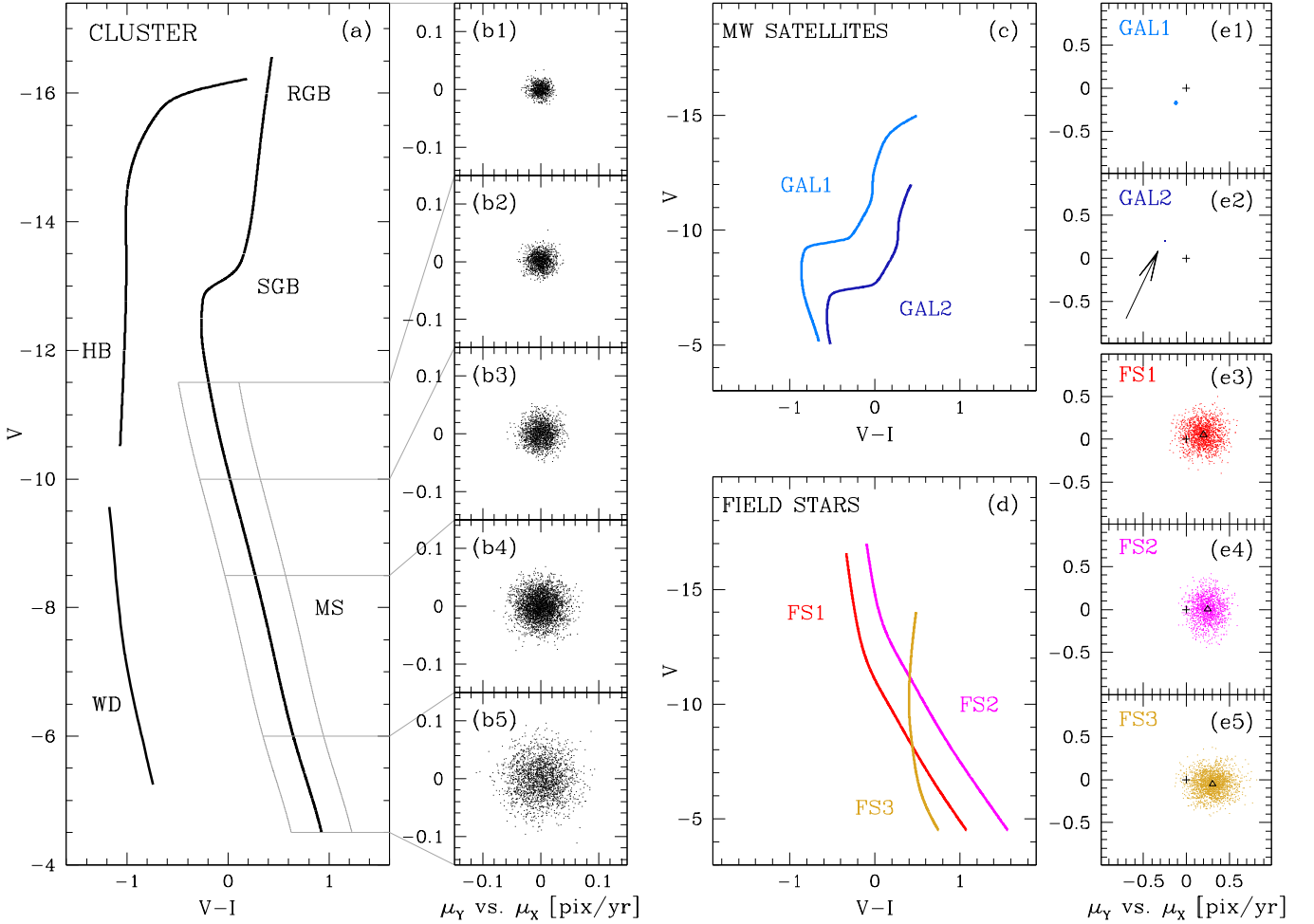}
\caption{Color-magnitude and vector point diagrams of the stars used
  for our comprehensive simulation. The CMD of cluster stars is in
  Panel (a). All the main evolutionary sequences have been
  included. We assigned to MS stars an increasing internal velocity
  dispersion at increasing magnitudes, to mimic some sort of energy
  equipartition. Panels (b2) to (b5) show the vector-point diagram of
  MS stars for 4 different values of the velocity dispersion. Bright
  MS stars and more evolved stars all have the same (smaller) velocity
  dispersion, as shown in Panel (b1). We also simulated 2 Milky-Way
  dwarf galaxies (GAL1 and GAL2, in azure and blue) and 3 components
  of field stars (FS1, FS2 and FS3, in red, magenta and yellow,
  respectively). Their CMDs are in Panel (c) and (d),
  respectively. We assigned a very small velocity dispersion (0.005
  \pixyr, 0.2 \masyr) to GAL1 stars (Panel (e1)), and no velocity
  dispersion at all to GAL2 stars (Panel (e2)). Field stars have the
  largest velocity dispersion. We assigned a bulk motion (black
  triangle) to field stars in such a way that they partially
  overlap to cluster stars in the vector-point diagram (Panels
  (e3), (e4) and (e5)).}
\label{f:inputpar}
\end{figure*}

\subsubsection{The Input Master Frame}

The spatial extension of the input master frame is $8000\times8000$
pixels, and allows us to fully populate single-exposure catalogs with
different dithers and roll-angle orientations. The CMD of cluster
stars resembles that of a real cluster, but it was drawn by hand
without aiming to be a reliable, physical representation of the real
CMD of any actual GC. Panel (a) of Fig.~\ref{f:inputpar} shows the
input CMD for cluster stars in instrumental magnitudes that for
simplicity are called V and I. As for the real data sets, we run the
simulation using instrumental-like magnitudes.  All the main
evolutionary sequences are traced.  We generated a total of 12074
cluster stars, divided as follows: 9964 main-sequence (MS) stars
(more numerous at increasing magnitudes), 350 sub-giant-branch (SGB)
stars, 651 red-giant-branch (RGB) stars, 1078 horizontal-branch (HB)
stars and 31 white-dwarf (WD) stars.

Cluster stars have a Gaussian-like distribution on the master frame
(centered at position $(5000,5000)$), to mimic the typical crowding
conditions of the center of GCs. Moreover, their positional dispersion
is larger at fainter magnitudes, to mimic some sort of mass
segregation.  The dispersion of MS stars grows from 344 to 600 pixels,
while evolved stars have the same 344-pixel spatial dispersion of the
bright MS stars.

The cluster's bulk motion is null by construction, as all measured
proper motions will be computed with respect to the bulk motion of the
cluster.  To resemble some sort of energy equipartition and test the
quality of measured PM errors we divided the MS into 5 groups, and
assigned to each of them an increasing velocity dispersion with
fainter magnitudes.  Velocity dispersions go from 0.01 \pixyr\ for the
brighter MS stars to 0.03 \pixyr\ at the faint end.  Evolved stars all
have the same velocity dispersion as the bright MS stars.  Panels (b1)
to (b5) of Fig.~\ref{f:inputpar} show the vector-point diagrams of
cluster stars for the 5 different values of input velocity dispersion.

Since it is not uncommon to have Milky-Way-satellite stars
superimposed on GC fields (e.g., Small-Magellanic-Cloud stars in
NGC~104 and NGC~362, or Sagittarius-Dwarf-Spheroidal stars in NGC~6681
and NGC~6715), we included the presence of two such nearby
galaxies. Panel (c) of Fig.~\ref{f:inputpar} shows their CMD. Galaxy
stars are placed randomly with a flat distribution on the master
frame. The brighter galaxy (GAL1) has 1126 stars and a bulk motion of
$(-0.12,-0.17)$ \pixyr. We set its internal velocity dispersion to be
small but still measurable:\ 5 milli-\pixyr (i.e., 0.2 \masyr).  The
faint galaxy (GAL2) has 685 stars and a bulk motion of $(-0.25,0.2)$
\pixyr. We assigned no internal velocity dispersion to its stars: this
way, we are able to obtain an external estimate of our measurement
errors.  Panel (e1) of Fig.~\ref{f:inputpar} shows the vector-point
diagram of GAL1 stars; the black cross marks the location of the
cluster's bulk motion.  An arrow in Panel (e2) points to the bulk
motion of GAL2.

We generated three sets of field stars, named FS1 (1516 stars), FS2
(1273 stars) and FS3 (2057 stars). Each set has its own ridge line
on the CMD (see Panel (d) of Fig.~\ref{f:inputpar}). While cluster and
galaxy stars do not have a color spread by construction (mimicking
single-stellar populations), we introduced a Gaussian scatter
($\sigma\sim0.5$ mag) to the color of field stars to resemble the fact
that they are not at the same distance, or do not have the same chemical
composition.

\begin{table*}[t!]
\begin{center}
\label{t:simpar}
\small{
\begin{tabular}{ccccccccc}
  \multicolumn{9}{c}{\textsc{Table 3}}\\
  \multicolumn{9}{c}{\textsc{Simulated single-exposure-catalog parameters}}\\
  \hline\hline
  Data set&$\Delta$time (yr)&Filter&Exposures&$\Delta$mag&Roll
  angle&Scale (\maspx)&X offset (pix)&Y
  offset (pix)\\
  \hline
 1&$-1.78$&$V$&5 long, 2 short&$-0.1$&$130^\circ$&50&2100&1900\\
    &             &$I$&5 long, 2 short&$+0.1$&$-190^\circ$&50&2200&1800\\
\hline
2&0.0&$V$&5 long, 2 short&$+0.05$&$20^\circ$&50&1900&2100\\
  &     &$I$&5 long, 2 short&$-0.5$&$85^\circ$&50&1800&2200\\
\hline
3&0.7&$V$&4 medium&$+1.5$&$80^\circ$&28.27&500&500\\
  &     &$I$&4 medium&$+1.5$&$80^\circ$&28.27&500&500\\
\hline
4&+1.3&$I$&5 long, 2 short&$-0.07$&$210^\circ$&40&2030&2020\\
\hline
5&+1.4&$V$&5 long, 2 short&$+0.1$&$60^\circ$&40&2020&2030\\
\hline\hline
\end{tabular}}
\end{center}
\end{table*}

The field FS1 has a bulk motion of $(0.2,0.05)$ \pixyr, with a round
velocity dispersion of 0.13 \pixyr. The bulk motion of field FS2 is
$(0.25,0.0)$ \pixyr, with a X-velocity dispersion of 0.12 \pixyr\ and
a Y-velocity dispersion of 0.14 \pixyr.  For the field FS3, these
three quantities are, respectively: $(0.3,-0.05)$ \pixyr, 0.14
\pixyr\ and 0.12 \pixyr.  The vector-point diagrams of field stars are
shown in Panels (e3), (e4) and (e5) of Fig.~\ref{f:inputpar}). The
bulk motion of each field component is marked by a triangle.

\begin{figure}[t!]
\centering
\includegraphics[width=7.5cm]{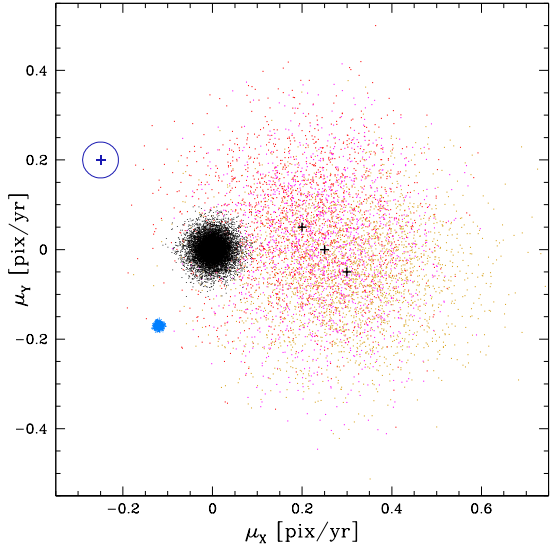}
\caption{The vector-point diagram of all the population components of
  our comprehensive simulation, color-coded as in
  Fig.~\ref{f:inputpar}. The GAL2 stars have zero PM dispersion, so
  they fall underneath the cross inside the blue circle.  The means of
  the three field components are marked by black crosses.}
\label{f:inputpmall}
\end{figure}

For clarity, Figure~\ref{f:inputpmall} shows the complete simulated
vector-point diagram. Each component is color-coded as in
Fig.~\ref{f:inputpar}. The location of the bulk motion of GAL2 stars
is highlighted by an open circle.

\subsubsection{Single-exposure catalogs}

Now that the input master frame has been defined, we can extract from
it single-exposure catalogs as follows.  We set up 5 data sets
spanning a total time baseline of 3.18 years. Each epoch has its own
orientation angle, offset (i.e., the center of the cluster is not
always at the center of the pointing), dither pattern, magnitude zero
point and pixel scale (to simulate the three cameras (ACS/WFC, ACS/HRC
and WFC3/UVIS). In addition, we added small random variations to all
these quantities:\ up to $0.2$\% variation for orientation angle,
scale (to mimic focus changes) and observing time (to mimic exposures
taken within a few days), and up to $\pm40$ pixels in ether direction
to resemble a dither pattern.

Table~3 lists the parameters adopted for each data set. The first two
data sets mimic ACS/WFC exposures (and the second one is designed to
be similar to GO-10775), the third refers to ACS/HRC exposures, while
WFC3/UVIS exposures are in data sets number 4 and 5. The magnitude
zero point $\Delta$mag listed in Table~3 is the difference in
instrumental magnitude between input master stars and deep-exposure
stars.  Stars in the short exposures are 2.2 mag fainter than those in
the deep ones.  Offsets are in units of pixels in the raw-coordinate
system of each catalog. We generated a total of 50 single-exposure
catalogs.

Stars of each single-exposure catalog are selected from the input
master frame according to their positional parameters (roll angle,
scale, offsets) and a magnitude zero point is applied.  Stars'
positions are then de-corrected for geometric distortion and put into
their raw-coordinate system. Finally, to resemble positional
uncertainties, an additional Gaussian-like shift in a random direction
is added to each star's position (with a dispersion equal to its
expected error; see Section~\ref{ss:experr}).  A similar method was
used to introduce scatter in the magnitudes."

\begin{figure*}[t!]
\centering
\includegraphics[width=17.1cm]{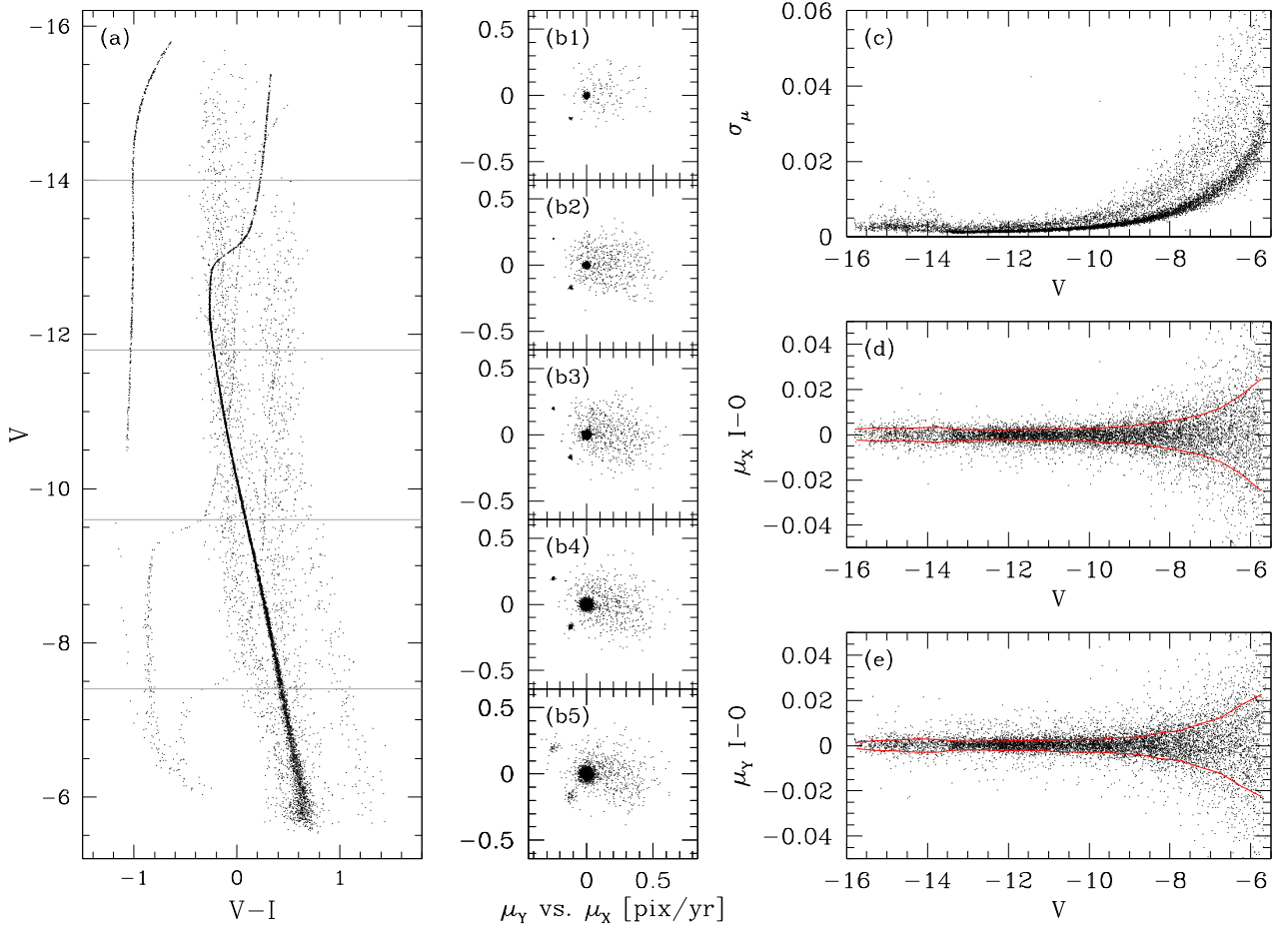}
\caption{Results of our comprehensive-data simulation. The recovered
  master-frame CMD is shown in Panel (a). Proper motions are divided
  into 5 magnitude bins (grey horizontal lines) and displayed in
  Panels (b1) to (b5), from the brighter to the fainter
  bin. Proper-motion errors as a function of the instrumental
  magnitude are shown in Panel (c). The input$-$output difference of
  stellar PMs along the X and the Y axes, as a function of the
  instrumental magnitude, are shown in Panels (d) and (e),
  respectively. Red lines in both panels mark the 68.27 percentile of
  the residuals around the median values.}
\label{f:outputpar}
\end{figure*}

\subsubsection{Results of the Full Simulation}
\label{sss:res}

We now have at our disposal single-exposure catalogs constructed as if
they were the result of reduced images. We derived from them an output
master frame using exposures of data set 2 for positions, and using
all the exposures for photometry. The recovered master frame is
necessarily different from the input master frame:\ it contains
uncertainties in the transformation parameters (because of the
position shift added to each star related to its PM plus measurement
error), and it has errors in the average position and errors in the
magnitude of its stars. The recovered master frame CMD is shown on
Panel (a) of Fig.~\ref{f:outputpar}. It contains only unsaturated
stars. Stars measured in deep exposures have a magnitude value up to
$\sim -13.5$, while brighter stars are measured only in short
exposures.

The input master frame was not used beyond this.  The recovered master
frame was the one used to compute proper motions. For simplicity,
hereafter we refer to the recovered master frame simply as the master
frame.

Because of the different pointings and orientation of each data set,
there will be master-frame stars present in some but not all of the
exposures. As a consequence, the time baseline available for some
stars will be shorter than 3.18 years.

We treated our master frame as if it came from the official GO-10775
release, and our simulated single-exposure catalogs as if they were
the output of our reduction routines. We measured PMs in the exact
same way that we do for real data sets.  Panels (b1) to (b5) of
Fig.~\ref{f:outputpar} show the recovered vector-point diagrams for 5
different magnitude bins, highlighted by grey horizontal lines in
Panel (a), from the bright bin to the faint one, respectively.

As expected, the velocity dispersion of GAL1 stars is found to be
larger than that of GAL2 stars (see, e.g., the different size of the
GAL1 and GAL2 clouds of points in Panels (b2) to (b5) of
Fig.\ref{f:outputpar}).  The one-dimensional velocity dispersion of
GAL2 stars, i.e. the estimate of our internal errors, goes from $\sim
3$ milli-\pixyr\ at $V=-11.5$ to $\sim 25$ milli-\pixyr\ at $V=-6.5$.
In the same magnitude interval, GAL1 stars have a measured velocity
dispersion (i.e., without subtracting the error in quadrature) ranging
from $\sim 7$ milli-\pixyr\ to $\sim 28$ milli-\pixyr, and is
systematically larger than that of GAL2 stars.  Table~4 lists
velocity-dispersion values for both galaxies in 6 magnitude ranges.

\begin{table}[t!]
\begin{center}
\label{t:g12t}
\small{
\begin{tabular}{ccc}
\multicolumn{3}{c}{\textsc{Table 4}}\\
\multicolumn{3}{c}{\textsc{Measured velocity dispersions of simulated}}\\
\multicolumn{3}{c}{\textsc{GAL1 and GAL2 stars}}\\
\hline\hline
Mag range&GAL1 $\sigma_\mu$&GAL2 $\sigma_\mu$\\
                & \pixyr&\pixyr\\
\hline
$(-12,-11)$&0.0068&0.0029\\
$(-11,-10)$&0.0066&0.0034\\
$(-10,-9)$&0.0071&0.0048\\
$(-9,-8)$&0.0102&0.0062\\
$(-8,-7)$&0.0226&0.0087\\
$(-7,-6)$&0.0282&0.0252\\
\hline\hline
\end{tabular}}
\end{center}
\end{table}

Panel (c) of Fig.~\ref{f:outputpar} illustrates the trend of PM errors
as a function of the instrumental magnitude. We can distinguish two
tails of errors at fainter magnitudes:\ a more populated, smaller
error trend, corresponding to stars with motions measured using the
full 3.18 years of time baseline, and a second, less populated tail
that corresponds to stars with a time baseline of 1.78 years.
Moreover, there is an increase in the PM errors for stars brighter
than $\sim-13.5$ mag. These stars are measured only in the short
exposures (8 out of 50), and therefore their PMs are less well
constrained.

Panels (d) and (e) show the difference (defined as input$-$output,
I$-$O) of each component of the motion. Red lines mark the $\pm68.27$
percentile (RMS) of the I-O values around the median values.  These
two plots provide another way to estimate the internal errors of our
procedure. For the particular simulation we set up, the $\mu_X$ I$-$O
RMS is about 0.0032 \pixyr\ (0.13 \masyr) for the short-exposure
regime, and goes from 0.0022 \pixyr\ (0.09 \masyr) at $V=-13$ to
0.0024 \pixyr\ (0.10 \masyr) at $V=-10$ to 0.006 \pixyr\ (0.24 \masyr)
at $V=-8$, and reaching 0.02 \pixyr\ (0.8 \maspx) at $V=-6$. The RMS
of $\mu_Y$ I$-$O has a similar behavior. These values are consistent
with the velocity dispersion of GAL2 stars.

The comparison of input and output PMs shows that our PM-measurement
algorithms are highly reliable. There are astrophysical applications
for which accurate error estimates are crucial. For instance, when we
want to measure the intrinsic velocity dispersion of cluster stars, we
have to subtract in quadrature the PM measurement errors from the
observed dispersion. When the errors contribute a large fraction of
the observed dispersion, a small over- or under-estimate of the errors
leads to biased results.

\begin{figure}[!t]
\centering
\includegraphics[width=\columnwidth]{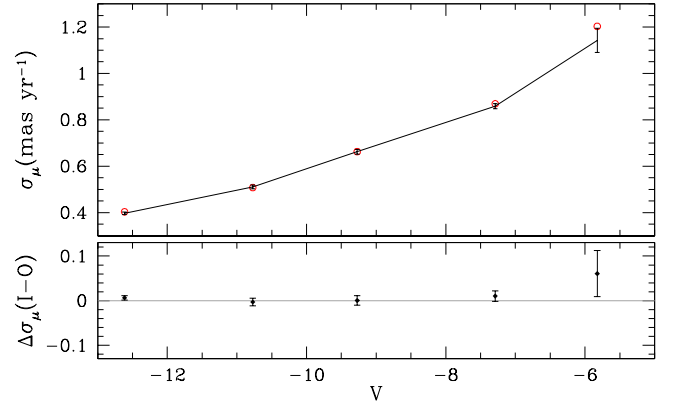}
\caption{The top panel shows the input (red open circles) and inferred
  (black with errorbars) velocity dispersion of cluster stars in our
  comprehensive simulation, as a function of the instrumental
  magnitude. The bottom panel shows the residuals between the input
  and the output values.}
\label{f:GHout}
\end{figure}

To test this, we compute the intrinsic velocity dispersion of cluster
stars from the PM catalog (as done in van der Marel \& Anderson 2010)
and check whether it is in agreement with the input values. The top
panel of Fig.~\ref{f:GHout} shows the inferred velocity dispersions
(in black, with errorbars) as a function of the instrumental magnitude
(0.4 \masyr\ corresponds to 0.01 \pixyr on the master frame). The real
(input) velocity dispersion of cluster stars is represented by red
open circles. The agreement between input and output velocity
dispersions (bottom panel) shows an absence of clear systematic
residuals, meaning that our quoted PM errors are accurate and
reliable.

There is perhaps a marginal discrepancy (at the $1.2\sigma$-level) at
the faint-end magnitude limit, where it seems that the PM errors have
been slightly overestimated, with the result that the inferred
velocity dispersion is lower than the input one. However, this should
not come as a surprise. The input velocity dispersion of faint GC
stars is 0.03 \pixyr, while their measured PM error is almost as large
($\sim 0.025$ \pixyr; see Panel (c) of Fig.~\ref{f:outputpar}). One
should always be careful in trusting results that come from the
quadrature difference of quantities of similar size, especially when
one of these quantities is an error estimate. The fact that even at
the faint limit of our simulated measurements input and output
velocity dispersions are still quite consistent (at the 1.2$\sigma$
level) is a further validation of our methodology.

\begin{figure}[!t]
\centering
\includegraphics[width=\columnwidth]{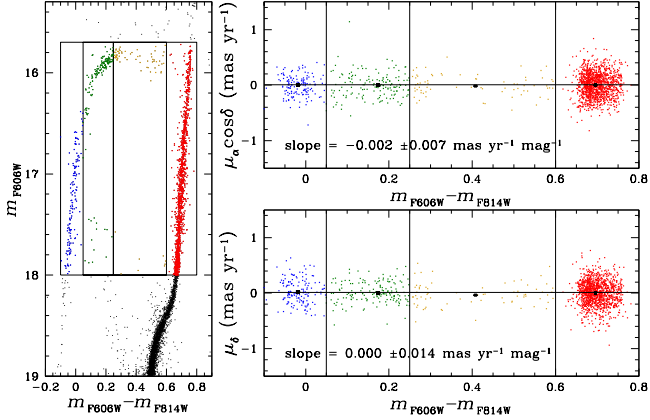}
\caption{The left panel shows the CMD of NGC~7078 around the
    HB and RGB regions, and the stars used to investigate the presence
    of chromatic-induced systematic effects. The right panels show the
    $\mu_\alpha cos\delta$ and $\mu_\delta$components of the motion of
    selected stars as a function of the star colors (top and bottom
    panels, respectively).  We divided and color-coded the selected
    stars into 4 groups according to their color, for clarity. We
    computed median motion and error for each group of stars, and
    fitted two lines to the median points (the size of the errors are
    comparable to, or smaller than, the median points). The slopes of
    the fitted lines, consistent with zero, imply no chromatic-induced
    systematic errors in our measurements.}
\label{f:dcr}
\end{figure}

\begin{figure*}[!t]
\centering
\includegraphics[width=\textwidth]{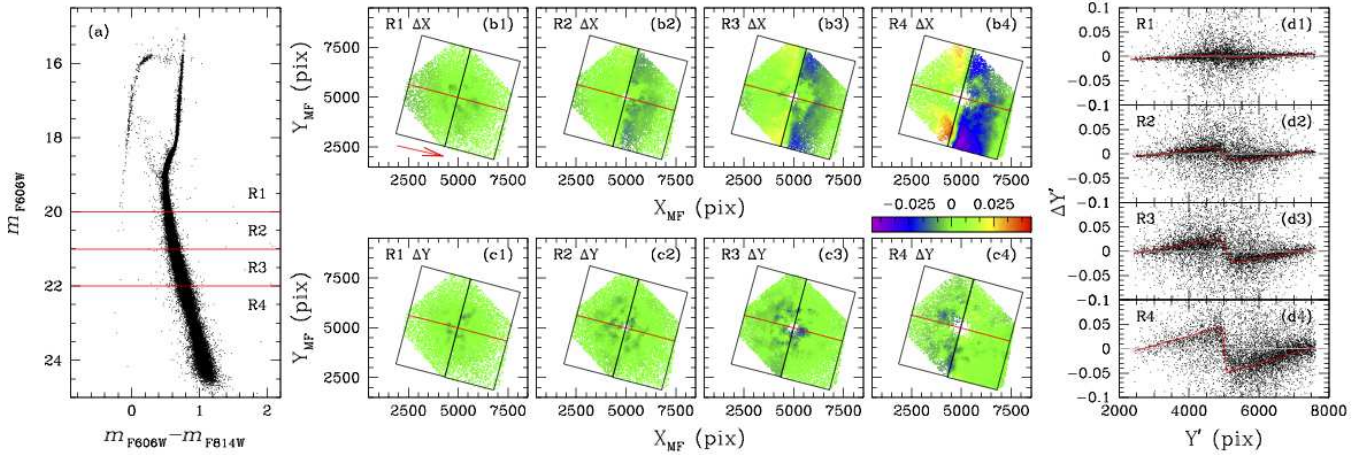}
\caption{Impact of uncorrected CTE effects on the GO-10775 NGC~7078
  master frame. Panel (a) shows the $m_{\rm F606W}$ vs. $m_{\rm
    F606W}-m_{\rm F814W}$ CMD. We divided the stars into 4 magnitude
  regions, labeled R1, R2, R3 and R4. For each region we computed the
  locally-averaged difference between the GO-10775 master-frame X and
  Y positions and those predicted by our PM fits at the epoch of the
  master frame. Panels (b1) to (b4) illustrate these differences for
  the X positions ($\Delta$X) as a function of the stellar location on
  the master frame, for the magnitude regions R1 to R4.  Panels (c1)
  to (c4) similarly show the differences in position along the Y axis
  ($\Delta$Y). Points are color coded according to the size of the
  differences. A footprint of the typical location of the GO-10775
  ACS/WFC chip placements is also shown in black, with individual
  amplifiers separated by a red line. A strong correlation between the
  pattern of position differences and the chip layout is evident.
  Panels (d1) to (d4) illustrate the position differences on a rotated
  reference system, so that the rotated Y$^\prime$ axis is parallel to
  the raw Y direction of the GO-10775 exposures. The averaged
  $\Delta$Y$^\prime$ residuals are highlighted by a red line. The fact
  that these residuals are strongly correlated with Y$^\prime$ and
  increase at fainter magnitudes is a clear signature of unaccounted
  for CTE losses.}
\label{f:CTEtrends}
\end{figure*}

\section{Mitigating Sources of Systematic Error}
\label{s:sys}

In the previous Section we demonstrated that our PM-measurement
algorithms are reliable when random errors and mild systematic effects
are taken into account.  Unfortunately, unaccounted for systematic
sources of error may also be present in real data. In this Section we
describe the methods we have adopted to mitigate their effects on our
PM measurements.

In what follows we will describe as an example the case of NGC~7078
(M~15). This is the cluster for which we will present the PM analysis
and catalog in Section~\ref{s:cat and res}. NGC~7078 is a typical case
among the 22 clusters in our study, in the sense that it has an
average time baseline and an average number of data sets.

\subsection{Chromatic effects}
\label{ss:dcr}

A systematic effect that is always present in ground-based PM
measurements is the so-called differential-chromatic refraction (DCR,
see., e.g., Anderson et al.\ 2006; Bellini et al.\ 2009). The DCR
effect shifts the photon positions on the CCD, and the displacement is
proportional to the photon wavelength and to the zenithal distance of
the observations. Space-based telescopes are obviously immune to DCR
effects. Nonetheless, as anticipated in Section~\ref{ss:gdc}, we found
a chromatic-dependent shift of blue and red stellar positions when UV
filters are used with the WFC3/UVIS camera (Bellini, Anderson \& Bedin
2011), and for this reason we decided not to include observations
taken with filters bluer than 330 nm.

A way to check whether or not our PM measurements are nonetheless
affected by some chromatic-induced systematic effects is to analyze
the behavior of the single components of the stellar motions as a
function of the star colors. The left panel of Figure~\ref{f:dcr}
shows the CMD of NGC~7078 around the HB and RGB regions. We selected
stars in the magnitude range $15.7<m_{\rm F606W}<18$ in order to cover
the largest available color baseline, and divided them into 4 color
bins (blue, green, yellow and red in the figure). The $\mu_\alpha
cos\delta$ component of their motions is shown in the top-right panel,
as a function of the star colors. We determined the median color and
motion, with error, for each of the four groups of stars (black full
squares). The same plot for the $\mu_\delta$ component of the stellar
motions is shown in the bottom-right panel.

The median motions in each of the two right-hand panels are fitted
with a weighted straight line (in black). Since we are using cluster
members for the test, in principle, the fitted lines should have no
slope. On the other hand, slopes that significantly differ from zero
would immediately reveal the possible presence of chromatic-induced
sytematic effects. The computed slopes and errors are:
$-0.002\pm0.007$\,mas\,yr$^{-1}$\,mag$^{-1}$ for $\mu_\alpha
cos\delta$, and $0.000\pm0.014$\,mas\,yr$^{-1}$\,mag$^{-1}$ for
$\mu_\delta$. These values are consistent with zero well within their
errors, and therefore we can rule out any presence of
chromatic-induced systematic effects in our PMs.

\subsection{CTE effects}
\label{ss:go10775-cte}

\begin{figure}[!t]
\centering
\includegraphics[width=\columnwidth]{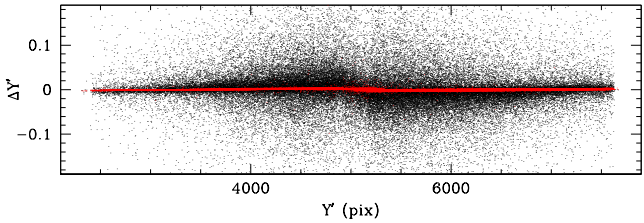}
\caption{Rotated $\Delta$Y$^\prime$ position offsets as a function of
  the Y$^\prime$ position using the original GO-10775 positions as the
  master frame (black, same as Panels (d) of Fig.~\ref{f:CTEtrends},
  but not binned in magnitude) and using the PM-predicted positions at
  t=0 (red). The latter are used for all our final PM catalogs.}
\label{f:iter1res}
\end{figure}

One problem not addressed by our simulations is that the GO-10775
master frame that was used for the real data is not really
astrometrically flat. At the time the GO-10775 catalogs were released
to the public, the pixel-based CTE correction for the ACS/WFC was yet
not available.  Stellar positions in the catalog thus suffer from this
systematic error. As a result, transformed single-exposure star
positions onto the master frame are affected by a systematic shift in
position that is a function of both the location of the stars on the
master frame and of their master-frame magnitude.

\begin{figure*}[!t]
\centering
\includegraphics[width=\textwidth]{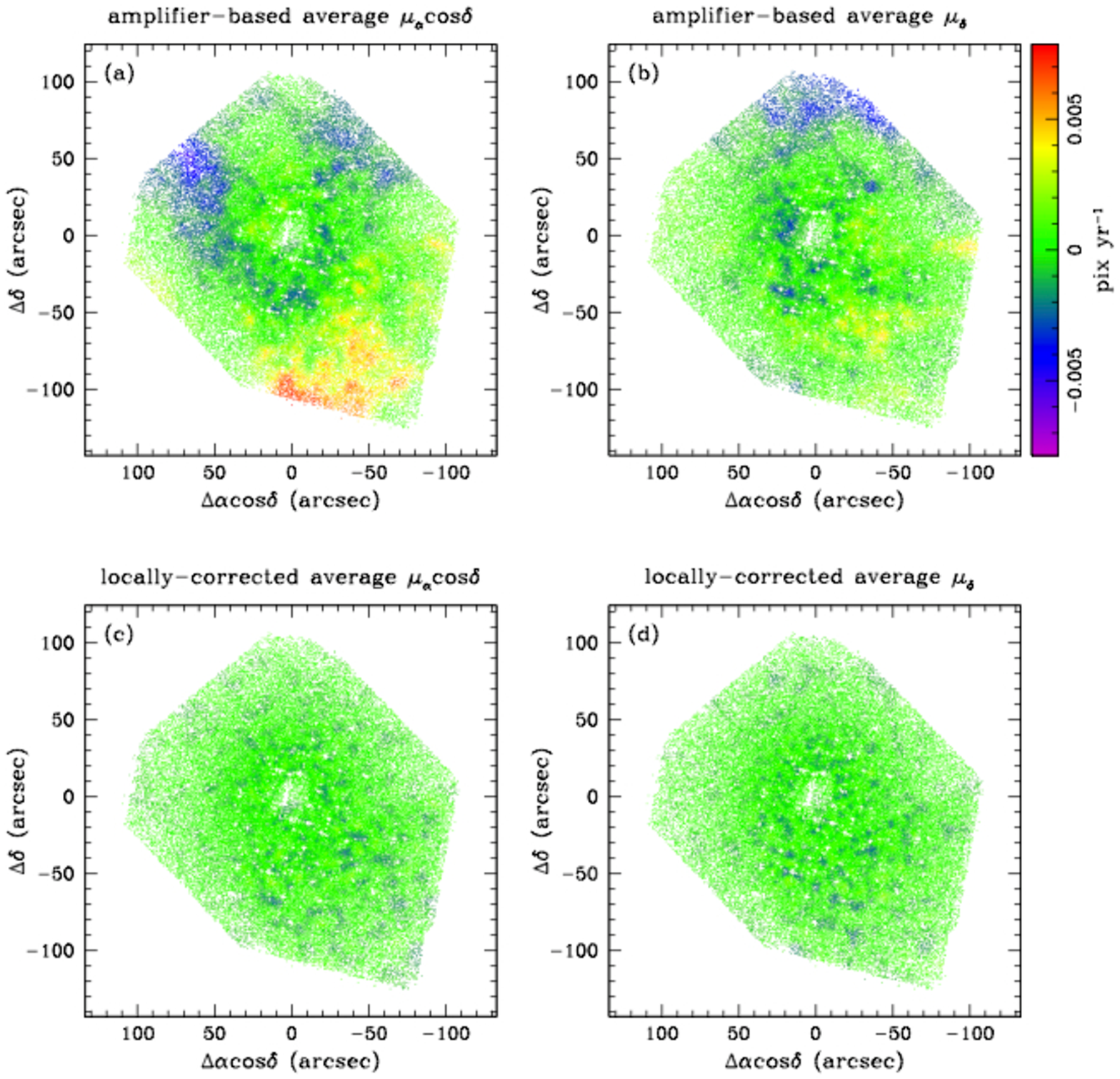}
\caption{The top panels show two-dimensional maps of the
  locally-averaged $\mu_{\alpha}\cos\delta$ (a) and $\mu_{\delta}$ (b)
  components of the PM, as a function of positions with respect to the
  cluster center (in units of arcsec). Stars are color-coded according
  to their locally-averaged PM, according to the color bar on the
  top-right. Bottom panels show the same after we applied our local
  correction described in Section~\ref{ss:loc_cor}.}
\label{f:loc_cor}
\end{figure*}

Our PM-measurement algorithms produce as output the predicted position
($\overline{x},\overline{y}$) of each star at the epoch of the master
frame (t=0), obtained as the intercept values of the least-squares
fits versus time.  This predicted position is based on a large number
of (CTE-corrected) exposures, and not just those from GO-10775, and
thus the new master frame should provide a better estimate of the true
star position at t=0.  A comparison between the GO-10775 master-frame
positions and the PM-based predicted positions
($\overline{x},\overline{y}$) should therefore reveal the signature of
uncorrected CTE effects in the GO-10775 master-catalog positions.

Panel (a) of Fig.~\ref{f:CTEtrends} shows the CMD of NGC~7078 for all
stars in our PM catalog. We divided the CMD into 4 magnitude regions,
from the brighter to the fainter, labeled R1 to R4.  For each star in
each magnitude region we computed the 3$\sigma$-clipped averaged
difference($\Delta$X, $\Delta$Y) (in pixels) between the master-frame
and the PM-predicted positions, locally averaged over its surrounding
200 stars.  Panels (b1) to (b4) show the map of the $\Delta$X
residuals for the magnitude regions R1 to R4, respectively.  Panels
(c1) to (c4) show the same for the $\Delta$Y residuals.  Stars are
colored according to the size of the residuals, following the
color-coded bar on top of Panel (c4).  In each of these middle panels
we overplotted the typical GO-10775 layout, in which ACS/WFC single
chips are drawn in black, while their amplifier subdivision is in red.

It is clear from Panels (b) and (c) of Fig.~\ref{f:CTEtrends} that the
pattern of residuals correlates with position on the master frame in
the manner expected for a master-frame not corrected for CTE losses.

CTE losses occur along the Y-axis direction of the raw GO-10775
exposures, highlighted by a red arrow in Panel (b1). By rotating the
master-frame in such a way that its rotated Y axis Y$^\prime$ is
parallel to the raw Y axis of GO-10775, the position residuals
$\Delta$Y$^\prime$ directly reveal the impact of CTE losses.  Panels
(d1) to (d4) of Fig.~\ref{f:CTEtrends} show the $\Delta$Y$^\prime$
residuals as a function of the Y$^\prime$ position for the 4 magnitude
regions. The red line in each panel indicates the average residual
trend. The results are remarkably similar to, e.g., Fig.~15 of
Anderson \& Bedin (2010), and leave no doubt that the source of the
systematic error is CTE losses.

To mitigate the impact of uncorrected CTE losses on the master-frame
positions, we re-measured all stellar PMs using
($\overline{x},\overline{y}$) values as the new master-frame
positions.  Figure~\ref{f:iter1res} shows the $\Delta$Y$^\prime$
residuals (not binned in magnitude) as a function of the Y$^\prime$
positions, obtained by using the original GO-10775 master frame (in
black) and the improved master frame (in red). This figure clearly
shows that our procedure successfully eliminates most of the impact of
uncorrected CTE losses in the GO-10775 master-frame positions.  We
therefore used this procedure for all final PM calculations.

\subsection{Other Residual Systematics}
\label{ss:second-order}

Even in the ideal case of a systematics-free master frame,
imperfectly-corrected geometric-distortion and CTE residuals are
always to be expected in our single-exposure star positions Depending
on how a given data set is oriented and dithered with respect to the
master frame, these uncorrected residuals may affect the measured PMs.

To assess the extent of any remaining systematic effects in our
catalogs, we considered two-dimensional maps of the mean PM of cluster
stars. To lowest order, no mean PM is expected. In the radial
direction, any contraction or expansion due to core collapse or
gravothermal oscillations is too slow to induce measurable PMs. The
same is true for any apparent contraction or expansion due to a
cluster's line-of-sight motion away from or towards us. In the
azimuthal direction, there may in principle be non-zero mean PMs due
to cluster rotation. However, clusters are generally close to
spherical, so any rotation is expected to be small.  Moreover, our
calibration procedure, using 6-parameter linear transformations to
align frames, removes any inherent solid-body rotation component from
the mean PM field (see discussion in van der Marel \& Anderson
2010). Therefore, the only mean PM components that may be in principle
present in our PM catalogs are small differential-rotation
components. Such components should be azimuthally aligned, with a
well-defined symmetry around the cluster center. Any other mean PM
component inherent in our catalogs is therefore a likely indication of
residual systematic errors.

\begin{table}[t!]
\begin{center}
\label{t:stat_preloc}
\footnotesize{
\begin{tabular}{ccccc}
\multicolumn{5}{c}{\textsc{Table 5}}\\
\multicolumn{5}{c}{\textsc{Amplifier-Based, Local Average PM Statistical Quantities}}\\
\hline\hline
Unit& Minimum & Median & Maximum & Semi-inter.\\
\hline
\multicolumn{5}{c}{$\mu_{\alpha}\cos\delta$}\\
\pixyr& $-$0.0049 & 0.0003 & 0.0079 & 0.0011\\
\masyr& $-$0.2017 & 0.0119 & 0.3143 & 0.0444\\
\kms  & $-$9.9487 & 0.5867 & 15.495 & 2.1914\\
\kms/$\sigma_{V_{\rm LOS}}$&
        $-$0.7368 & 0.0405 & 1.1478 & 0.1623\\
\hline
\multicolumn{5}{c}{$\mu_{\delta}$}\\
\pixyr& $-$0.0042 & 0.0003 & 0.0049 & 0.0008\\
\masyr& $-$0.1737 & 0.0111 & 0.1948 & 0.0322\\
\kms  & $-$8.5683 & 0.5472 & 9.6037 & 1.5875\\
\kms/$\sigma_{V_{\rm LOS}}$&
        $-$0.6346 & 0.0405 & 0.7114 & 0.1176\\
\hline\hline
\end{tabular}}
\end{center}
\end{table}

We constructed a two-dimensional map for each component of the average
motion by color-coding each star in our NGC~7078 PM catalog according
to average motion of its surrounding 200 stars.  We used
3$\sigma$-clipping to remove any influence from non-cluster members.
The top panels of Fig.~\ref{f:loc_cor} show the so-derived 2D maps for
the X (left) and the Y (right) component of the motion.  The color
scale is shown in the top-right panel of the figure, in units of
\pixyr. The panels reveal the presence of systematic
errors. Transitions between lower and higher average PM values happen
in proximity to the detector/amplifier edges of the adopted data sets,
namely: GO-10401, GO-10775, GO-11233, and GO-12605 (see Table~A21 for
the full list of exposures we used).  To quantify the size of these
systematic trends, we computed for each component of the
locally-averaged motion the minimum, median, maximum and
semi-interquartile values in four different PM units:\ \masyr, \pixyr,
\kms\ and \kms/$\sigma_{V_{\rm LOS}}$, where $\sigma_{V_{\rm LOS}}$ is
from Table~1. Table~5 collects these values.

In an absolute sense, the systematic trends are generally very
small. In fact, 50\% of the stars in our catalog have locally-averaged
PMs smaller than 0.0011 and 0.0004 \pixyr\ for the X and the Y
component, respectively. As a reference, we recall that we can measure
the position of bright, unsaturated stars in each exposure with an
average precision of $\sim 0.01$ pixel. Nevertheless, there are
locations on the master frame where the systematic trends are as large
as $\sim 0.008$ \pixyr. The available time baseline for these
locations is about 5.5 years, giving a total displacement of more than
0.04 pixels.

These systematic trends have the potential to significantly affect
specific scientific studies. Even though the systematic trends are
typically only as large as $\sim 15$\% of the the quoted velocity
dispersion $\sigma_{V_{\rm LOS}}$ (at least for NGC~7078), there are
locations on the master frame where the systematic effects are even
larger than $\sigma_{V_{\rm LOS}}$, so this may affect dynamical
studies of the spatially-dependent kinematics. By contrast, other
scientific studies, e.g. those focusing on differences in kinematics
between different sub-populations of the cluster, won't be affected by
these systematic trends. Locally the PM of stars of different
populations will be biased in the same way.

The user of the catalogs can decide to simply not include stars in any
high-mean PM regions in the analysis, but it can be tricky to
carefully choose which stars are good and which stars are not. The
choice depends on the specific scientific needs. In order to make our
PM catalogs useful for a wide range of scientific investigations, the
PMs in our catalogs are offered in two ways:\ the amplifier-based PM
measurements discussed so far, and the locally-corrected PM
measurements obtained as described in the following Section.

\begin{table}[t!]
\begin{center}
\label{t:stat_loccor}
\footnotesize{
\begin{tabular}{ccccc}
\multicolumn{5}{c}{\textsc{Table 6}}\\
\multicolumn{5}{c}{\textsc{Locally-Corrected, Local Average PM Statistical Quantities}}\\
\hline\hline
Unit& Minimum & Median & Maximum & Semi-inter.\\
\hline
\multicolumn{5}{c}{$\mu_{\alpha}\cos\delta$}\\
\pixyr& $-$0.0024 & 0.0000 & 0.0028 & 0.0004\\
\masyr& $-$0.0992 & 0.0007 & 0.1100 & 0.0149\\
\kms  & $-$4.8954 & 0.0345 & 5.4230 & 0.7345\\
\kms/$\sigma_{V_{\rm LOS}}$&
        $-$0.3625 & 0.0026 & 0.4017 & 0.0544\\
\hline
\multicolumn{5}{c}{$\mu_{\delta}$}\\
\pixyr& $-$0.0026 & 0.0000 & 0.0027 & 0.0004\\
\masyr& $-$0.1063 & 0.0010 & 0.1063 & 0.0151\\
\kms  & $-$5.2454 & 0.0493 & 5.2406 & 0.7444\\
\kms/$\sigma_{V_{\rm LOS}}$&
        $-$0.3885 & 0.0037 & 0.3882 & 0.0551\\
\hline\hline
\end{tabular}}
\end{center}
\end{table}

\subsection{Local Corrections}
\label{ss:loc_cor}

Local PM corrections can be obtained in two ways:\ (1) ``A-priori'',
by using a local sample of reference stars to compute the linear
transformations from each single-exposure catalog on to the master
frame (the so-called local-transformation approach, see e.g. Anderson
et al.\ 2006; Bellini et al.\ 2009); (2) ``A-posteriori'', by locally
correcting the PM of each star by the net motion of its surrounding
neighbors. Our adopted local PM correction is of the latter kind.

Surrounding neighbors are chosen as follows. For each star in the PM
catalog, we identify surrounding cluster stars within 600 pixels and
within $\pm 0.5$ $m_{\rm F606W}$ magnitudes from the target star (to
mitigate the impact of both uncorrected geometric-distortion and
uncorrected CTE residuals). Then, we compute the 3.5$\sigma$-clipped
median value of each component of the motion for these neighbors:
$\overline{\mu_{\alpha}\cos\delta}$ and $\overline{\mu_{\delta}}$. We
correct the motion of the target star by subtracting these values.  If
there are less than 50 neighbor stars, no correction is applied. If
there are more than 150 neighbor stars, we compute
$\overline{\mu_{\alpha}\cos\delta}$ and $\overline{\mu_{\delta}}$
values using only the closest 150 stars.

Panels (c) and (d) of Fig.~\ref{f:loc_cor} show the locally-averaged
PMs after our local correction is applied. Points are color-coded in
the same way as for the amplifier-based average motions. As expected,
all systematic spatial PM trends have been removed. Table~6 collects
the same statistical quantities as Table~5, but now for the
local-corrected PMs. The improvement offered by the local correction
with respect to the amplifier-based PMs is evident in all values
listed in Table~6.  

\begin{figure}[!t]
\centering
\includegraphics[width=\columnwidth]{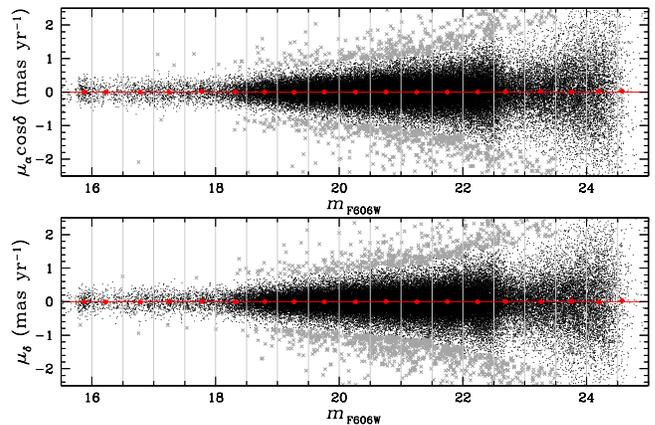}
\caption{PM components as a function of the $m_{\rm F606W}$ magnitude:
  $\mu_{\alpha} cos \delta$ (top), $\mu_\delta$ (bottom). Motions are
  divided into magnitude bins and their 3.5$\sigma$-clipped median are
  shown in red, for each bin. The size of the median errors are
  comparable to, or even smaller than, the median points. Rejected
  points are marked with grey crosses.  The red horizontal line shows
  the absence of any magnitude trend, and is not a fit to the points.}
\label{f:magsys}
\end{figure}

Because uncorrected CTE residuals are a function of both stellar
positions and magnitudes, a further proof that our local corrections
are able to properly remove any systematic-error residual would be the
absence of trends in the PM versus magnitude plane. The two panels of
Figure~\ref{f:magsys} show each component of the locally-corrected PMs
as a function of the stellar magnitude. We computed
3.5$\sigma$-clipped median motions and errors binning every 0.5 mag
(red points. Errorbars are comparable to, or smaller than, the median
points). Rejected points are marked with grey crosses. The red
horizontal lines indicate the absence of any systematic trend, and are
not a fit to the points, which all lie on the lines well within their
errors.

It is clear from Figures~\ref{f:loc_cor} and \ref{f:magsys} that
locally-corrected proper motions succesfully correct any spatially-
and magnitude-dependent systematic trends. However, users should
carefully consider whether it is best to use the amplifier-based PMs
or the locally-corrected PMs. The latter have fewer systematics, so
they may be best for studies of, e.g., cluster velocity dispersion
profiles. However, locally-corrected PMs have any intrinsic mean
motion removed by brute force. Therefore, they are not suitable for
studies of, e.g., cluster rotation.

\subsection{Selections based on Data-Quality Parameters}
\label{ss:qfit+}

\begin{figure*}[!t]
\centering
\includegraphics[width=\textwidth]{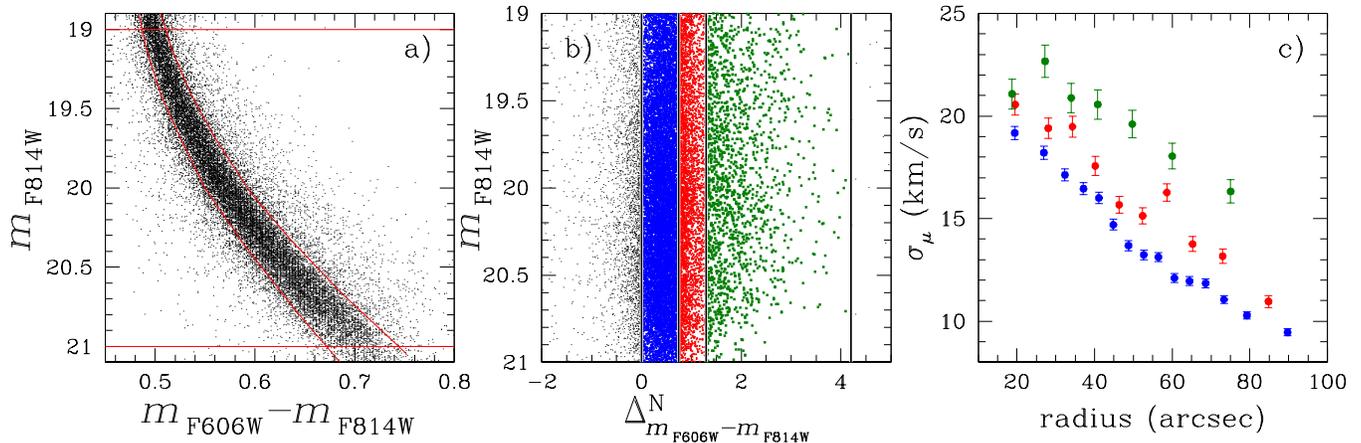}
\caption{Panel a) the upper MS of NGC~7078. Stars in the magnitude
  range $19<m_{\rm F814W}<21$ (horizontal red lines) are selected for
  measurement of the velocity dispersion. The two red lines along the
  MS are used for the rectification of MS stars shown in panel b),
  where we define 3 samples of stars according to their color:\ bMS
  (blue), rMS (red), and very-red objects (vrO, in green). The radial
  velocity dispersion profile of the three components is shown in
  panel c), where we can see the effects of crowding/blending on
  $\sigma_\mu$ as described in the text.}
\label{qfit_7078_pre}
\end{figure*}

\begin{figure}[!t]
\centering
\includegraphics[width=\columnwidth]{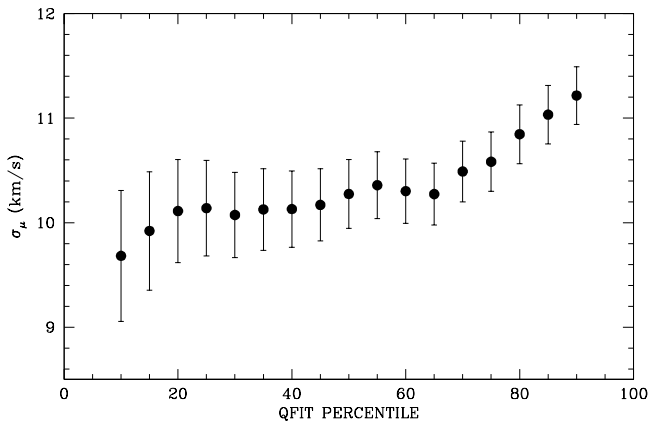}
\caption{Sensitivity of the inferred velocity dispersion of NGC~7078
  stars with similar kinematical properties to different
  \texttt{QFIT} selection cuts on the PM catalog. NGC~7078 is the
  prototype of high central-density clusters with unbalanced filters
  in the different epochs, for which use of appropriate cuts based on
  data-quality parameters is important.}
\label{f:qfit}
\end{figure}

In the previous sections we discussed systematic effects that impact
all our PM catalogs. Other sources of systematic errors, e.g. those
caused by crowding, affect some clusters more than others. Moreover,
such systematics are relevant to only some of the scientific
investigations listed in the Introduction (e.g., internal motions). As
part of the PM analysis, we derive several data-quality parameters
that are reported in our catalogs. These parameters can serve as
diagnostics to determine which stars to include or exclude from a
particular analysis, depending on the specific scientific needs.

We do not include in our catalogs stars with obvious neighbors (see
Section~\ref{s:master frame}). Nonetheless, some stars in our catalogs
will be affected by (faint) neighbors, even when not explictly
recognized as such. The resulting crowding-induced systematic effects
are among the most subtle sources of error. In clusters with a very
dense core, the measured position of sources with neighbors are shifted
away from its true position. This causes a systematic PM error, if the
shift is not the same at different epochs. This can happen if the
sources have a high relative motion, or if the sources are observed
with different filters at different epochs. To illustrate the latter
case, consider the case of two close sources: a red and a blue
star. When observed through a red filter, the apparent shift induced
by one star on the position of the other is different than when
observed through a blue filter. If we have only two epochs of
observations, one based on red and one based on blue exposures, then
this will induce systematic PM errors. The situation is obviously
worse the closer the stars are (and especially when dealing with
complete blends), or when there are multiple close neighbors of
different colors.

The \texttt{QFIT} parameter included in the GO-10775 catalogs (which
is also replicated in our PM catalogs) is an important diagnostic to
assess crowding effects. This parameter quantifies how well a source
has been fit with the PSF model. This, in turn, this correlates with
the amount of light contamination from neighbor stars that fell within
the region over which the stellar profile was fitted. Lower
\texttt{QFIT} values correspond to more isolated, less
systematic-affected stars.

Another parameter that helps in assessing crowding effects on the PM
measurements is the reduced $\chi^2$. For position measurements with
only Gaussian random errors, the least-square linear fits we used to
measure PMs should generate $\chi^2 \approx N$, where $N$ is the
number of degrees of freedom.  Hence, it should result in a reduced
$\chi^2 \approx 1$. Instead, when the position measurements also
contain systematic errors, then the reduced $\chi^2$ tends to be
larger. Rejecting stars with large \texttt{QFIT} and/or large reduced
$\chi^2$ values therefore helps to minimize the impact of
crowding-induced systematics on the PM catalog.

A third diagnostic worth mentioning is N$_R$, defined as the ratio
$N_{\rm used}/N_{\rm found}$. Here, N$_{\rm found}$ is the total
number of data points initially available for the PM straight-line
fits, and N$_{\rm used}$ is the final number of data points actually
used after the one-point-at-a-time rejection algorithm (see
Section~\ref{ss:linfit}). If N$_R$ is low, then a high fraction of
data points are rejected in the PM fit of a given star, and one should
be suspicious about the quality of the resulting PM measurement.

\begin{figure*}[!t]
\centering
\includegraphics[width=\textwidth]{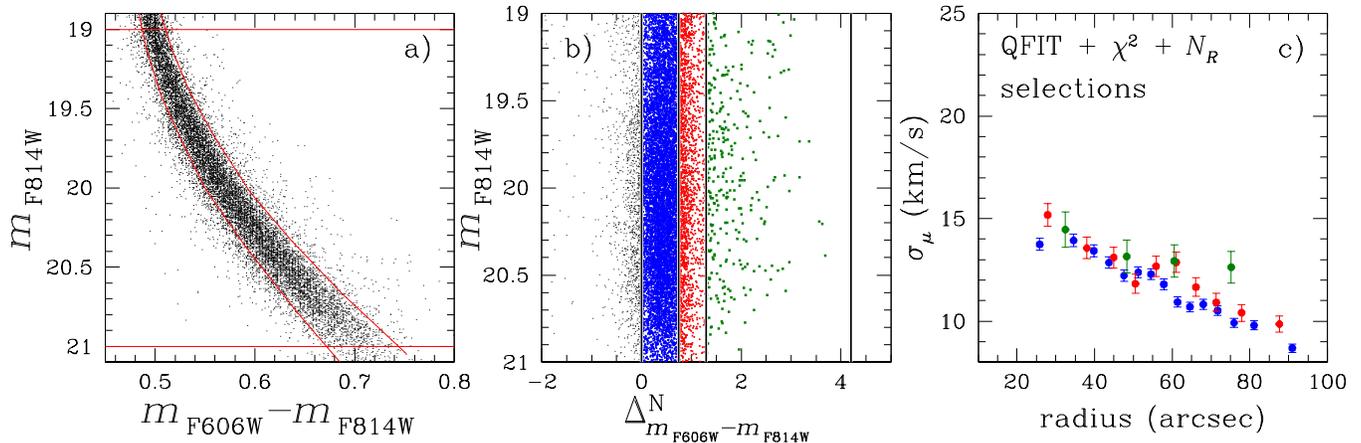}
\caption{Similar to Figure~\ref{qfit_7078_pre}, but for the subset of
  NGC~7078 catalog stars with high-quality PMs. There is now no
  disagreement between the velocity dispersions of the three MS
  samples in panel c, and the $\sigma_\mu$ values are reduced across
  the board compared to Figure~\ref{qfit_7078_pre}.}
\label{qfit_7078_after}
\end{figure*}

As a practical example, let us again consider NGC~7078. Since this is
a post core-collapse cluster, its level of crowding is very high even
at \textit{HST} resolution. Therefore, crowding/blending-driven
systematics are expected to play an important role. When two stars on
the MS are blended, their blended sum typically shows up as source on
the red side of the MS (this is because the fainter star that
`perturbs' the brighter star is redder, owing to the MS slope in the
CMD). So to look for a possible signature of systematic PM errors, we
studied in NGC~7078 the dependence of the PM kinematics as function of
color within a given magnitude range.

We selected NGC~7078 stars along the MS in the magnitude interval
between $19<m_{\rm F814W}<21$ (panel a of Figure~\ref{qfit_7078_pre}).
We drew by hand two fiducial lines on the blue and on the red side of
the MS (in red in the panel), and used them to `rectify' the MS so
that the blue-side and red-side fiducials have a $\Delta^N$color of 0
and 1, respectively, on the rectified plane (panel b). We then defined
3 subsamples of stars:\ the blue MS (bMS, in blue), the red MS (rMS,
in red) and a sample containing very-red objects (vrO, in green). The
vrO sample should mostly contain blends, since the binary fraction of
NGC~7078 is less than 4\% (Milone et al. 2012) and the photometry is
corrected for differential reddening. The velocity dispersion profiles
for the three PM subsamples (determined as described in
Section~\ref{ss:vdp} below) is shown as a function of the radial
distance in panel (c).

It is evident that the velocity dispersion is systematically higher
for redder stars. This can be explained by assuming that the redder
stars are affected by blending, and that this blending induces a
systematic component of PM scatter that is observed in addition to the
actual random motions of the stars in the cluster. To test this
hypothesis, one can repeat the analysis using only stars with smaller
values of \texttt{QFIT} and reduced $\chi^2$, and higher N$_R$. One
would expect this to reduce the difference in velocity dispersion
between the bMS, rMS and vrO stars.

Choosing the optimal cuts for the \texttt{QFIT}, reduced $\chi^2$ and
N$_R$ selections is a delicate issue. In principle, one can use an
iterative approach in which one gradually rejects stars using
increasingly stringent cuts, and then measures the velocity dispersion
for each progressive cut. Convergence in the measured velocity
dispersions might occur if at some cut level all blended sources have
been removed from the sample. In practice though, the selections
(especially those based on \texttt{QFIT} and reduced $\chi^2$)
preferentially remove fainter stars close to the cluster center, and
these stars have intrinsically a higher velocity dispersion than other
stars because of hydrostatic equilibrium and energy
equipartition. This means that every time a sharper cut is applied to
the sample, a counteracting bias is also applied to the surviving
sample of stars. Hence, there may be no convergence in the inferred
velocity dispersions as stronger cuts are applied.

For these reasons, the best way to choose cuts without introducing
excessive selection biases is to select stars of similar luminosity
(e.g., mass) and distance from the cluster center. As an example, we
selected NGC~7078 stars in an annulus between 60 and 70 arcsec from
the cluster center, and between $m_{\rm F606W}=20.3$ and 20.6 (about 1
mag below the turnoff)\footnote{If we were to use stars that are
  fainter or closer to the center, then low-number statistics would
  have become a problem.}. We chose fixed cuts for the reduced
$\chi^2$ ($<1.25$) and $N_R$ ($>0.85$), and applied various
\texttt{QFIT} cuts to show how this impacts the measured velocity
dispersion. The initial total number of selected stars is 510.  We
measured the stellar velocity dispersion $\sigma_\mu$ by keeping the
best 90, 85, 80, \dots, 10 percentile of the \texttt{QFIT} values in
both the $m_{\rm F606W}$ and $m_{\rm F814W}$ magnitudes.

Figure~\ref{f:qfit} shows the velocity dispersions thus derived for
different \texttt{QFIT} cuts. Stars with high \texttt{QFIT} values are
those with a higher chance of being affected by crowding/blending
effects. As expected, going from right to left in the figure, more
stringent \texttt{QFIT} cuts produce a smaller velocity dispersion for
the surviving sample. Below the 65th percentile, the velocity
dispersions converge and stay constant to within the errors. From this
we infer that a 65th percentile cut is able to remove most of the
blended objects from the sample. The small decrease of $\sigma_\mu$ as
a function of the \texttt{QFIT} below the 65th percentile is likely
due to the fact that even in the small magnitude and radial range
under consideration, progressively stronger cuts induce a kinematical
bias in the surviving sample as described above.

Based on these considerations, we reanalyzed the bMS, rMS and vrO
samples of NGC~7078 as in Figure~\ref{qfit_7078_pre}, but now
including only stars that have $\chi^2<1.25$, $N_R>0.85$, and that
survive a 50th-percentile \texttt{QFIT} cut. The results are shown in
Figure~\ref{qfit_7078_after}. The velocity dispersions of the three MS
components are now comparable. This supports the hypothesis that the
kinematical differences evident in Figure~\ref{qfit_7078_pre}c were
entirely due to blending-induced PM systematics. It also supports the
notion that the cuts applied here are necessary and sufficient for
this particular PM catalog. It should be noted that even for the bMS
stars, for which the observed color provides no indication of
blending, the velocity dispersion drops significantly after
application of the cuts.  Therefore, for dynamical studies of clusters
such as NGC~7078, it is critical to use the data-quality parameters
provided in our catalogs to compose an optimal sample. This is due to
the combination of several effects, including the fact that NGC 7078
is post core-collapse, the fact that it is relatively distant, the
fact that we only have a few epochs of data for this cluster, and the
fact that the data at different epochs were taken in different
filters. Faint stars and stars at small radii are most sensitive to
these effects, because they tend to be most affected by crowding.

\begin{figure}[!t]
\centering
\includegraphics[width=\columnwidth]{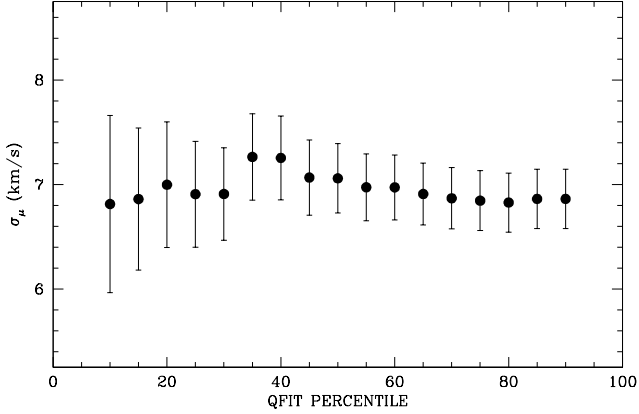}
\caption{Similar to Figure~\ref{f:qfit}, but for NGC~6752, a closer
  and less massive cluster and with a more homogeneous filter/epoch
  coverage than NGC~7078. In this case, cuts based on data-quality
  parameters do not significantly affect the inferred velocity
  dispersion.}
\label{f:qfit2}
\end{figure}

Other less-crowded clusters, or clusters for which a large number of
exposures is available (even when taken through a variety of different
filters), are far less affected by crowding/blending-induced PM
systematics. As an example, we repeated the same selection test shown
in Figure~\ref{f:qfit} on the PM catalog of NGC~6752. This cluster has
near 300 exposures of its core taken with nine different filters
spanning from F390W to F814W (see Table~A19), and it is much closer
than NGC~7078 (4.0 kpc instead of 10.4).  The test was performed on MS
stars with magnitudes $18.3<m_{\rm F606W}<18.6$ (about 1 mag below the
turnoff), and between 50 and 60 arcsecs from the cluster center.
Figure~\ref{f:qfit2} shows the results of this second test.  In this
case, the measured velocity dispersions all agree within the
uncertainties regardless of the applied \texttt{QFIT} cut.

\subsection{Caveats}
\label{ss:caveats}

In Section~\ref{s:sim} we showed that the techniques we developed to
measure high-precision PMs with \textit{HST} are highly reliable, and
our PM errors are a very good representation of the true errors. In
this Section we showed that we are able to identify and correct
systematic errors introduced by the use of non-optimal master frames
(Section~\ref{ss:go10775-cte}), by uncorrected geometric-distortion
and uncorrected CTE residuals in the single-exposure catalogs
(Sections~\ref{ss:second-order} and Section~\ref{ss:loc_cor}), and by
crowding and blending (Section~\ref{ss:qfit+}). We believe that with
the corrections described in these sections, our PM measurements are
as good as they can be, given the limitations of the data available in
the \textit{HST} archive (which are extremely heterogeneous, and were
rarely obtained for the purpose of astrometry). Nevertheless, several
more issues need to be kept in mind when using our PM catalogs.

\begin{figure*}[!t]
\centering
\includegraphics[width=\textwidth]{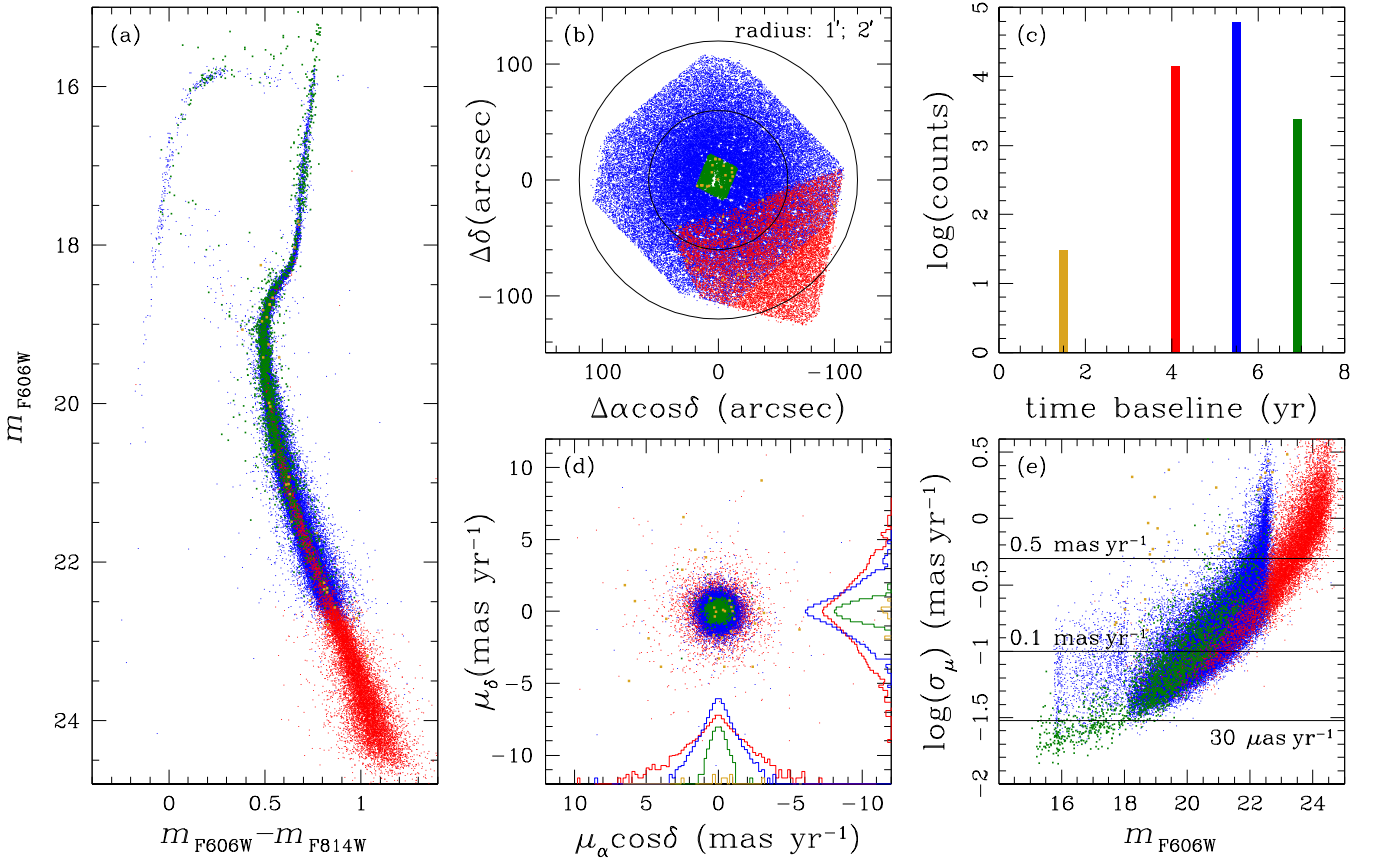}
\caption{Panel (a):\ The CMD of all stars in the NGC~7078 PM
  catalog. Panel (b):\ Stellar spatial distribution with respect to
  the cluster center, in arcsec. Two circles at $1^\prime$ and
  $2^\prime$ are shown for reference. Panel (c):\ Logarithmic
  histogram of the time baseline used to compute PMs. ``Counts'' refer
  to the number of stars.  Panel (d):\ PM diagram, together with
  histograms of the two PM components for each available time
  baseline. Panel (e):\ PM errors as a function of the $m_{\rm F606W}$
  magnitude. In each panel stars are color-coded according to their
  time baseline.}
\label{f:n7078intro}
\end{figure*}

Our catalogs are necessarily incomplete, and in different ways for
different clusters. For instance, in the most-crowded central regions
of each cluster we can measure PMs for only the brightest
stars. Specific dynamical studies, like the search for
intermediate-mass black holes, require a large number of stars with
high-quality PMs in the very proximity of the cluster center. This
does not mean that our PM catalogs are not suitable for these kind of
studies in general, but some clusters will be more appropriate than
others, and it depends on the crowding conditions of their centers. A
better way to measure high-quality PMs for a large number of stars in
the cluster centers would be to have used a master frame based on
higher-spatial-resolution ACS/HRC exposures (when available) rather
than on the ACS/WFC data, but this goes beyond the scope of the
present work.

We saw in Section~\ref{sss:res} that, at the faint limit, there might
be some non-negligible systematic errors in the measured velocity
dispersion.  Estimation of the velocity dispersion requires, in
essence, that the PM-measurement uncertainties be subtracted in
quadrature from the observed PM scatter. At the faint end, the PM
uncertainties become comparable to (or exceed) the velocity dispersion
of the cluster.  Very accurate estimates of the PM-measurement
uncertainties are then required in order to obtain reliable
results. In our somewhat idealized simulations of
Figure~\ref{f:GHout}, PM uncertainties can be fairly reliably
estimated at all magnitudes. But in practice, there is always the
potential of low-level unidentified systematic errors. The random
errors estimated by our algorithms are then at best only an
approximation to the true uncertainties. For this reason, it is
advised to restrict any dynamical analysis to stars for which the PM
uncertainties are well below the cluster velocity dispersion. This is
particularly important for studies of energy equipartition (e.g.,
Anderson \& van der Marel 2010; Trenti \& van der Marel 2014), which
rely on quantifying the increase of the velocity dispersion with
decreasing stellar mass. It is then particularly important to reliably
understand how the PM-measurement errors increase towards fainter
magnitudes.

Errors in our catalogs are not homogeneously distributed. Some
locations of the master frame will have larger time baselines and/or
more single-exposure measurements. Taking special care in selecting
high-quality PMs is therefore always crucial --and a delicate
matter--, regardless of the specific scientific needs (unless PMs are
only used to select a cleaned sample of cluster stars for photometric
studies).

\section{Proper-Motion Kinematics of NGC~7078}
\label{s:cat and res}

Our PM catalog for NGC~7078 is described in Appendix~B, and is
distributed electronically as part of this paper (Table~B2).

\subsection{Overview}
\label{ss:overv}

Figure~\ref{f:n7078intro} provides a visual overview of the
information contained in the catalog. Panel (a) shows the GO-10775
CMD, corrected for differential reddening, for all stars with a PM
measurement. We measured PMs from just above the HB region down to
$\sim 5$ magnitudes below the MS turn-off. The total spatial coverage
of the catalog is shown in Panel (b), with respect to the cluster
center.  We added two circles of radius $1^\prime$ and $2^\prime$ for
reference.  The histogram of the time baseline used to compute each
star's motion is shown in Panel (c). The Y axis of the plot is in
logarithmic units, to properly show all histogram bins using the same
scale.  Panel (d) shows the PM vector-points diagram, in units of
\masyr. Histograms of the PM distribution for each component of the
motion, and for each time-baseline bin, are also shown, again on a
logarithmic Y-axis scale. Finally, PM errors as a function of the
$m_{\rm F606W}$ magnitude are shown in Panel (e). In each panel, stars
are color-coded according to their time baseline. The figure gives an
immediate sense of the PM distribution, quality and respective
magnitude range in each location of the available FoV.  Proper-motion
errors are smaller than 30 $\mu$as$\,$yr$^{-1}$ for the brightest
stars with the longest time baseline, and increase up to $\sim 3$
\masyr\ for the faintest stars in the catalog.  There are 32 stars in
the catalog with a time baseline of less than 2 years. Although the PM
of these stars is poorly constrained, they are included in the catalog
for completeness.

\subsection{Comparison with other Published PM Catalogs}
\label{ss:comparison}

\begin{figure}[!t]
\centering
\includegraphics[width=\columnwidth]{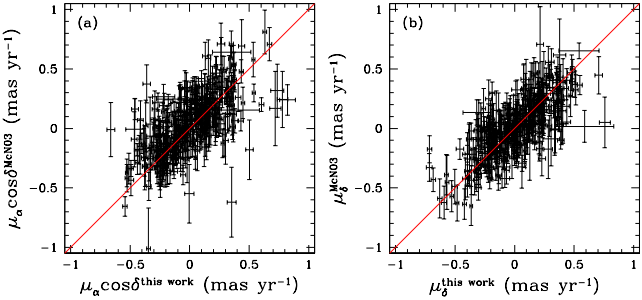}
\caption{PM component-to-component comparison between our catalog and
  that of McN03. Most of the scatter is due to the larger errorbars of
  the McN03 catalog.}
\label{f:mcnamara}
\end{figure}

\begin{figure}[!t]
\centering
\includegraphics[width=\columnwidth]{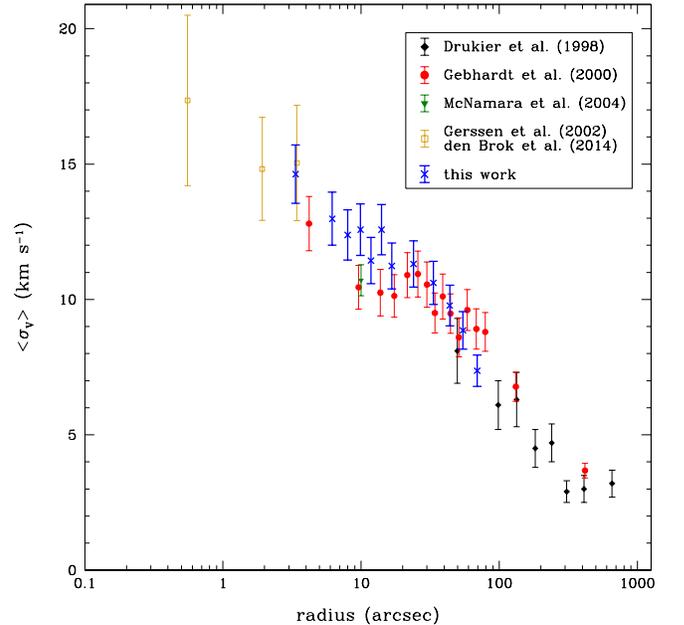}
\caption{Velocity-dispersion profiles in the literature (black, red
  and green points) and that of obtained with RGB stars in our catalog
  (in blue), assuming a cluster distance of 10.4 kpc (see Table~1).}
\label{f:los}
\end{figure}

\begin{figure*}[!t]
\centering
\includegraphics[height=7.5cm]{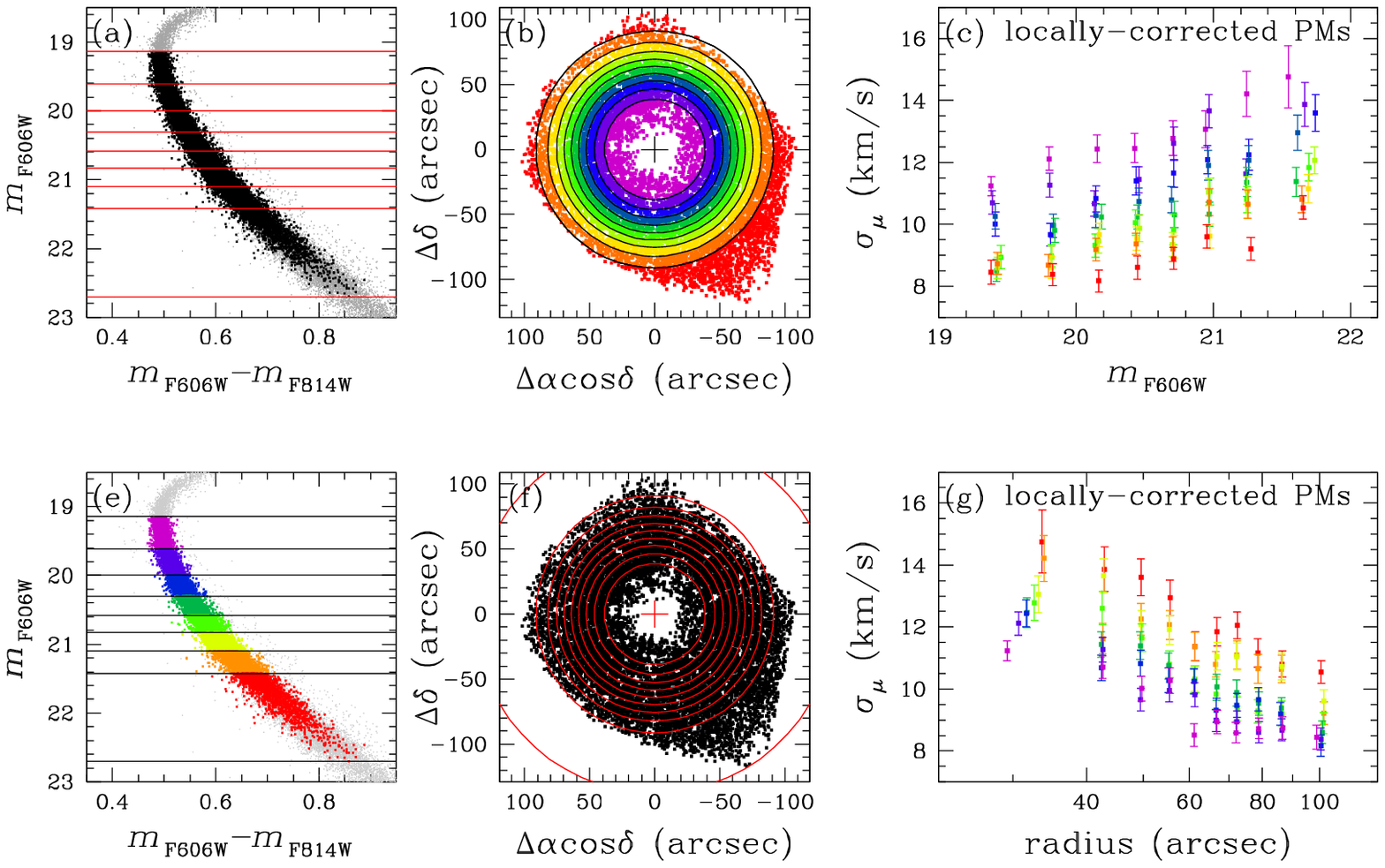}
\includegraphics[height=7.5cm]{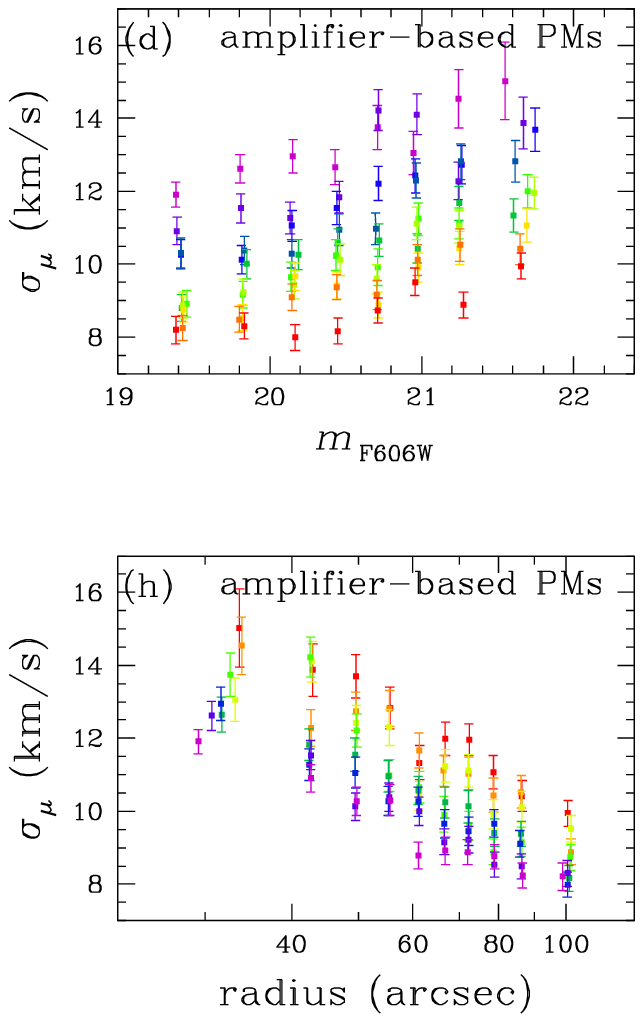}
\caption{Top panels show the velocity-dispersion profiles
  $\sigma_{\mu}$ of MS stars in different radial intervals as a
  function of the $m_{\rm F606W}$ magnitude. (a) The CMD of NGC~7078
  around its MS for all selected stars (grey) and for those with PM
  errors smaller than half the local velocity dispersion (black). The
  red lines define 8 magnitude bins with the same number of stars. (b)
  The spatial distribution of high-quality-PM MS stars. The black
  circles define 10 radial intervals with the same number of stars.
  Panels (c) and (d) show $\sigma_{\mu}$ values as a function of
  $m_{\rm F606W}$ for the locally-corrected and amplifier-based PMs,
  respectively. Points and errorbars are color-coded according to
  their radial interval.  Bottom panels show the $\sigma_{\mu}$
  profiles for the same stars in different magnitude intervals as a
  function of their distance from the cluster center. The magnitude
  and radial bins are the same as in the top panels. This time points
  are color-coded according to their magnitude bin.}
\label{f:sigv_r_mag}
\end{figure*}

The internal PM dispersion of NGC~7078, based on 210 bright RGB stars,
was first (barely) detected by Cudworth (1976), using photograhic
plates spanning over 70 years of time baseline.  The first
high-quality PM catalog of NGC~7078 was published by McNamara et al.\
(2003, hereafter McN03) for 1764 stars in the core, obtained with the
\textit{HST}'s WFPC2 detector. The authors computed proper motions
using 4 first-epoch and 12 second-epoch exposures taken $\sim 8$ years
apart. Their catalog includes positions in the
geometric-distortion-corrected frame of their first exposure, in
pixels, and proper motions as displacements in pixels over the
available time baseline\footnote{Note that McN03 quoted displacements
  are to be intended as first-epoch positions minus second-epoch
  positions and not vice-versa.}.

We applied general 6-parameter linear transformations to translate the
McN03 WFPC2 positions into our master frame, and cross-identified
their stars with the closest stars in our catalog within 2.5 pixels. A
total of 686 stars were found in common, 323 of which were used in
their internal PM analysis. Among them, there are 26 stars in
proximity of McN03 FoV edges that exhibit a significant offset in
position with respect to our master frame, probably due to
unaccounted-for WFPC2 geometric-distortion residuals. These 26 stars
are not included in the PM comparison.  Finally, we transformed quoted
McN03 PMs and errors into ($\mu_\alpha \cos\delta, \mu_\delta$)
units\footnote{In order to convert McN03 quoted PMs into
  \masyr\ units, we applied a scaling factor of 5.69 instead of their
  suggested 5.75 (1\% difference). This difference is due to the
  different pixel scale adopted for WFPC2:\ they use a 46
  \maspx\ scale value, while we directly measured their plate scale on
  our master frame to be 45.46 \maspx.}.

In Figure~\ref{f:mcnamara} we illustrate the comparison between our
PMs and those of McN03, with $\mu_{\alpha}\cos\delta$ in Panel (a),
and $\mu_{\delta}$ in Panel (b).  Most of the scatter is due to the
uncertainties of the McN03 PM measurements, which are significantly
larger than those in our catalog (our catalog is also superior in that
it has 40 times as many stars).  The fact the the points are mostly
aligned along the red line implies that our PMs are consistent with
those of McN03. The scatter of the points along the direction
perpendicular to the red line (which is not a fit to the data but just
the plane bisector) reveals a small but marginal (within the errors)
disagreement. The fact that our PMs are consistent with those of McN03
is a further indication of the reliability of our measurements.

\subsection{Velocity-Dispersion Profiles}
\label{ss:vdp}

In 1989, Peterson, Seitzer \& Cudworth (1989) first measured the
line-of-sight velocity-dispersion profile of NGC~7078, based on 120
spectra of individual stars in the centermost $4\farcm6$.  In
subsequent years, many authors have analyzed the line-of-sight
velocity-dispersion profile of NGC~7078 with various telescopes and
techniques. High signal-to-noise spectra are generally obtained from
only the brightest stars in a GC (i.e., RGB stars). In
Fig.~\ref{f:los} we therefore compare literature high-quality
velocity-dispersion profiles (in black, red, green and yellow for
Drukier et al.\ 1998, Gebhardt et al.\ 2000, McNamara et al.\ 2004,
and Gerssen et al. 2002/den Brok et al. 2014,
respectively\footnote{Gerssen et al. (2002) published individual star
  velocities and unparametrized profiles of $V$ and $\sigma_V$ of
  stars in the core of the NGC~7078, obtained with the \textit{HST}
  STIS spectrograph. den Brok et al. (2014) combined Gerssen et
  al. (2002) velocities with those of Gebhardt et al. (2000) to
  compute radial-binned profiles. Here we consider only the innermost
  three data points of den Brok et al. (2014) profile (their Fig.~1),
  which are mostly (if not completely) derived using the Gerssen et
  al. (2002) data.}) with that obtained from the stars in our catalog
brighter than the SGB (in blue), using 10.4 kpc as the cluster
distance (see Table~1). There is excellent agreement between our
values and those obtained from spectra, as expected for a cluster with
an isotropic velocity distribution and a correctly-estimated distance.
This once again confirms the high-quality and reliability of our PM
catalog.

Here and henceforth, velocity dispersions were estimated from the PM
catalog using the same method as in van der Marel \& Anderson (2010).
This corrects the observed scatter for the individual stellar PM
uncertainties. Unless stated otherwise, we quote the average
one-dimensional velocity dispersion $\sigma_{\mu}$, based on the
combined $x$ and $y$ PM measurements. Moreover, we adopted an
appropriate sample of high-quality PM stars for the analysis.

Satisfied that our PM measurements appear to be solid both internally
(see. Section~\ref{s:sim}) and externally
(see. Section~\ref{ss:comparison}), we proceed by analyzing more in
detail the MS velocity dispersion profile of NGC~7078.  In order to
select the best-measured stars we proceeded as follows.  First of all,
we selected likely cluster members on the basis of their positions on
the CMD. In addition, we kept only those stars with the
\texttt{QFIT}-percentile values below 50\%, reduced $\chi^2$ values
below 1.25, and $N_R>0.85$, that proved to remove crowding/blending as
a source of systematic effects (see Section~\ref{ss:qfit+}).

Then, we adopted an iterative procedure that further identifies and
rejects stars for which the measurement error is larger than F times
the local $\sigma_\mu$, where F is a certain threshold value, and the
local $\sigma_\mu$ is computed for each star using the 100 stars
closest in radial distance and magnitude to the target star. We
iterated this procedure until we obtained convergence of the
dispersion profiles. We found that F=0.5 provides the best compromise
between accuracy and sample size.  After these procedures were
applied, there were no remaining candidate field stars with
highly-discrepant ($>5\sigma$) PMs. Our final sample consists of
18$\,$136 stars, of which 15$\,$456 are MS stars with $m_{\rm F606W}$
magnitudes between 19.15 (which here defines the turnoff) and 22.7,
and between $11\farcs6$ and $136\farcs6$ from the cluster center.

We divided this sample into 8 magnitude bins each having approximately
the same number of stars, and into 10 radial intervals, again each
having approximately the same number of stars. These subdivisions
define 80 regions in the magnitude-radius space, with each containing
on average 193 stars. Obviously, the innermost radial intervals have
fewer faint stars on average than the outermost ones, because of
crowding-driven incompleteness. The number of stars in each region
ranges from 72 to 342.  For each region we computed the velocity
dispersion $\sigma_{\mu}$ and its error for both amplifier-based and
locally-corrected PMs.

Figure~\ref{f:sigv_r_mag} collects the results of the
velocity-dispersion analysis. We show the results in two ways:\ (1)
$\sigma_{\mu}$ as a function of the magnitude for different radial
intervals (top panels); and (2) $\sigma_{\mu}$ as a function of the
radial distance for different magnitude bins (bottom panels).  Panels
(a) and (e) show the CMD of selected stars around the MS of
NGC~7078. Horizontal lines delimit the magnitude bins. Panels (b) and
(f) show the spatial distribution of the selected stars. The circles
define the radial intervals. Panels (c) and (d) show the
$\sigma_{\mu}$ profiles as a function of the $m_{\rm F606W}$ magnitude
for locally-corrected and amplifier-based PMs, respectively. Points
and errorbars are color-coded according to their radial
intervals. Panels (g) and (h) show the $\sigma_{\mu}$ profiles as a
function of the radial distance from the cluster center, with points
and errorbars color-coded according to their magnitude bin.

Figure~\ref{f:sigv_r_mag} reveals a complex behavior of $\sigma_{\mu}$
as a function of both magnitude and radius. Bright, more massive stars
are kinematically colder than faint, less massive stars at all radii.
This behavior is a direct consequence of the effects of energy
equipartition.  Moreover, stars at larger radii are colder than stars
closer to the cluster center for each magnitude bin, which is a direct
consequence of hydrostatic equilibrium.  There is little
(statistically insignificant) difference between amplifier-based and
locally-corrected velocity-dispersion profiles, with the latter being
on average only slightly lower than the former.
Figure~\ref{f:sigv_r_mag} also tells us that LOS velocity dispersions
quoted in the literature based on RGB stars are to be considered as
lower limits.  The vast majority of stars are less massive than RGB
stars and move faster.

\subsection{Anisotropy}
\label{ss:iso}

\begin{figure}[!t]
\centering
\includegraphics[width=\columnwidth]{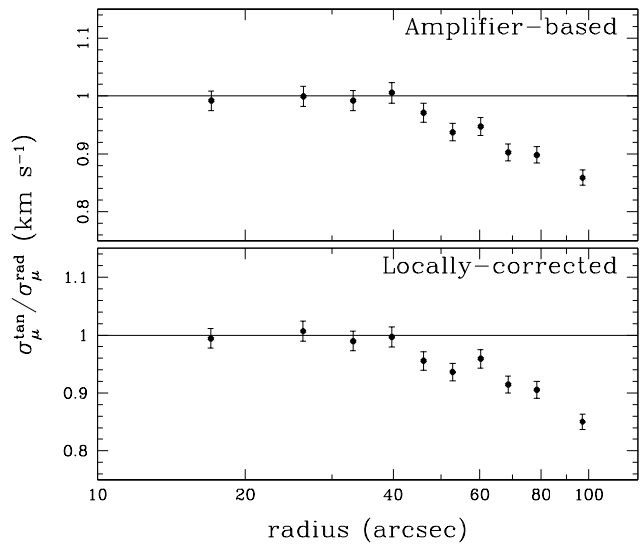}
\caption{Anisotropy in the proper-motion Velocity-dispersion as a
  function of the radial distance. Amplifier-based PMs are in the top
  panel, while locally-corrected PMs are in the bottom one.}
\label{f:iso}
\end{figure}

A direct estimate of the degree of velocity anisotropy of the cluster
is obtained by studying the ratio between tangential and radial
proper-motion dispersions as a function of the radial distance. We
measured the velocity dispersion in each direction, using the full
sample of 15$\,$546 high-quality stars, in order to map the
velocity-anisotropy profile.  Moreover, velocity-dispersions are
computed using both amplifier-based and locally-corrected PM values.

The results are summarized in Fig.~\ref{f:iso}, using amplifier-based
PMs in the top panel, and locally-corrected PMs in the bottom panel.
As before, there is only a small difference between the two ways of
computing PMs. The velocity distribution of NGC~7078 in the central
$\sim 45^{\prime\prime}$ (comparable to the half-light radius $r_{\rm
  h} = 60^{\prime\prime}$; Harris 1996, 2010 edition) is close to
isotropic. This is consistent with what might be expected given the
short two-body relaxation time of NGC~7078.  There is evidence of
motions that are preferably oriented radially rather than tangentially
at distances greater than $45^{\prime\prime}$.

\section{Conclusions and Discussion}
\label{s:conclusions}

Our understanding of the internal kinematics of globular clusters is
based largely on studies of modest samples of stellar LOS
velocities. PM studies with \textit{HST} have the potential to
significantly advance our understanding, by extending the measurements
to two or three-dimensional velocities, lower stellar masses, and
larger sample sizes. We have presented here the first study of
\textit{HST} PMs for a large sample of globular clusters, based on
heterogenous data assembled from the \textit{HST} Archive. This first
paper in a series has focused on the data-reduction procedures, data
quality, and new kinematical quantities inferred for NGC~7078
(M~15). Subsequent papers will explore a range of applications,
including the many scientific topics of interest highlighted in
Section~\ref{sec:1}.

We identified clusters in the \textit{HST} Archive with suitable
exposures spread over multiple epochs, resulting in a sample of 22
clusters. For these clusters we analyzed a total of 2510 different
exposures, obtained over the past decade with the ACS/WFC, ACS/HRC,
and WFC3/UVIS instruments. We created photometric, astrometric, and PM
catalogs from these data. For this we used and extended the software
developed in the context of our previous globular-cluster studies, and
in the context of our HSTPROMO collaboration. The data reduction also
folded in and improved many of the single-epoch catalogs previously
obtained in the context of the \textit{HST} Globular Cluster Treasury
Program GO-10775. Significant effort was invested to develop a
reduction procedure that can be used in a homogeneous way for all
clusters to obtain high-quality PM measurements, despite the very
heterogeneous nature of the Archival data (which were not generally
obtained for high-precision astrometry).

We demonstrated the quality of the PM measurements through extensive
Monte-Carlo simulations for single stars and comprehesive data sets.
These show that input PM distributions and dispersions can be reliably
recovered for realistic observational setups and random errors. In
practice we also have to contend with various sources of systematic
errors. We have discussed in detail the impact on the PM measurements
due to charge-transfer-inefficiency effects, uncorrected
geometric-distortion residuals, and crowding and blending. We have
developed and discussed techniques to remove systematic PM errors due
to these effects to the extent possible. We have presented various
tests that have shown that with these corrections, our PM data quality
is excellent.

From our analyses we were able to measure the PM of over 1.3 million
stars in the central regions of the target clusters, with a median
number of $\sim$60,000 stars per cluster. Most of the PM catalogs will
be disseminated as part of future papers in this series. Here we focus
on, and release, the catalog for NGC~7078, which consists of 77,837
stars. The number of stars with measured velocities is $\sim$ 40 times
larger than in the best catalogs of NGC~7078 PMs and LOS velocities
previously available (Gebhardt \etal 2000; McNamara \etal 2004). Our
measurements are consistent with these previous catalogs in the areas
of overlap. For the PMs we demonstrated this on a star-by-star basis,
and for the LOS velocities we demonstrated this by comparison of the
velocity-dispersion profiles for bright stars under the assumption of
isotropy.

We present a preliminary analysis of the PM kinematics of NGC~7078
that demonstrates the potential of our data. The large number of
measurements allows detailed studies of the velocity dispersion as a
function of radius, as a function of stellar magnitude (or mass) along
the main sequence, and as a function of direction in the plane of the
sky (radial or tangential). The velocity dispersion increases towards
the center as expected from hydrostatic equilibrium, and it increases
towards lower masses as expected from energy equipartition. The
velocity dispersion is isotopic near the center, as expected from
two-body relaxation. There is evidence of motions that are preferably
oriented radially rather than tangentially outside the half-light
radius.

Although this work represents the most detailed study of globular
cluster PMs to date, there continues to be room for significant
improvement in the observations and measurements. New observations of
the cores of globular clusters are taken each \textit{HST} observing
cycle. This makes it possible to construct PM catalogs for more
clusters, and to extend the time baselines (and reduce the
uncertainties) for clusters with existing PM catalogs. Also, the
measurements presented here were not optimized to deal with very
crowded fields. Some clusters have deep ACS/HRC observations of their
cores. These have higher spatial resolution than the ACS/WFC
observations that were used to build the GO-10775 master frames used
for our analysis. Moreover, these ACS/HRC observations are often taken
in bluer filters, will yield less crowding (since the brightest stars
tend to be {\it red} giants). New photometric reduction techniques for
the WFC3 detector (Anderson et al., in preparation) can measure
stellar positions and fluxes after prior subtraction of surrounding
neighbors (deblending, cf.~Anderson et al. 2008 for ACS/WFC). Master
frames based on the ACS/HRC observations, combined with data-reduction
techniques that explictly deblend, have the potential to yield
catalogs with more stars with more accurately measured PMs and better
characterized errors. This is especially relevant close to the cluster
centers, which are dominated by crowding/blending issues. These
central regions are crucial for studies of intermediate-mass black
holes in globular clusters.

\acknowledgments \noindent \textbf{Acknowledgments.}  We thank the
anonymous referee for comments that helped improve the presentation of
our results.  Support for this work was provided by grants for
\textit{HST} programs GO-9453, GO-10335, GO-10401, GO-11664, GO-11801,
GO-12274, AR-12656 and AR-12845, provided by the Space Telescope
Science Institute, which is operated by AURA, Inc., under NASA
contract NAS 5-26555.  FRF and DM acknoledge the support from the
Cosmic-Lab project (web site:\ \url{http://www.cosmic-lab.eu}) funded
by the European Research Council (under contract ERC-2010-AdG-267675).
PB acknowledges financial support from the International Max Planck
Research School for Astronomy and Cosmic Physics (IMPRS) at the
University of Heidelberg and travel support from the Heidelberg
Graduate School for Fundamental Physics (HGSFP).  The authors are
grateful to Laura Ferrarese, Rodrigo Ibata and Carlton Pryor for help
with the proposal preparation for some of these projects, and to
Michele Trenti for helpful discussions.

\appendix

\section{Appendix A. Complete list of the data sets used for each cluster}
\label{a1}

Tables A1 through A22 provide the full list of used exposures for each
cluster, ordered by program number, camera and filter. These tables
are available only in the electronic version of the article.

\vskip -5mm
\begin{table*}[th!]
\begin{center}
\label{tab:A1}
\scriptsize{
\begin{tabular}{cccccc}
\multicolumn{6}{c}{\textsc{Table A1}}\\
\multicolumn{6}{c}{\textsc{List of observations of NGC~104}}\\
\hline\hline
GO&PI&Intr./Cam.&Filter&$N\times$Exp. time& Epoch\\
\hline
 9019& Bohlin & ACS/HRC & F330W & 18{\x}66s                                & Apr 2002\\
     &        &         & F435W & 2$\times$5s, 2{\x}20s, 17$\times$60s, 2$\times300$s&\\          
     &        &         & F475W &10$\times$60s                             &         \\          
     &        &         & F555W &14$\times$60s                             &         \\           
     &        &         & F606W &10$\times$60s                             &         \\          
     &        &         & F625W &10$\times$60s                             &         \\           
     &        &         & F775W &13$\times$60s                             &         \\           
     &        &     & F814W &2$\times$5s, 2$\times$20s, 14$\times$60s, 2$\times$300s\\           
     &        &         & F850LP&10$\times$60s                            &         \\
\hline                                                                                            
 9028&Meurer  & ACS/HRC & F475W & 40$\times$60s                            & Apr 2002\\           
     &        & ACS/WFC & F475W & 20$\times$60s                            &         \\           
\hline                                                                                            
 9281&Grindlay& ACS/WFC & F435W & 1$\times$10s, 6$\times$100s, 3$\times$115& Sep-Oct 2002\\       
     &        &         & F625W & 2$\times$10s, 20$\times$65s              &             \\       
     &        &         & F658N & 6$\times$350s, 6$\times$370s, 8$\times$390s&           \\       
\hline                                                                                            
 9575& Sparks  & ACS/WFC & F475W & 3{\x}700s                              & Apr 2002\\            
     &         &         & F775W & 1{\x}578s, 5{\x}700s                   &         \\            
     &         &         & F850LP& 6{\x}700s                              &         \\            
\hline                                                                                            
 9443& King   & ACS/HRC &F330W & 1{\x}350s                                & Jul 2002\\            
     &        &        & F435W & 1{\x}350s                               & \\                     
     &        &        & F475W & 20{\x}60s, 1{\x}350s          &         \\                       
     &        &        & F555W & 1{\x}350s                     &         \\                       
     &        &        & F606W & 1{\x}350s                     &         \\                       
     &        &        & F814W & 1{\x}350s                     &         \\                       
     &        & ACS/WFC &F435W & 1{\x}150s                                &         \\            
     &        &        & F475W & 5{\x}60s, 1{\x}150s                         &         \\         
     &        &        & F555W & 1{\x}150s                                &         \\            
     &        &        & F606W & 1{\x}100s                                &         \\            
     &        &        & F814W & 1{\x}150s                                &         \\            
\hline                                                                                            
 9453& Brown  & ACS/WFC & F606W & 1{\x}6s, 1{\x}70s                      & Jul 2002\\             
     &        &         & F814W & 1{\x}5s, 1{\x}72s                      &         \\             
\hline                                                                                         
 9662& Gilliland& ACS/HRC & F606W &2{\x}1s                                & Sep 2002\\
\hline
 9503 & Nagar  & ACS/WFC & F475W & 1{\x}60s                                & Jan 2003\\
      &        &         & F658N & 1{\x}340s                               &         \\
\hline
10055& Biretta & ACS/HRC & F330W & 2{\x}40s, 6{\x}150s                     & Feb 2004\\               
     &         &         & F435W & 2{\x}20s, 6{\x}60s                      &\\                        
     &         &         & F606W & 2{\x}10s                       &         \\                        
     &         &         & F775W & 2{\x}10s                       &         \\                        
\hline                                                                                                
10375& Mack    & ACS/HRC & F435W & 4{\x}60s                                & 2004--2005\\             
     &         &         & F475W & 4{\x}60s                                &         \\               
     &         &         & F555W & 4{\x}60s                                &         \\               
     &         &         & F606W & 4{\x}60s                                &         \\               
      &         &         & F625W & 4{\x}60s                                &         \\              
     &         &         & F775W & 4{\x}60s                                &         \\               
      &         &         & F814W & 4{\x}60s                                &         \\              
      &         &         & F850LP& 4{\x}60s                                &         \\              
 \hline                                                                                               
 10737& Mack    & ACS/HRC & F330W & 2{\x}66s                                &2005--2006\\             
      &         &         & F435W & 6{\x}60s                               &\\                        
      &         &         & F475W & 6{\x}60s                                &         \\              
      &         &         & F555W & 6{\x}60s                                &         \\              
      &         &         & F606W & 6{\x}60s                                &         \\              
      &         &         & F625W & 6{\x}60s                                &         \\              
      &         &         & F775W & 6{\x}60s                                &         \\              
      &         &         & F814W & 6{\x}60s                                &         \\              
      &         &         & F850LP& 6{\x}60s                                &         \\              
 \hline                                                                                               
 10775&Sarajedini& ACS/WFC& F606W & 1{\x}3s, 4{\x}50s                        & Mar 2006\\             
      &          &        & F814W & 1{\x}3s, 4{\x}50s                        &         \\             
 \hline                                                                                               
 11664& Brown   &WFC3/UVIS& F390W & 2{\x}10s, 2{\x}348s, 2{\x}940s          & Sep 2010\\              
      &         &         & F555W & 1{\x} 1s, 1{\x}30s, 2{\x}665s           &         \\              
      &         &         & F814W & 1{\x}30s, 2{\x}565s                     &         \\              
 \hline                                                                                               
 11729& Holtzman&WFC3/UVIS& F336W & 1{\x}30s, 2{\x}580s                     & Sep 2010\\              
      &         &         & F390W & 1{\x}10s                                &\\                       
      &         &         & F467M & 1{\x}40s, 2{\x}450s                     &\\                       
 \hline                                                                                               
 12116&Dalcanton& ACS/WFC & F475W & 2{\x}7s                                 &Jul 2012\\               
\hline\hline
\end{tabular}}
\end{center}
\end{table*}
\vskip -15mm

\begin{table*}[th!]
\scriptsize{
\begin{center}
\label{tab:A2}
\begin{tabular}{cccccc}
\multicolumn{6}{c}{\textsc{Table A2}}\\
\multicolumn{6}{c}{\textsc{List of observations of NGC~288}}\\
\hline\hline
GO&PI&Intr./Cam.&Filter&$N\times$Exp. time& Epoch\\
\hline
10120& Anderson& ACS/WFC & F435W & 1{\x}60s, 2{\x}340s                    & Sep 2004\\
     &         &         & F625W & 1{\x}10s, 1{\x}75s, 1{\x}115s, 1{\x}120s&        \\
     &         &         & F658N & 2{\x}340, 2{\x}540x                     &        \\
\hline
10775&Sarajedini& ACS/WFC & F606W & 2{\x}10s, 8{\x}130s                    & Jul 2006\\
     &          &         & F814W & 2{\x}10s, 8{\x}150s                    &         \\
\hline
12193& Lee      &WFC3/UVIS& F467M & 1{\x}964s, 1{\x}1055s                  &Nov 2010\\
\hline\hline
\end{tabular}
\end{center}}
\end{table*}
\vskip -5mm

\begin{table*}[th!]
\scriptsize{
\begin{center}
\label{tab:A3}
\begin{tabular}{cccccc}
\multicolumn{6}{c}{\textsc{Table A3}}\\
\multicolumn{6}{c}{\textsc{List of observations of NGC~362}}\\
\hline\hline
GO&PI&Intr./Cam.&Filter&$N\times$Exp. time& Epoch\\
\hline
10005& Lewin& ACS/WFC & F435W & 4{\x}340s                               & Dec 2003\\
     &      &         & F625W & 2{\x}110s, 2{\x}120s                    &         \\
     &      &         & F658N & 2{\x}440s, 2{\x}500s                    &         \\
\hline
10401& Chandar & ACS/HRC & F435W & 17{\x}85s                            & Dec 2004\\
\hline
10615&Anderson & ACS/WFC & F435W & 5{\x}70s, 30{\x}340s                 & Sep 2005\\
\hline
10775&Sarajedini& ACS/WFC & F606W & 1{\x}10s, 4{\x}150s                 & Jun 2006\\
     &          &         & F814W & 1{\x}10s, 4{\x}170s                 &         \\
\hline\hline
\end{tabular}
\end{center}}
\end{table*}
\vskip -5mm

\begin{table*}[th!]
\scriptsize{
\begin{center}
\label{tab:A4}
\begin{tabular}{cccccc}
\multicolumn{6}{c}{\textsc{Table A4}}\\
\multicolumn{6}{c}{\textsc{List of observations of NGC~1851}}\\
\hline\hline
GO&PI&Intr./Cam.&Filter&$N\times$Exp. time& Epoch\\
\hline
10458& Biretta  & ACS/HRC & F555W & 12{\x}10s, 4{\x}100s, 2{\x}500s       & Aug2005\\
\hline
10775&Sarajedini& ACS/WFC & F606W & 1{\x}20s, 5{\x}350s                 & May 2006\\
     &          &         & F814W & 1{\x}20s, 5{\x}350s                 &         \\
\hline
12311&  Piotto  &WFC3/UVIS& F814W & 7{\x}100s                 & 2010--2011\\
\hline\hline
\end{tabular}
\end{center}}
\end{table*}
\vskip -5mm

\begin{table*}[th!]
\scriptsize{
\begin{center}
\label{tab:A5}
\begin{tabular}{cccccc}
\multicolumn{6}{c}{\textsc{Table A5}}\\
\multicolumn{6}{c}{\textsc{List of observations of NGC~2808}}\\
\hline\hline
GO&PI&Intr./Cam.&Filter&$N\times$Exp. time& Epoch\\
\hline
9899& Piotto  & ACS/WFC & F475W & 6{\x}340s                           & May 2004\\
\hline
10335& Ford   & ACS/HRC & F435W & 24{\x}135s                          & Jun 2006\\
     &        &         & F555W &  4{\x}50s                           &         \\
\hline
10775&Sarajedini& ACS/WFC & F606W & 1{\x}23s, 4{\x}360s                 & Mar 2006\\
     &          &         & F814W & 1{\x}23s, 4{\x}370s                 &         \\
\hline
10922& Piotto    & ACS/WFC & F475W & 1{\x}20s, 2{\x}350s, 2{\x}360s    & Aug--Nov 2006\\
     &           &         & F814W & 1{\x}10s, 3{\x}350s, 4{\x}360s    &  \\
\hline
11801& Ford       &WFC3/UVIS& F438W & 7{\x}20s, 9{\x}160s              & Dec 2009\\
\hline\hline
\end{tabular}
\end{center}}
\end{table*}
\vskip -15mm

\begin{table*}[th!]
\scriptsize{
\begin{center}
\label{tab:A6}
\begin{tabular}{cccccc}
\multicolumn{6}{c}{\textsc{Table A6}}\\
\multicolumn{6}{c}{\textsc{List of observations of NGC~5139}}\\
\hline\hline
GO&PI&Intr./Cam.&Filter&$N\times$Exp. time& Epoch\\
\hline
9442 & Cool   & ACS/WFC & F435W & 9{\x}12s, 27{\x}340s                   & Jun 2002\\
     &        &         & F625W & 8{\x}8s, 27{\x}340s                   &         \\
     &        &         & F658N & 36{\x}440s                            &         \\
\hline
10252& Anderson & ACS/WFC & F606W & 1{\x}15s, 5{\x}340s                 & Dec 2004\\
     &          &         & F814W & 1{\x}15s, 5{\x}340s                 &         \\
\hline
10775&Sarajedini& ACS/WFC & F606W & 1{\x}4s, 4{\x}80s                 & Mar--Jul 2006\\
     &          &         & F814W & 1{\x}4s, 4{\x}80s                 &         \\
\hline
11452& Kim Quijano   &WFC3/UVIS& F336W & 9{\x}350s                         & Jul 2009\\
     &               &         & F606W & 1{\x}35s                         &\\
     &               &         & F814W & 1{\x}35s                         &\\
\hline
11911& Sabbi         &WFC3/UVIS& F336W & 19{\x}350s                         & Jan--Jul 2010\\
     &               &         & F390W & 15{\x}350s                         &\\
     &               &         & F438W & 25{\x}350s                         &\\
     &               &         & F555W & 18{\x}40s                         &\\
     &               &         & F606W & 22{\x}40s                         &\\
     &               &         & F775W & 16{\x}350s                         &\\
     &               &         & F814W & 24{\x}40s                         &\\
     &               &         & F850LP& 17{\x}60s                         &\\
\hline
12094& Petro         &WFC3/UVIS& F606W & 9{\x}40s                          &Apr 2010\\
\hline
12339& Sabbi         &WFC3/UVIS& F336W & 9{\x}350s                         & Feb--Mar 2011\\
     &               &         & F438W & 9{\x}350s                         &\\
     &               &         & F555W & 9{\x}40s                         &\\
     &               &         & F606W & 9{\x}40s                         &\\
     &               &         & F814W & 9{\x}40s                         &\\
     &               &         & F850LP& 9{\x}60s                         &\\
\hline
12353&Kozhurina-Platais&WFC3/UVIS& F606W & 11{\x}40s                      & 2010--2011\\
\hline
12694& Long         &WFC3/UVIS& F467M & 3{\x}400s, 3{\x}450s                      & Feb--Apr 2012\\
\hline
12700& Riess        &WFC3/UVIS& F775W & 2{\x}450s                            & Jun 2012\\
\hline
12714&Kozhurina-Platais&WFC3/UVIS& F606W & 4{\x}40s                                 & Mar 2012\\
\hline
13100&Kozhurina-Platais&WFC3/UVIS& F606W & 6{\x}48s                         & 2012--2013\\
\hline\hline
\end{tabular}
\end{center}}
\end{table*}
\vskip -15mm

\begin{table*}[th!]
\scriptsize{
\begin{center}
\label{tab:A7}
\begin{tabular}{cccccc}
\multicolumn{6}{c}{\textsc{Table A7}}\\
\multicolumn{6}{c}{\textsc{List of observations of NGC~5904}}\\
\hline\hline
GO&PI&Intr./Cam.&Filter&$N\times$Exp. time& Epoch\\
\hline
10120 & Anderson & ACS/WFC & F435W & 1{\x}70s, 2{\x}340s              & Aug 2004\\
      &          &         & F625W & 1{\x}10s, 1{\x}70s, 2{\x}110s    &\\
      &          &         & F658N & 2{\x}340s, 2{\x}540s             &\\
\hline
10615 & Anderson & ACS/WFC & F435W & 1{\x}130s, 3{\x}215s, 25{\x}240s & Feb 2006\\
\hline
10775&Sarajedini& ACS/WFC & F606W & 1{\x}7s, 4{\x}140s                 & Mar 2006\\
     &          &         & F814W & 1{\x}7s, 4{\x}140s                 &         \\
\hline
11615& Ferraro  &WFC3/UVIS& F390W & 6{\x}500s                          & Jul 2010\\
     &          &         & F606W & 4{\x}150s                          &\\
     &          &         & F814W & 4{\x}150s                          &\\
\hline\hline
\end{tabular}
\end{center}}
\end{table*}
\vskip -15mm

\begin{table*}[th!]
\scriptsize{
\begin{center}
\label{tab:A8}
\begin{tabular}{cccccc}
\multicolumn{6}{c}{\textsc{Table A8}}\\
\multicolumn{6}{c}{\textsc{List of observations of NGC~5927}}\\
\hline\hline
GO&PI&Intr./Cam.&Filter&$N\times$Exp. time& Epoch\\
\hline
9453 & Brown   & ACS/WFC & F606W & 1{\x}2s, 1{\x}30s, 1{\x}500s              & Aug 2002\\
     &         &         & F814W & 1{\x}15s, 1{\x}340s                       &\\
\hline
10775&Sarajedini& ACS/WFC & F606W & 1{\x}30s, 5{\x}350s                 & Apr 2006\\
     &          &         & F814W & 1{\x}25s, 5{\x}360s                 &         \\
\hline
11664& Brown    &WFC3/UVIS& F390W & 2{\x}40s, 2{\x}348s, 2{\x}800s      & Aug 2010\\
     &          &         & F555W & 1{\x}50s, 2{\x}665s                 &\\
     &          &         & F814W & 1{\x}50s, 2{\x}455s                 &\\
\hline
11729& Holtzman &WFC3/UVIS& F336W & 2{\x}475s                 &Sep 2010\\
     &          &         & F467M & 2{\x}365s                 &\\
\hline\hline
\end{tabular}
\end{center}}
\end{table*}
\vskip -15mm

\begin{table*}[th!]
\scriptsize{
\begin{center}
\label{tab:A10}
\begin{tabular}{cccccc}
\multicolumn{6}{c}{\textsc{Table A9}}\\
\multicolumn{6}{c}{\textsc{List of observations of NGC~6266}}\\
\hline\hline
GO&PI&Intr./Cam.&Filter&$N\times$Exp. time& Epoch\\
\hline
10120& Anderson & ACS/WFC & F435W & 1{\x}200s, 2{\x}340s              & Aug 2004\\
     &          &         & F625W & 1{\x}30s, 1{\x}120s, 3{\x}340s    &\\
     &          &         & F658N & 1{\x}340s, 3{\x}350s, 3{\x}365s, 3{\x}375s&\\
\hline
11609& Chanam\'{e}  &WFC3/UVIS& F390W & 4{\x}35s, 5{\x}393s, 5{\x}421s    & Jun 2010\\
\hline\hline
\end{tabular}
\end{center}}
\end{table*}
\vskip -15mm

\begin{table*}[th!]
\scriptsize{
\begin{center}
\label{tab:A11}
\begin{tabular}{cccccc}
\multicolumn{6}{c}{\textsc{Table A10}}\\
\multicolumn{6}{c}{\textsc{List of observations of NGC~6341}}\\
\hline\hline
GO&PI&Intr./Cam.&Filter&$N\times$Exp. time& Epoch\\
\hline
9453 & Brown   & ACS/WFC & F606W & 1{\x}5s, 1{\x}90s                   & Aug 2002\\
     &         &         & F814W & 1{\x}6s, 1{\x}100s                       &\\
\hline
10120& Anderson& ACS/WFC & F435W & 1{\x}90s, 2{\x}340s                 &Aug 2004\\
     &         &         & F625W & 1{\x}10s, 3{\x}120s                 &\\
     &         &         & F658N & 2{\x}350s, 2{\x}555s                 &\\
\hline
10335& Ford    & ACS/HRC & F435W & 36{\x}85s                           & 2004--2006\\
     &         &         & F435W & 15{\x}40s                           &\\
\hline
10443& Biretta & ACS/HRC & F330W & 8{\x}100s, 4{\x}500s               &Feb 2005\\
     &         &         & F555W & 78{\x}10s, 33{\x}100s, 18{\x}500s&\\
     &         &         & F606W & 14{\x}357                          &\\
\hline
10455& Biretta & ACS/HRC & F555W & 12{\x}10s, 41{\x}100s, 2{\x}500s   &Feb 2005\\
\hline
10505& Gallart& ACS/WFC& F475W & 1{\x}3s, 1{\x}20s, 1{\x}40s            & Jan 2006\\
     &        &        & F814W & 1{\x}7s, 1{\x}10s, 1{\x}20s            &\\
\hline
10615&Anderson& ACS/WFC & F435W & 30{\x}340s                            & Jan 2006\\
\hline
10775&Sarajedini& ACS/WFC & F606W & 1{\x}7s, 5{\x}140s                 & Apr 2006\\
     &          &         & F814W & 1{\x}7s, 5{\x}150s                 &         \\
\hline
11664& Brown    &WFC3/UVIS& F390W & 2{\x}348s, 2{\x}795s      & Oct 2009\\
     &          &         & F555W & 1{\x}30s, 2{\x}665s                 &\\
     &          &         & F814W & 1{\x}30s, 2{\x}415s                 &\\
\hline
11801& Ford     &WFC3/UVIS& F438W & 6{\x}10s, 11{\x}110s      &Nov 2009\\
\hline
11729& Holtzman &WFC3/UVIS& F336W & 1{\x}30s, 2{\x}425s                 &Oct 2010\\
     &          &         & F390W & 1{\x}10s                            &\\
     &          &         & F467M & 1{\x}40s, 2{\x}350s                 &\\
\hline\hline
\end{tabular}
\end{center}}
\end{table*}
\vskip -15mm

\begin{table*}[th!]
\scriptsize{
\begin{center}
\label{tab:A12}
\begin{tabular}{cccccc}
\multicolumn{6}{c}{\textsc{Table A11}}\\
\multicolumn{6}{c}{\textsc{List of observations of NGC~6362}}\\
\hline\hline
GO&PI&Intr./Cam.&Filter&$N\times$Exp. time& Epoch\\
\hline
10775&Sarajedini& ACS/WFC & F606W & 1{\x}10s, 4{\x}130s         & May 2006\\
     &          &         & F814W & 1{\x}10s, 4{\x}150s         &         \\
\hline
12008& Kong     &WFC3/UVIS& F336W & 1{\x}368s, 5{\x}450s        & Aug 2010\\
\hline\hline
\end{tabular}
\end{center}}
\end{table*}
\vskip -15mm

\clearpage
\clearpage

\begin{table*}[th!]
\scriptsize{
\begin{center}
\label{tab:A13}
\begin{tabular}{cccccc}
\multicolumn{6}{c}{\textsc{Table A12}}\\
\multicolumn{6}{c}{\textsc{List of observations of NGC~6388}}\\
\hline\hline
GO&PI&Intr./Cam.&Filter&$N\times$Exp. time& Epoch\\
\hline
9821 & Pritzl    & ACS/WFC & F435W & 6{\x}11s                       & 2003--2004\\
     &           &         & F555W & 6{\x}7s                        &\\
     &           &         & F814W & 6{\x}3s                        &\\
\hline
9835 & Drukier   & ACS/HRC & F555W & 48{\x}155s                     &Oct 2003\\
     &           &         & F814W & 5{\x}25s, 2{\x}469s, 10{\x}505s&\\
\hline
10350& Cohn      & ACS/HRC & F330W & 2{\x}1266s, 4{\x}1314s         &Apr 2006\\
     &           &         & F555W & 3{\x}155s                      &\\
\hline
10474& Drukier   & ACS/HRC & F555W & 48{\x}155s                     &Apr 2006\\
     &           &         & F814W & 4{\x}25s, 8{\x}501s, 4{\x}508s &\\
\hline
10775&Sarajedini& ACS/WFC & F606W & 1{\x}40s, 5{\x}340s         & Apr 2006\\
     &          &         & F814W & 1{\x}40s, 5{\x}350s         &         \\
\hline
11739& Piotto   &WFC3/UVIS& F390W & 6{\x}880s                   &Jun--Jul 2010\\
\hline\hline
\end{tabular}
\end{center}}
\end{table*}

\begin{table*}[th!]
\scriptsize{
\begin{center}
\label{tab:A14}
\begin{tabular}{cccccc}
\multicolumn{6}{c}{\textsc{Table A13}}\\
\multicolumn{6}{c}{\textsc{List of observations of NGC~6397}}\\
\hline\hline
GO&PI&Intr./Cam.&Filter&$N\times$Exp. time& Epoch\\
\hline
10257& Anderson & ACS/WFC & F435W & 5{\x}13s, 5{\x}340s                & 2004--2005\\
     &          &         & F625W & 5{\x}10s, 5{\x}340s                & \\
     &          &         & F658N & 20{\x}390s, 20{\x}395s             & \\
\hline
10775&Sarajedini& ACS/WFC & F606W & 1{\x}1s, 4{\x}15s         & May 2006\\
     &          &         & F814W & 1{\x}1s, 4{\x}15s         &         \\
\hline
11633& Rich     &WFC3/UVIS& F336W & 6{\x}620s                 & Mar 2010\\
     &          &         & F606W & 6{\x}360s                 & \\
\hline\hline
\end{tabular}
\end{center}}
\end{table*}

\begin{table*}[th!]
\scriptsize{
\begin{center}
\label{tab:A15}
\begin{tabular}{cccccc}
\multicolumn{6}{c}{\textsc{Table A14}}\\
\multicolumn{6}{c}{\textsc{List of observations of NGC~6441}}\\
\hline\hline
GO&PI&Intr./Cam.&Filter&$N\times$Exp. time& Epoch\\
\hline
9835 & Drukier   & ACS/HRC & F555W & 36{\x}240s                     &Sep 2003\\
     &           &         & F814W & 5{\x}40s, 2{\x}413s, 10{\x}440s&\\
\hline
10775&Sarajedini& ACS/WFC & F606W & 1{\x}45s, 5{\x}340s         & May 2006\\
     &          &         & F814W & 1{\x}45s, 5{\x}350s         &         \\
\hline
11739& Piotto   &WFC3/UVIS& F390W & 2{\x}880s, 2{\x}884s, 8{\x}885s&2010--2011\\
\hline\hline
\end{tabular}
\end{center}}
\end{table*}

\begin{table*}[th!]
\scriptsize{
\begin{center}
\label{tab:A16}
\begin{tabular}{cccccc}
\multicolumn{6}{c}{\textsc{Table A15}}\\
\multicolumn{6}{c}{\textsc{List of observations of NGC~6535}}\\
\hline\hline
GO&PI&Intr./Cam.&Filter&$N\times$Exp. time& Epoch\\
\hline
10775&Sarajedini& ACS/WFC & F606W & 1{\x}12s, 4{\x}130s         & Mar 2006\\
     &          &         & F814W & 1{\x}12s, 4{\x}150s         &         \\
\hline
12008& Kong     & ACS/WFC & F625W & 1{\x}100s, 1{\x}148s        & Sep 2010\\
     &          &         & F658N & 1{\x}588s, 1{\x}600s        &\\
     &          &WFC3/UVIS& F336W & 1{\x}253s, 5{\x}400s        &\\
\hline\hline
\end{tabular}
\end{center}}
\end{table*}

\begin{table*}[th!]
\scriptsize{
\begin{center}
\label{tab:A17}
\begin{tabular}{cccccc}
\multicolumn{6}{c}{\textsc{Table A16}}\\
\multicolumn{6}{c}{\textsc{List of observations of NGC~6624}}\\
\hline\hline
GO&PI&Intr./Cam.&Filter&$N\times$Exp. time& Epoch\\
\hline
10401& Chandar & ACS/HRC & F435W &20{\x}200s                       &Feb 2005\\
\hline
10775&Sarajedini& ACS/WFC & F606W & 1{\x}15s, 5{\x}350s         & Apr 2006\\
     &          &         & F814W & 1{\x}15s, 5{\x}350s         &         \\
\hline
10573& Mateo & ACS/WFC & F435W & 1{\x}360s                       & Jun 2006\\
     &       &         & F555W & 1{\x}160s                       & \\
     &       &         & F814W & 1{\x}65s                        & \\
\hline\hline
\end{tabular}
\end{center}}
\end{table*}

\vskip -15mm

\begin{table*}[th!]
\scriptsize{
\begin{center}
\label{tab:A18}
\begin{tabular}{cccccc}
\multicolumn{6}{c}{\textsc{Table A17}}\\
\multicolumn{6}{c}{\textsc{List of observations of NGC~6656}}\\
\hline\hline
GO&PI&Intr./Cam.&Filter&$N\times$Exp. time& Epoch\\
\hline
10775&Sarajedini& ACS/WFC & F606W & 1{\x}3s, 4{\x}55s         & Apr 2006\\
     &          &         & F814W & 1{\x}3s, 4{\x}65s         &         \\
\hline
11558 & De Marco  & ACS/WFC & F502N & 2{\x}441s, 1{\x}2102x, 1{\x}2322s& Mar 2010\\
\hline
12311& Piotto &WFC3/UVIS& F814W & 4{\x}50s                    &2010--2011\\
\hline\hline
\end{tabular}
\end{center}}
\end{table*}

\vskip -15mm

\begin{table*}[th!]
\scriptsize{
\begin{center}
\label{tab:A19}
\begin{tabular}{cccccc}
\multicolumn{6}{c}{\textsc{Table A18}}\\
\multicolumn{6}{c}{\textsc{List of observations of NGC~6681}}\\
\hline\hline
GO&PI&Intr./Cam.&Filter&$N\times$Exp. time& Epoch\\
\hline
9019 & Bohlin & ACS/HRC & F330W & 4{\x}170s & Apr 2002\\
\hline 
9010 & Tran & ACS/HRC & F330W & 6{\x}70s & May--June 2002\\
\hline
9565 & De Marchi & ACS/HRC & F330W & 16{\x}70s &Jun-Sep 2002\\
\hline
9566 & De Marchi & ACS/HRC & F330W & 17{\x}70s & Feb 2003\\
\hline
9655 & Giavalisco & ACS/HRC & F330W & 16{\x}70s & Feb--Sep 2003\\
\hline
10047 &Giavalisco & ACS/HRC & F330W & 6{\x}70s & Mar--Sep 2004\\
\hline 
10401& Chandar & ACS/HRC & F435W &26{\x}125s                     & Feb 2005\\
\hline
10373 & Giavalisco & ACS/HRC & F330W & 4{\x}70s  & 2005--2006\\
\hline
10736&Maiz-Apellaniz& ACS/HRC & F330W & 8{\x}20s                & Mar 2006\\
     &              &         & F435W & 4{\x}2s                 &\\
     &              &         & F555W & 4{\x}2s                 &         \\
     &              &         & F625W & 4{\x}1s                 &         \\
     &              &         & F814W & 4{\x}1s                 &         \\
\hline
10775&Sarajedini& ACS/WFC & F606W & 1{\x}10s, 4{\x}140s         & May 2006\\
     &          &         & F814W & 1{\x}10s, 4{\x}150s         &         \\
\hline
12516& Ferraro  &WFC3/UVIS& F390W & 12{\x}348s                  & Nov 2011\\
     &          &         & F555W & 2{\x}127s, 8{\x}150s        &         \\
     &          &         & F814W & 13{\x}348s                  &         \\
\hline\hline
\end{tabular}
\end{center}}
\end{table*}
\vskip -15mm

\begin{table*}[th!]
\scriptsize{
\begin{center}
\label{tab:A20}
\begin{tabular}{cccccc}
\multicolumn{6}{c}{\textsc{Table A19}}\\
\multicolumn{6}{c}{\textsc{List of observations of NGC~6752}}\\
\hline\hline
GO&PI&Intr./Cam.&Filter&$N\times$Exp. time& Epoch\\
\hline
9453 & Brown    & ACS/WFC & F606W & 1{\x}4s, 1{\x}40s           & Sep 2002\\
     &          &         & F814W & 1{\x}4s, 1{\x}45s           &\\
\hline
9899 & Piotto   & ACS/WFC & F475W & 6{\x}340s                   & Jul 2004\\
\hline
10121& Bailyn   & ACS/WFC & F555W & 12{\x}80s, 11{\x}435s       & Sep 2004\\
     &          &         & F814W & 12{\x}40s                   &\\
\hline
10335& Ford     & ACS/HRC & F435W & 24{\x}35s                   &2004--2006\\
     &          &         & F555W & 13{\x}10s                   &\\
\hline
10458& Biretta  & ACS/HRC & F555W & 12{\x}10s, 4{\x}100s, 2{\x}500s&Aug 2005\\
     &          &         & F606W & 2{\x}357s                   &\\
\hline
10459& Biretta  & ACS/WFC & F606W & 8{\x}450&                 Oct 2005\\
\hline
10335& Ford     & ACS/HRC & F435W & 24{\x}35s                    &Jun 2004\\
     &          &         & F555W & 13{\x}10s                    &\\
\hline
10775&Sarajedini& ACS/WFC & F606W & 1{\x}2s, 4{\x}35s         & May 2006\\
     &          &         & F814W & 1{\x}2s, 4{\x}40s         &         \\
\hline
11801& Ford     &WFC3/UVIS& F438W & 4{\x}5s, 18{\x}45s        &Nov 2009\\
\hline
11664& Brown    &WFC3/UVIS& F390W & 2{\x}348s, 2{\x}880s      & May 2010\\
     &          &         & F555W & 1{\x}30s, 2{\x}665s       &\\
     &          &         & F814W & 1{\x}30s, 2{\x}495s       &\\
\hline
11904& Kalirai  &WFC3/UVIS& F555W & 15{\x}550s                &Jul--Aug 2010\\
     &          &         & F814W & 15{\x}550s                &\\
\hline
12254& Cool     & ACS/WFC & F435W & 6{\x}10s, 12{\x}380s      &May--Nov 2011\\
     &          &         & F625W & 18{\x}10s, 12{\x}360s      &\\
     &          &         & F658N & 12{\x}724s, 12{\x}820s     &\\
\hline
12311& Piotto   &WFC3/UVIS& F814W & 2{\x}50s                     &Mar--Apr 2011\\
\hline\hline
\end{tabular}
\end{center}}
\end{table*}
\vskip -15mm

\begin{table*}[th!]
\scriptsize{
\begin{center}
\label{tab:A21}
\begin{tabular}{cccccc}
\multicolumn{6}{c}{\textsc{Table A20}}\\
\multicolumn{6}{c}{\textsc{List of observations of NGC~6715}}\\
\hline\hline
GO&PI&Intr./Cam.&Filter&$N\times$Exp. time& Epoch\\
\hline
10775&Sarajedini& ACS/WFC & F606W & 2{\x}30s, 10{\x}340s         & May 2006\\
     &          &         & F814W & 2{\x}30s, 10{\x}350s         &         \\
\hline
12274& van der Marel&WFC3/UVIS& F438W & 10{\x}30s, 5{\x}234s, 5{\x}256s&Sep 2011\\
\hline\hline
\end{tabular}
\end{center}}
\end{table*}
\vskip -15mm

\begin{table*}[th!]
\scriptsize{
\begin{center}
\label{tab:A22}
\begin{tabular}{cccccc}
\multicolumn{6}{c}{\textsc{Table A21}}\\
\multicolumn{6}{c}{\textsc{List of observations of NGC~7078}}\\
\hline\hline
GO&PI&Intr./Cam.&Filter&$N\times$Exp. time& Epoch\\
\hline
10401& Chandar   & ACS/HRC & F435W & 13{\x}125s               &Dec 2004\\
\hline
10775&Sarajedini & ACS/WFC & F606W & 1{\x}15s, 4{\x}130s         & May 2006\\
     &           &         & F814W & 1{\x}15s, 4{\x}150s         &         \\
\hline
11233& Piotto    &WFC3/UVIS& F390W & 6{\x}827s                   & May 2010\\
\hline
12605& Piotto    &WFC3/UVIS& F336W & 6{\x}350s                   & Oct 2011\\
     &           &         & F438W & 6{\x}65s                    &\\
\hline\hline
\end{tabular}
\end{center}}
\end{table*}
\vskip -15mm

\begin{table*}[th!]
\scriptsize{
\begin{center}
\label{tab:A23}
\begin{tabular}{cccccc}
\multicolumn{6}{c}{\textsc{Table A22}}\\
\multicolumn{6}{c}{\textsc{List of observations of NGC~7099}}\\
\hline\hline
GO&PI&Intr./Cam.&Filter&$N\times$Exp. time& Epoch\\
\hline
10401& Chandar     & ACS/HRC & F435W & 13{\x}125s                & Dec 2004\\
\hline
10775&Sarajedini & ACS/WFC & F606W & 1{\x}7s, 4{\x}140s         & May 2006\\
     &           &         & F814W & 1{\x}7s, 4{\x}140s         &         \\
\hline\hline
\end{tabular}
\end{center}}
\end{table*}
\vskip -15mm

\newpage
\clearpage

\section{Appendix B. Proper-Motion catalog of NGC~7078}
\label{a2}

Our procedures generate a large number of parameters for each star,
but most users will need only the high-level data. The PM catalog of
NGC~7078 contains 91 lines of header information, followed by one line
for each star with a PM measurement, for a total of 77$\,$837
stars. Stars in the catalog are sorted according to their distance
from the cluster center, as given in Table~1.

The header starts with some general information about the cluster,
such as the reference time of the master frame and the adopted cluster
center position, in both equatorial and master-frame units. Then
follows a column-by-column description of the catalog.  The columns
contain:\ the reference-frame positions and distance from the cluster
center, calibrated and differential-reddening-corrected F606W and
F814W magnitudes with errors and some photometric-quality information,
PMs with errors derived using both the expected errors as a weight and
the actual residuals around the PM least-squares fits (see
Section~\ref{ss:MC}, some additional astrometric-quality information,
and finally the differences between local-corrected and
amplifier-based PMs (see Section~\ref{ss:loc_cor}). A description of
each column of the catalog is given in Table~B1, while the first 10
lines of the NGC~7078 PM catalog are shown in Table~B2.

\begin{table}[!th]
\begin{center}
\scriptsize{
\begin{tabular}{ccl}
  \multicolumn{3}{c}{\textsc{Table B1}}\\
  \multicolumn{3}{c}{\textsc{Column-by-column information contained in the catalog}}\\
  \hline\hline
  Col&Name (unit)&Explanation\\
  \hline
  \multicolumn{3}{c}{Astrometric information}\\
\hline
  1&$r$ $(^{\prime\prime})$& Distance from the cluster center\\
  2&$\Delta x_0$ $(^{\prime\prime})$& GO-10775 x-position in the rectified Cartesian system with respect to the adopted center\\
  3&$\Delta y_0$ $(^{\prime\prime})$& GO-10775 y-position in the rectified Cartesian system with respect to the adopted center\\
  4&$\mu_{\alpha}\cos \delta$ (\masyr)& PM along the x axis (parallel to and increasing as R.A.)\\
  5&$\mu_{\delta}$ (\masyr)& PM along the y axis (parallel to and increasing as Dec.)\\
  6&$\sigma_{\mu_{\alpha}\cos \delta}$ (\masyr)& 1-$\sigma$ uncertainty in $\mu_{\alpha} \cos \delta$ computed using actual residuals\\
  7&$\sigma_{\mu_{\delta}}$ (\masyr)& 1-$\sigma$ uncertainty in $\mu_{\delta}$ computed using actual residuals\\
  8&$x_{\rm M}$ (pixel)& x-position on the master frame\\
  9&$y_{\rm M}$ (pixel)& y-position on the master frame\\
  10&$\Delta x$ (pixel)& difference between $x_{\rm M}$ and the PM-predicted position at the reference time ($\overline{x}$)\\
  11&$\Delta y$ (pixel)& difference between $y_{\rm M}$ and the PM-predicted position at the reference time ($\overline{y}$)\\
  12&err$_{\mu_{\alpha} \cos \delta}$ (\masyr)& 1-$\sigma$ uncertainty in $\mu_{\alpha} \cos \delta$ computed using expected errors\\
  13&err$_{\mu_{\delta}}$ (\masyr)& 1-$\sigma$ uncertainty in $\mu_{\delta}$ computed using expected errors\\
\hline 
 \multicolumn{3}{c}{Photometric information}\\
\hline 
 14&$m_{\rm F606W}$ (mag)&Differential-reddening-corrected GO-10775 F606W Vega-mag photometry\\
  15&$m_{\rm F814W}$ (mag)&Differential-reddening-corrected GO-10775 F814W Vega-mag photometry\\
16&$\sigma_{m_{\rm F606W}}$ (mag)& Photometric error in F606W (from GO-10775)\\
17&$\sigma_{m_{\rm F814W}}$ (mag)& Photometric error in F814W (from GO-10775)\\
18&\texttt{QFIT}$_{\rm F606W}$& Quality of F606W PSF-fit (from GO-10775)\\
19&\texttt{QFIT}$_{\rm F814W}$& Quality of F814W PSF-fit (from GO-10775)\\
\hline
\multicolumn{3}{c}{Proper-motion quality information}\\
\hline
20&$\chi^2_{\mu_{\alpha}\cos \delta}$& Reduced $\chi^2$ of the fit of the x-component of the motion\\
21&$\chi^2_{\mu_\delta}$& Reduced $\chi^2$ of the fit of the y-component of the  motion\\
22&$\sigma_{\overline{x}}$ (pix)&1-$\sigma$ uncertainty in the intercept of the PM fit for the x-component using actual residuals\\
23&$\sigma_{\overline{y}}$ (pix)&1-$\sigma$ uncertainty in the intercept of the PM fit for the y-component using actual residuals\\
24& time (yr)& Time baseline, in Julian years\\ 
25&err$_{\overline{x}}$ (pix)&1-$\sigma$ uncertainty in the intercept of the PM fit for the x-component  using expected errors\\
26&err$_{\overline{y}}$ (pix)&1-$\sigma$ uncertainty in the intercept of the PM fit for the y-component  using expected errors\\
27&U$_{\rm ref}$& Flag: 1 if used as reference bona-fide cluster star for the linear transformations, 0 otherwise\\
28&N$_{\rm found}$&Initial number of data points for the PM fits\\
29&N$_{\rm used}$&Final number of data points used for the PM fits\\
30&ID&ID number for each star (not the GO-10775 ID)\\
\hline 
 \multicolumn{3}{c}{Local PM corrections}\\
\hline 
31&$\Delta \mu_{\alpha}\cos \delta$ (\masyr)& Difference in $\mu_{\alpha}\cos \delta$ between locally-corrected and amplifier-based PMs. Add to column 4\\
& &to obtain locally-corrected PMs.\\
32&$\Delta \mu_{\delta}$ (\masyr) &  Difference in $\mu_{\delta}$ between locally-corrected and amplifier-based PMs. Add to column 5 to \\
&&obtain locally-corrected PMs.\\
\hline\hline
\end{tabular}}
\end{center}
\end{table}

\begin{sidewaystable}
\centering
\scriptsize{
\begin{tabular}{cccccccccccccccccccccccccccccccc}
\multicolumn{17}{c}{\textsc{Table B2}}\\
\multicolumn{17}{c}{\textsc{First ten lines of the NGC~7078 PM catalog}}\\
\hline\hline
$r$ $(^{\prime\prime})$&$\Delta x_0$ $(^{\prime\prime})$&$\Delta y_0$ $(^{\prime\prime})$&$\mu_{\alpha}\cos \delta$&$\mu_{\delta}$&
$\sigma_{\mu_{\alpha}\cos \delta}$&$\sigma_{\mu_{\delta}}$&$x_{\rm M}$&$y_{\rm M}$&$\Delta x$&$\Delta y$&err$_{\mu_{\alpha}\cos \delta}$&err$_{\mu_{\delta}}$&
$m_{\rm F606W}$&$m_{\rm F814W}$&$\sigma_{m_{\rm F606W}}$&$\rightarrow$\\
(1)&(2)&(3)&(4)&(5)&(6)&(7)&(8)&(9)&(10)&(11)&(12)&(13)&(14)&(15)&(16)&\\
\hline
0.22148 &   0.19883 &   0.09756 &$-$0.203 &   0.249 & 0.039 & 0.030 & 4984.312 & 5019.940 &   0.024 &   0.014 & 0.032 & 0.030 & 17.015 & 16.276 & 9.900 &\dots\\
0.50339 &    0.24141 &   0.44172 &$-$3.057 &   9.266 & 0.367 & 2.021 & 4983.246 & 5028.540 &$-$0.078 &$-$0.241 & 0.418 & 0.957 & 18.253 & 17.774 & 9.900 &\dots\\
1.13357 &    0.84530 &   0.75528 &   0.201 &   0.245 & 0.045 & 0.054 & 4968.149 & 5036.379 &   0.030 &   0.003 & 0.042 & 0.037 & 15.508 & 15.113 & 9.900 &\dots\\
1.24526 &    1.18454 &   0.38412 &$-$0.283 &   0.055 & 0.020 & 0.038 & 4959.674 & 5027.097 &$-$0.014 &   0.052 & 0.021 & 0.034 & 15.985 & 15.801 & 9.900 &\dots\\
1.32849 &    0.86993 &   1.00404 &   0.001 &$-$0.192 & 0.023 & 0.021 & 4967.535 & 5042.601 &   0.012 &$-$0.008 & 0.024 & 0.018 & 16.974 & 16.193 & 9.900 &\dots\\
1.33293 &    0.59227 &   1.19412 &   0.321 &$-$0.101 & 0.027 & 0.031 & 4974.479 & 5047.344 &   0.004 &$-$0.013 & 0.027 & 0.035 & 17.419 & 16.724 & 9.900 &\dots\\
1.46104 & $-$1.44918 &   0.18576 &   0.176 &$-$0.084 & 0.022 & 0.023 & 5025.506 & 5022.140 &   0.016 &   0.012 & 0.018 & 0.028 & 16.686 & 15.977 & 9.900 &\dots\\
1.62112 & $-$0.24352 &   1.60272 &   0.054 &$-$0.045 & 0.034 & 0.019 & 4995.371 & 5057.557 &$-$0.030 &   0.022 & 0.029 & 0.023 & 15.478 & 15.406 & 9.900 &\dots\\
1.77721 & $-$1.39604 &$-$1.09980 &$-$0.403 &   0.109 & 0.025 & 0.046 & 5024.188 & 4990.005 &   0.022 &   0.005 & 0.024 & 0.036 & 17.375 & 16.719 & 9.900 &\dots\\
1.90239 & $-$1.31299 &   1.37664 &   0.387 &$-$0.474 & 0.015 & 0.031 & 5022.109 & 5051.913 &   0.021 &$-$0.021 & 0.019 & 0.029 & 17.443 & 16.765 & 9.900 &\dots\\
\dots&\dots&\dots&\dots&\dots&\dots&\dots&\dots&\dots&\dots&\dots&\dots&\dots&\dots&\dots&\dots&\dots\\
\hline
$\rightarrow$&$\sigma_{m_{\rm F814W}}$&\texttt{QFIT}$_{\rm F606W}$&\texttt{QFIT}$_{\rm F814W}$&$\chi^2_{\mu_{\alpha}\cos\delta}$&$\chi^2_{\mu_\delta}$&$\sigma_{\overline{x}}$&$\sigma_{\overline{y}}$&time&err$_{\overline{x}}$&err$_{\overline{y}}$&
U$_{\rm ref}$&N$_{\rm found}$&N$_{\rm used}$&ID&$\Delta \mu_{\alpha}\cos \delta$&$\Delta \mu_{\delta}$\\
&(17)&(18)&(19)&(20)&(21)&(22)&(23)&(24)&(25)&(26)&(27)&(28)&(29)&(30)&(31)&(32)\\
\hline
\dots& 9.900 & 0.080 & 0.056 & 2.412 &  2.116 & 0.0018 &  0.0017 & 6.96206 & 0.0018 & 0.0016 & 1 & 30 & 24 & 86023 &   0.004&   0.005\\
\dots& 9.900 & 0.331 & 0.347 & 2.328 & 11.882 & 0.0134 &  0.0749 & 1.48741 & 0.0153 & 0.0350 & 0 & 26 & 15 & 86021 &   0.023&$-$0.079\\
\dots& 9.900 & 0.084 & 0.049 & 3.502 &  2.750 & 0.0020 &  0.0022 & 6.96206 & 0.0019 & 0.0017 & 0 & 20 & 18 & 86020 &   0.047&   0.013\\
\dots& 9.900 & 0.062 & 0.043 & 1.706 &  4.652 & 0.0011 &  0.0019 & 6.96206 & 0.0011 & 0.0018 & 0 & 25 & 23 & 86022 &   0.049&   0.014\\
\dots& 9.900 & 0.118 & 0.063 & 1.279 &  0.796 & 0.0014 &  0.0011 & 6.96206 & 0.0013 & 0.0010 & 1 & 24 & 21 & 86019 &   0.017&$-$0.002\\
\dots& 9.900 & 0.115 & 0.117 & 1.475 &  2.553 & 0.0016 &  0.0021 & 6.96206 & 0.0015 & 0.0020 & 1 & 25 & 23 & 86018 &$-$0.006&   0.017\\
\dots& 9.900 & 0.080 & 0.084 & 1.045 &  2.306 & 0.0010 &  0.0017 & 6.96206 & 0.0010 & 0.0016 & 1 & 27 & 26 & 86483 &   0.022&$-$0.028\\
\dots& 9.900 & 0.046 & 0.042 & 1.746 &  1.067 & 0.0015 &  0.0011 & 6.96195 & 0.0014 & 0.0011 & 0 & 16 & 14 & 86228 &   0.033&   0.001\\
\dots& 9.900 & 0.098 & 0.068 & 1.140 &  2.624 & 0.0013 &  0.0020 & 6.96206 & 0.0013 & 0.0021 & 1 & 25 & 25 & 86481 &$-$0.010&   0.011\\
\dots& 9.900 & 0.146 & 0.096 & 0.895 &  2.024 & 0.0013 &  0.0016 & 6.96206 & 0.0012 & 0.0018 & 1 & 27 & 26 & 86485 &$-$0.010&   0.005\\
\dots&\dots&\dots&\dots&\dots&\dots&\dots&\dots&\dots&\dots&\dots&\dots&\dots&\dots&\dots&\dots&\dots\\
\hline\hline
\end{tabular}}
\end{sidewaystable}

\end{document}